\documentclass[aps,prb,twocolumn,groupedaddress,showpacs,superscriptaddress,amssymb,amsmath,floatfix,longbibliography]{revtex4-1}
\usepackage{physics}
\usepackage[T1]{fontenc}
\usepackage[latin9]{inputenc}
\usepackage{float}
\usepackage{amsmath}
\usepackage{amssymb}
\usepackage{comment}
\usepackage{graphicx}
\usepackage{wasysym}
\usepackage{color}
\usepackage{physics}
\usepackage{color}
\usepackage{amsmath}
\usepackage{comment}
\usepackage{dcolumn}
\usepackage{hyperref}
\usepackage{bm}
\usepackage{epsf}
\newcommand{\bea}{\begin{eqnarray}}
\newcommand{\eea}{\end{eqnarray}}

\hypersetup{
  colorlinks=true,
  citecolor=blue,
  linkcolor=blue,
  urlcolor=blue}
  
\makeatletter
\usepackage{babel}

\makeatother
\begin{document}
\title{Open quantum systems with noncommuting coupling operators: An analytic approach}


\author{Jakub Garwo\l a}
\affiliation{Department of Physics, 60 Saint George St., University of Toronto, Toronto, Ontario, M5S 1A7, Canada}

\author{Dvira Segal}
\affiliation{Department of Chemistry and Centre for Quantum Information and Quantum Control,
University of Toronto, 80 Saint George St., Toronto, Ontario, M5S 3H6, Canada}
\affiliation{Department of Physics, 60 Saint George St., University of Toronto, Toronto, Ontario, M5S 1A7, Canada}
\email{dvira.segal@utoronto.ca}

\begin{abstract}
We present an analytic approach to treat open quantum systems strongly coupled to multiple environments via noncommuting system operators, a prime example is a qubit concurrently coupled to both decoherring and dissipative baths. Our approach, which accommodates strong system-bath couplings, generalizes the recently developed reaction-coordinate polaron transform method [PRX Quantum {\bf 4}, 020307 (2023)] to handle couplings to baths via noncommuting system operators. Our approach creates an effective Hamiltonian that reveals the cooperative effect of the baths on the system. For a spin impurity coupled to both dissipative and decoherring environments, the effective Hamiltonian predicts the suppression of relaxation by decoherence---a phenomenon previously observed in simulations but lacking so far a theoretical foundation. We also apply the method to an ensemble of spins coupled to local baths through noncommuting operators, demonstrating the engineering of the Kitaev XY spin chain interaction. 
Noncommutativity is a feature of quantum systems; future prospects of our approach include the study of thermal machines that leverage such genuine quantum effects.
\end{abstract}

\maketitle

\date{\today}


\section{Introduction}


Open quantum systems (OQS) often couple to multiple independent environments. For example, electron tunneling in quantum dots may be coupled to substrate lattice phonons, 
to radiation or cavity fields, 
and to the electron sea in metal electrodes \cite{witczak_flexible_2023,Petta17}. 
Qubits in quantum processors are subject to both decoherence and dissipation, often assumed to be uncorrelated.
In qubits designed based on neutral atom arrays, which are trapped by optical tweezers, errors emerge due to excited state decay, atoms scatterings and losses, hyperfine transitions, power fluctuations, and other effects, resulting in both decay and depolarizing effects of the state \cite{Lukin22}. In the context of quantum thermal transport, superconducting qubits have been designed to couple to two independent ohmic resistors, representing separate bosonic thermal baths \cite{PekolaR}.

From the simulation side, different types of noise processes are often modelled phenomenologically, by using independent jump operators in a Lindblad equation formalism \cite{Nielsen2010,noise21,noise03,Francesco24}. 
However, the Lindblad equation, and similarly, the Redfield quantum master equation \cite{Nitzan} are rooted in the weak system-bath coupling approximation and as such they miss cooperative effects between different baths.
Numerous tools have been developed to treat OQS that are {\it strongly and simultaneously} coupled to multiple independent baths. A very partial list includes stochastic and deterministic path integral approaches \cite{MCGuy13,SegalSimine13,makri2024quantumdynamics,PRXQuantum.3.010321,Keeling24}, polaronic quantum master equations \cite{SegalSB1,SegalSB2,SegalSB3}, 
Hierarchical Equations of Motion \cite{HEOMThoss16,Tanimura20}, wavefunction methods \cite{MCTDH08,MCTDH09},
and Markovian embedding techniques \cite{MarlonTN,GooldME,RC_termodynamics}.

Treating OQS that are {\it strongly} coupled to multiple baths through {\it noncommuting system operators} is particularly complex. 
Studies of such systems based on analytical and numerical renormalization group calculations \cite{Afflek03,Zarand05,Vojta12,flow13},
path integral
\cite{non-commuting_PI,Nalbach19,PRXQuantum.3.010321,Richter_2022,Keeling24},
polaron transform \cite{Cao_unusual_transport,AhsanNC19,Dahai24},
and reaction coordinate \cite{Schaller16,Janet3B} equations of motion 
demonstrated a plethora of unique phenomena, e.g., effects that can be referred to as ``quantum  frustration''. 
In the path integral formalism, handling noncommuting interaction terms required generalizing the standard quasi-adiabatic path integral algorithm \cite{non-commuting_PI,Nalbach19,PRXQuantum.3.010321,Richter_2022,Keeling24}.
Interestingly, simulations reported in Ref. \cite{Suppressing_relaxation_through_dephasing} revealed an intriguing phenomenon: the suppression of spin relaxation with increased decoherence, which takes place due to the action of an independent heat bath. 
Although this observation has been demonstrated in simulations, it lacks a theoretical explanation. 
A related observation, the enhancement of entanglement between a spin and its environments---when coupled isotropically to three independent baths---was recently discussed in Ref. \citenum{Janet3B}. 

Providing an analytical treatment, thus fundamental insights for an OQS that couples to multiple baths via noncommuting system operators, especially for arbitrary coupling strengths, remains a challenge to-date. In this study, we address this theoretical challenge and develop such a treatment. Using our approach, we (i) explain the nontrivial effect of relaxation suppression by decoherence, and (ii) demonstrate bath engineering of desired interactions in spin chains. 
More broadly, our analytical approach can tackle equilibrium and nonequilibrium dynamics of quantum frustration as arising from coupling a system through noncommuting components to independent baths \cite{Afflek03,flow13}. 

The recently developed Effective Hamiltonian (EFFH) method, building on the reaction coordinate and the polaron transformation \cite{Nick_PRX}, provides a semi-analytic approach towards studying the dynamics \cite{brenes2024bathinduced}, equilibrium state \cite{Anto_Sztrikacs_2023}, and the nonequilibrium steady state \cite{Anto-Sztrikacs_2021, non-equalibrium_strong_coupling} of OQS that are coupled to multiple harmonic environments at arbitrary coupling strength.
The method results in building an Effective Hamiltonian, a powerful starting point for analysis and simulations. The EFFH method consists of the extraction of a collective ``reaction coordinate'' mode from each harmonic bath, followed by a polaron transform, and then truncation of the shifted reaction coordinate modes. This approach has proven itself effective for understanding the characteristics of quantum thermal machines based on qubits, qtrits, and spin chains \cite{Nick_PRX,Anto-Sztrikacs_2021,min2024bathengineering}. This approximate analytical tool complements numerically-exact methods such as the Time Evolving Density matrices method that uses Orthogonal Polynomials (TEDOPA) \cite{TEDOPA}, or influence functional path integral-based methods \cite{Makri_PI,non-commuting_PI,10.1063/5.0066891,PATHSUM,Keeling24}.

The main advantage of the EFFH approach is that it produces a closed-form expression for the effective system Hamiltonian. This so-called Hamiltonian of mean force is conjectured to construct the equilibrium state and the nonequilibrium steady state of the system, handling system-bath coupling strengths from ultraweak to ultrastrong \cite{Nick_PRX}.
However, as we illustrate in Sec. \ref{sec:polaron1}, the EFFH method as formulated in Ref. \citenum{Nick_PRX}, cannot handle OQS that are coupled to multiple baths through noncommuting system operators. 

In this work, we address this deficiency by developing a theory that allows for the study of the equilibrium state and dynamics of spin impurities and lattices coupled to multiple bosonic baths, even when those couplings take place through {\it noncommuting operators of the system}.
Our method extends the EFFH technique \cite{Nick_PRX} by generalizing the polaron transform step within that procedure. The outcome is an effective Hamiltonian that captures the intertwined impact of the different environments on the OQS, revealing cooperative effects between them. 

\begin{figure}[htbp]
\centering{}
\includegraphics[width=0.4\textwidth]{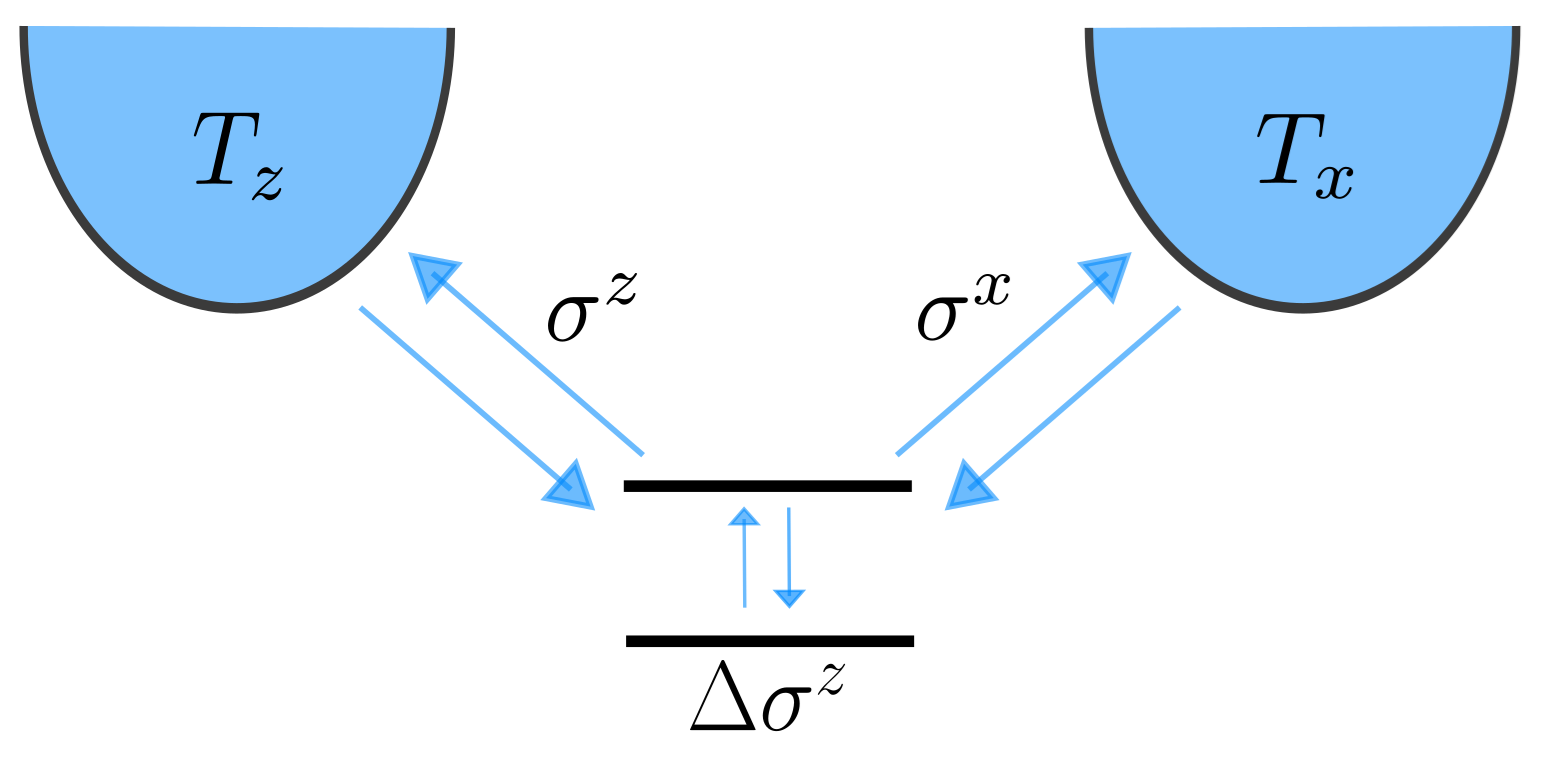}
\caption{Spin impurity coupled simultaneously to independent baths through noncommuting system operators. 
$\Delta$ is the spin splitting, $T_{z}$ and $T_{x}$ are the temperatures of the two heat baths coupled to the system via two Pauli matrices $\sigma^{z}$ and $\sigma^{x}$, respectively. In simulations we assume that the baths have the same temperatures.}
\label{fig:impurity_sys}
\end{figure}

\begin{figure}
\begin{centering}
\includegraphics[width=0.45\textwidth]{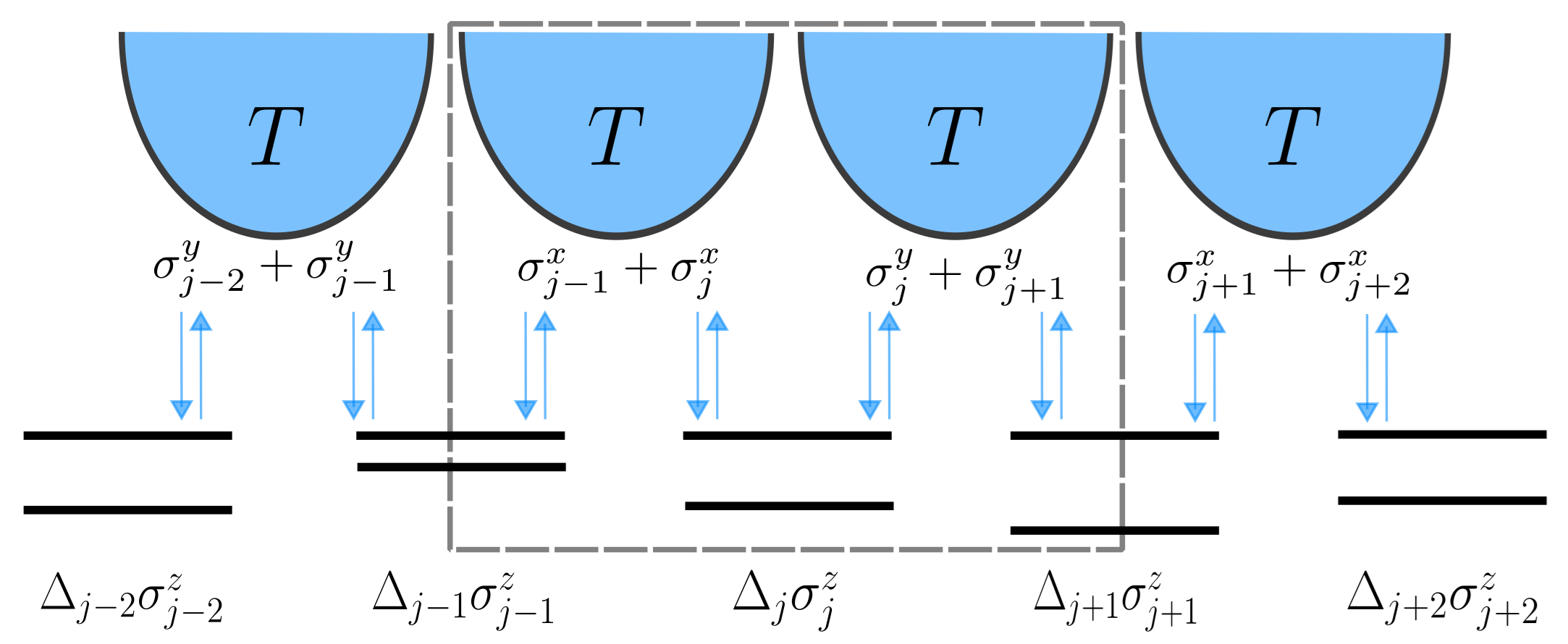} 
\par\end{centering}
\caption{Creating the Kitaev XY spin chain model through bath engineering, with each bath coupled to neighboring spins and alternating between $\sigma^{x}$ and $\sigma^{y}$ coupling operators.}
\label{fig:spin_chain}
\end{figure}

This theoretical advance allows us to examine two nontrivial problems and discover the following:
(i) Suppression of relaxation by decoherence:
We study the relaxation dynamics and equilibrium state of a qubit coupled simultaneously to both dissipative and decoherring baths, as illustrated in Fig. \ref{fig:impurity_sys}. The effective Hamiltonian emerging from the procedure provides a clear explanation for the intriguing phenomenon where relaxation is suppressed by decoherence. Our result, a relaxation rate dressed by both baths' couplings, constitutes the main result of this paper and it is given by Eq. (\ref{eq:GammaXM}).
(ii) Bath Engineering spin chain interactions: We construct the Kitaev XY spin chain through bath engineering by designing a coupling pattern where independent baths interact with the spins in the model. This design is depicted in Fig. \ref{fig:spin_chain}  with the main result reported in Eq. (\ref{eq:chain_Hseff}).

The paper is organized as follows. In Section \ref{sec:ModelMethod}, we present our model and summarize the EFFH procedure. In Section \ref{sec:polaron1}, we illustrate the nonuniqueness problem of the standard EFFH method when applied to noncommuting operators. We explain in Sec. \ref{sec:polaron2} how to resolve this ambiguity using a more general, non factorized polaron transform. 
This generalization leads to a closed-form formula for the effective system Hamiltonian when considering two noncommuting coupling operators.
In Section \ref{sec:TLS}, we apply our method to investigate the suppression of relaxation by decoherence in a spin qubit. 
We discuss the realization of the Kitaev XY spin chain through bath engineering 
in Section \ref{sec:lattice}, and
we conclude in Section \ref{sec:summary}. Details of derivations are delegated to appendices \ref{app:1}, \ref{app:2}, \ref{app:3}, and \ref{app:4}.


\section{Model and the basic EFFH method}
\label{sec:ModelMethod}

\subsection{Model: OQS with noncommuting couplings}

We begin by presenting a generic OQS of interest.
We consider a quantum system coupled to multiple (at least two) bosonic baths through different, possibly noncommuting system operators. The model is described by the following Hamiltonian,
\begin{align}
\hat{H} & =\hat{H}_{S}+\sum_{n}\sum_{k}\Big(t_{n,k}\hat{S}_{n}(\hat{c}_{n,k}^{\dagger}+\hat{c}_{n,k})+\nu_{n,k}\hat{c}_{n,k}^{\dagger}\hat{c}_{n,k}\Big).
\label{eq:H_original}
\end{align}
Here, $\hat{H}_{S}$ is the Hamiltonian of the system and $\hat{S}_{n}$ is an operator of the system, which is coupled to the $n$th bath. In this work, we allow that $[\hat{S}_{n},\hat{S}_{n'}]\neq0$ for all $n \neq n'$. 
For example, in Fig. \ref{fig:impurity_sys} a spin impurity in a magnetic field couples to two independent environments, dissipative and decoherring, with $\hat{H}_{S} = \Delta \sigma^z$, $\hat S_z=\sigma^z$ and $\hat S_x=\sigma^x$.
As for the environments, $\hat{c}^{\dagger}_{n,k}$ 
($\hat{c}_{n,k}$) are bosonic creation (annihilation) operators of modes with momentum $k$ in the $n$-th bath, where $\nu_{n,k}$ are the  frequencies of the modes.  We denote by $t_{n,k}$  the system-bath coupling energies. For simplicity, throughout this paper we assume that these coupling coefficients are real-valued. The spectral density function for each bath is given by 
$J_{n}(\omega)=\sum_{k}t_{n,k}^{2}\delta(\nu_{n,k}-\omega)$ .

As for the initial state of the baths, for simplicity, we assume that the different baths are maintained at thermal equilibrium characterized by the same temperature, $T$. As such, the system eventually relaxes into an equilibrium state that depends on $T$. However, because the system couples to its environments at arbitrary coupling strength, the equilibrium state of the system may deviate from the canonical Gibbs state as shown in multiple studies, e.g., Refs. \citenum{Thoss09, CaoFP,Miller2018, Nick_PRX,
Anto_Sztrikacs_2023,Anton22, strong_limit_MFGS, AndersAVS}.
More generally, we are interested in the state of the system as it evolves in time from a certain initial condition to thermal equilibrium. For a spin system, observables of interest include the spin polarization, $\langle \sigma^z\rangle(t) = {\rm Tr}_B [\rho_S(t) \sigma^z]$ and its
coherences, $\langle \sigma^x\rangle(t) = {\rm Tr}_B [\rho_S(t) \sigma^x]$,  with $\rho_S(t)$ the state of the system at time $t$. 
 We work in units in which the Boltzmann constant is set as $k_B=1$ and $\hbar =1$.
 
\subsection{Principles of the EFFH mapping}

We focus here on systems that are made of a single spin (impurity model) or a collection of spins (lattice models), with examples presented in Figs. \ref{fig:impurity_sys} and \ref{fig:spin_chain}. The two nontrivial elements of our study are the coupling of the system to the baths {\it beyond the weak coupling limit} and through {\it noncommuting} system operators. 
The first challenge in solving the dynamics of the system described by Eq.~(\ref{eq:H_original}) arises from its potentially strong coupling to the environments, which precludes a low-order perturbative treatment. We address this problem by employing the EFFH method \cite{Nick_PRX}, a Markovian embedding technique. The second challenge is the focus of our work, and it will be resolved in Sec. \ref{sec:polaron2}.

The objective of the EFFH mapping is to derive an approximate, effective system Hamiltonian that is weakly coupled to its environments, in contrast to the original system, which is coupled to the baths at arbitrary strength. 
This is accomplished through three steps ({\bf A}, {\bf B}, {\bf C}), which we now summarize. 
For simplicity, we focus on a system coupled to two baths only.

In the present discussion, we describe the EFFH approach in general terms. In Sec. \ref{sec:polaron1}, we show that the standard factorized polaron used in Ref. \cite{Nick_PRX} cannot handle noncommuting coupling operators. In Sec. \ref{sec:polaron2}, we resolve this issue by generalizing step {\bf B}.

{\bf A.} The reaction coordinate (RC) transformation \cite{RC_termodynamics} is applied to Eq. (\ref{eq:H_original}) to extract a single collective mode (the `reaction coordinate') from each bath $n$, with a corresponding creation (annihilation) operator $\hat{a}_{n}^{\dagger}$ ($\hat{a}_{n}$) and frequency $\Omega_{n}$.
Reaction coordinates may strongly couple to the system, depending on the original spectral functions. However, the transformation is defined such that the extracted reaction coordinates only weakly couple to the residual baths.  
From Eq. (\ref{eq:H_original}), the resulting RC Hamiltonian is (for the case of two baths),
\begin{equation}
\begin{aligned} & \hat{H}_{\rm RC}  =\hat{H}_{S}+\sum_{n=1,2}\Big(\lambda_{n}\hat{S}_{n}(\hat{a}_{n}^{\dagger}+\hat{a}_{n})+\Omega_{n}\hat{a}_{n}^{\dagger}\hat{a}_{n}\Big)\\
 & +\sum_{n=1,2}\tilde{\sum_{k}}\Big(f_{n,k}(\hat{a}_{n}^{\dagger}+\hat{a}_{n})(\hat{b}_{n,k}^{\dagger}+\hat{b}_{n,k})+\omega_{n,k}\hat{b}_{n,k}^{\dagger}\hat{b}_{n,k}\Big).
\end{aligned}
\label{eq:HRC}
\end{equation}
The RC transformation defines how to build the creation and annihilation bosonic operators 
$\hat b_{n,k}^{\dagger}$, $\hat b_{n,k}$ of the residual baths
from the original set of operators $\hat c^{\dagger}_{n,k}$, $\hat c_{n,k}$,
as well as how to construct
the spectral density functions, $J_n^{{\text RC}}(\omega)= \sum_k f_{n,k}^2 \delta(\omega-\omega_{n,k})$, from the original function, $J_n(\omega)$ \cite{Irene1,Irene2,RC_termodynamics}.
In Eq. (\ref{eq:HRC}), the sum with a tilde covers an infinite set of modes, consisting of linear combinations of the original bath modes, with the exclusion of the reaction coordinate, itself a collective mode of the original set. The parameters $\lambda_n$ and $\Omega_n$ characterize the $n$th (original) bath, representing the system-bath interaction energy and the characteristic frequency of the bath, respectively. Before the RC mapping, these parameters define the spectral density function $J_n(\omega)$, see for example Eq. (\ref{eq:Brownian}). After the RC mapping, these parameters appear explicitly in the model Hamiltonian \cite{Nick_PRX,RC_termodynamics,Anto-Sztrikacs_2021}.  

We now identify the extended system Hamiltonian, which includes the original system along with the reaction coordinates, 
\begin{equation}
    \begin{aligned}
        \hat{H}_S^{\rm RC} = \hat{H}_{S}+\sum_{n=1,2}\Big(\lambda_{n}\hat{S}_{n}(\hat{a}_{n}^{\dagger}+\hat{a}_{n})+\Omega_{n}\hat{a}_{n}^{\dagger}\hat{a}_{n}\Big).
    \end{aligned}
\end{equation}
Assuming weak coupling of this system to its residual baths, a Redfield equation of motion can be employed to follow the dynamics of the extended system. Such simulations, which we present below, are referred to as ``reaction coordinate simulations'' and denoted as RC.

{\bf B.} A polaron transformation, enacted by $\hat U_P$, is applied to Eq. (\ref{eq:HRC}), to decouple the RCs from the system \cite{variational_polaron}. This step imprints the system-bath interaction energies, $\lambda_n $, into the system's Hamiltonian. Additionally, the polaron transformation generates coupling terms between the transformed system and the residual baths \cite{Nick_PRX}. Formally, the polaron mapping can be summarized as
\bea
\hat H_{\rm RC-P} =\hat U_P\hat H_{\rm RC} \hat U_P^{\dagger}.
\label{eq:HRCP}
\eea
Previously, the EFFH method was applied to multi-bath problems as in the nonequilibrium spin-boson model  \cite{Anto-Sztrikacs_2021,Nick_PRX} 
and dissipative spin chains \cite{min2024bathengineering}. 
However, in those studies the system's operators, $\hat S_n$, commuted, which was essential for reaching a unique EFFH. 

{\bf C.} The reaction coordinates (two for the case of two baths) are eliminated by retaining only the low-energy states of the Hamiltonian, truncating each RC to its ground state. This step is approximate and it is generally justified under the assumption that the frequencies of the reaction coordinates represent the largest energy scale in the model. A generalization based on a variational polaron mapping was developed in Ref. \citenum{Anto_Sztrikacs_2023}.

Concisely, steps {\bf B} and {\bf C} can be written as
\bea
\hat H^{\rm eff} = {\rm Tr_{\rm RC}} \left(\Pi_0\hat U_P\hat H_{\rm RC} \hat U_P^{\dagger} \Pi_0\right),
\label{eq:HEFF}
\eea
creating the effective Hamiltonian of the original model. The operator $\Pi_0$ projects to the subspace of the ground states of the RCs involved. 

In the original EFFH method, the polaron transform $\hat U_P$ was factorized to terms from different baths. In Sec. \ref{sec:polaron1} we show that when the operators $\hat{S}_{n}$ do not commute, the EFFH approach of Ref. \citenum{Nick_PRX} encounters an ambiguity, resulting in effective Hamiltonians that depend on the order of the application of the polaron transforms. To handle noncommuting operators, we introduce in Sec. \ref{sec:polaron2} a non-factorized polaron transformation. The revised transformation introduces technical complexities---but it resolves the ambiguity and provides a well-defined effective Hamiltonian, which serves as the starting point for simulating the dynamics and steady state of systems with noncommuting coupling operators.

Before continuing with the generalization of Step {\bf B}, we provide additional comments on the EFFH method. We performed so far a single approximation: Truncating the manifold of the RC mode to retain only its ground state. This approximation is justified at low temperatures and when energy scales of the system ($\Delta$) and the interaction Hamiltonian ($\lambda$) are small relative to $\Omega$, the RC frequency. 
Another approximation that we make later in simulations is that the residual bath only weakly couples to the system. This allows us to describe the impact of the residual bath on the system  within a perturbatuve approach, specifically the Redfield quantum master equation.

The EFFH method can be extended beyond these two limitations. 
The polaron step can be exercised in a variational manner, allowing the treatment of lower frequency modes, $\Omega$ order of temperature and $\Delta$ \cite{Anto_Sztrikacs_2023}. 
Another approach for handling low frequency baths is to retain more levels in the RC manifold. It is also notable that the EFFH method may be valid (model dependent) when the interaction Hamiltonian is very strong, with $\Omega$ and $\lambda$ of the same order, and even in the ultrastrong coupling limit \cite{Nick_PRX}.
As for the assumption of weak coupling between the effective system and the residual bath, one can simulate the EFFH model using
more accurate techniques than the Redfield equation, such as with the influence functional path integral method \cite{Makri23}, particularly allowing more feasible convergence in the new effective representation, compared to the original one.
One can also operate the EFFH method after a chain mapping or in an iterative manner. 
In this paper, as we focus on method development and the elucidation of the mechanism of suppression of relaxation by decoherence,  we follow the original EFFH technique to present our contribution.

\vspace{2mm}

\section{Factorized polaron transforms: The Uniqueness problem}
\label{sec:polaron1}

Step {\bf B} of the EFFH procedure involves a polaron shift of the reaction coordinate. If the system-bath coupling operators $\hat S_n$ all commute, we factorize the polaron transform into two shift operators, one for each bath, indexed by $n\in\{1,2\}$, 
\bea
\hat U_P=\hat U_{1,P} \hat U_{2,P},\,\,\,\, \hat{U}_{n,P}=\exp\left[\frac{\lambda_{n}}{\Omega_{n}}\left(\hat{a}_{n}^{\dagger}-\hat{a}_{n}\right)\hat{S}_{n}\right].
\label{eq:polaron12}
\eea
These transformations are applied sequentially on $\hat{H}_{\text {RC}}$, Eq. (\ref{eq:HRC}). 
The third step (C) in the procedure involves truncating each RC to its ground state. Focusing on the system's Hamiltonian, we get from equation (\ref{eq:HEFF}),
\begin{align}
\hat{H}_{S}^{\text{eff}} & =\langle0|\hat{U}_{2,P}\hat{U}_{1,P}\hat{H}_{S}\hat{U}_{1,P}^{\dagger}\hat{U}_{2,P}^{\dagger}|0\rangle,
\label{eq:Heff_2polaron_1}
\end{align}
where we used the short notation $|0\rangle=|0_{1}\rangle\otimes|0_{2}\rangle$ for the ground states of the two reaction coordinates. 
To simplify this expression, we note that the polaron operator has a similar structure to the displacement operator, $D(\alpha) = e^{\alpha \hat{a}^{\dagger} - \alpha \hat{a}}$; in our study, $\alpha$ is an operator of the system, $\alpha \equiv \frac{\lambda}{\Omega}\hat{S}$ and $\hat S=\hat S^{\dagger}$.
We now use the following properties of the displacement operator: 
$D(-\alpha) = D^{\dagger}(\alpha)$,  $D(\alpha)\ket{0} = \ket{\alpha}$, implying that $D^{\dagger}(\alpha)\ket{0} = \ket{-\alpha}$. Furthermore, coherent states can be represented by eigenstates of the harmonic oscillator, $|n\rangle$, as $\ket{\alpha}=e^{-\frac{|\alpha|^2}{2}}
\sum_{n=0}^{\infty} \frac{\alpha^n}{\sqrt{n!}}|n\rangle$.
Since in the models examined here all the $\hat S_n$ operators are hermitian with real-valued elements (symmetric matrices), we ignore the absolute value symbol. Altogether, we simplify Eq. (\ref{eq:Heff_2polaron_1}) and arrive at the effective system Hamiltonian,
\begin{widetext}
\begin{equation}
\begin{aligned}\hat{H}_{S}^{\text{eff}} & =\langle0_{2}|\hat{U}_{2,P}\langle0_{1}|\hat{U}_{1,P}\hat{H}_{S}\hat{U}_{1,P}^{\dagger}|0_{1}\rangle\hat{U}_{2,P}^{\dagger}|0_{2}\rangle\\
 & =\langle0_{2}|\hat{U}_{2,P}e^{-(\lambda_{1}^{2}/2\Omega_{1}^{2})\hat{S}_{1}^{2}}\Big(\sum_{n=0}^{\infty}\frac{\lambda_{1}^{2n}}{\Omega_{1}^{2n}n!}\hat{S}_{1}^{n}\hat{H}_{S}\hat{S}_{1}^{n}\Big)e^{-(\lambda_{1}^{2}/2\Omega_{1}^{2})\hat{S}_{1}^{2}}\hat{U}_{2,P}^{\dagger}|0_{2}\rangle\\
 & =e^{-(\lambda_{2}^{2}/2\Omega_{2}^{2})\hat{S}_{2}^{2}}\sum_{p=0}^{\infty}\frac{\lambda_{2}^{2p}}{\Omega_{2}^{2p}p!}\Big(\hat{S}_{2}^{p}e^{-(\lambda_{1}^{2}/2\Omega_{1}^{2})\hat{S}_{1}^{2}}\sum_{n=0}^{\infty}\frac{\lambda_{1}^{2n}}{\Omega_{1}^{2n}n!}\Big(\hat{S}_{1}^{n}\hat{H}_{S}\hat{S}_{1}^{n}\Big)e^{-(\lambda_{1}^{2}/2\Omega_{1}^{2})\hat{S}_{1}^{2}}\hat{S}_{2}^{p}\Big)e^{-(\lambda_{2}^{2}/2\Omega_{2}^{2})\hat{S}_{2}^{2}}.
\end{aligned}
\label{eq:Heff_2polaron_2}
\end{equation}
\end{widetext}
If $\hat S_1$ and $\hat S_2$ indeed commute, the order of application of $\hat{U}_{n,P}$ does not impact the resulting effective Hamiltonian. However, if they do not commute, the resulting effective Hamiltonian depends on which polaron transform was (arbitrarily) selected to be applied first, $\hat U_{P,1}$ or $\hat U_{P,2}$.

For example, consider $\hat H_S=\Delta \sigma^z$,
$\hat S_1= \cos (\pi/4)  \sigma^z + \sin(\pi/4) \sigma^x$
and $\hat S_2= \sigma^x$.  Applying $\hat S_1$ first (at the centre) and then $\hat S_2$ yields \cite{Anto_Sztrikacs_2023}
\bea
\hat{H}_S^{\rm eff} =  \frac{\Delta}{2}\left(1+e^{-\frac{2\lambda_1^2}{\Omega_1^2}}\right) e^{-\frac{2\lambda_2^2}{\Omega_2^2}}\sigma^z 
+ \frac{\Delta}{2} \left(1-e^{-\frac{2\lambda_1^2}{\Omega_1^2}}\right) \sigma^x.
\nonumber\\
\eea
However, switching the operation $1\leftrightarrow2$ yields the symmetric structure,
\bea
\hat{H}_S^{\rm eff} &=&  \frac{\Delta}{2}\left(1+e^{-\frac{2\lambda_2^2}{\Omega_2^2}}\right) e^{-\frac{2\lambda_1^2}{\Omega_1^2}}\hat{\sigma}^z 
\nonumber\\
&+& \frac{\Delta}{2} \left(1-e^{-\frac{2\lambda_2^2}{\Omega_2^2}}\right) e^{-\frac{2\lambda_1^2}{\Omega_1^2}}\sigma^x.
\eea
This ambiguity also affects the system-bath interaction Hamiltonian; recall that  we need to apply the polaron transformations of Eq. (\ref{eq:polaron12}) on all other terms in the Hamiltonian Eq. (\ref{eq:HRC}).
Once again, the order $(1,2)$ of the polaron transformation will change the result, as we can get either ($n\neq m$)
\begin{equation}
\begin{aligned}\hat{U}_{n,P}\hat{U}_{m,P}\hat{a}_{n}\hat{U}_{m,P}^{\dagger}\hat{U}_{n,P}^{\dagger} & =\hat{U}_{n,P}\hat{a}_{n}\hat{U}_{n,P}^{\dagger}=\hat{a}_{n}-\frac{\lambda_{n}}{\Omega_{n}}\hat{S}_{n},
\end{aligned}
\label{eq:a_2polaron_1}
\end{equation}
or
\begin{equation}
\begin{aligned}\hat{U}_{m,P}\hat{U}_{n,P}\hat{a}_{n}\hat{U}_{n,P}^{\dagger}\hat{U}_{m,P}^{\dagger} & =\hat{U}_{m,P}\Big(\hat{a}_{n}-\frac{\lambda_{n}}{\Omega_{n}}\hat{S}_{n}\Big)\hat{U}_{m,P}^{\dagger}\\
 & =\hat{a}_{n}-\frac{\lambda_{n}}{\Omega_{n}}\hat{U}_{m,P}\hat{S}_{n}\hat{U}_{m,P}^{\dagger}.
\end{aligned}
\label{eq:a_2polaron_2}
\end{equation}
Explicitly, by expanding the exponential inside the polaron transform and using the identity  (\ref{eq:Com_identity}) we find that
\bea
\hat{U}_{m,P}\hat{S}_{n}\hat{U}_{m,P}^{\dagger} & =&\sum_{p=0}^{\infty}\frac{1}{p!}\frac{\lambda_{m}^{p}}{\Omega_{m}^{p}}(\hat{a}_{m}^{\dagger}-\hat{a}_{m})^{n}\text{Com}^{\ocircle p}(\hat{S}_{p},\hat{S}_{m})
\nonumber\\
&\neq&\hat{S}_{n},
\label{eq:Seff_naive}
\eea
see Appendix \ref{app:1} for details.
Here, $\text{Com}^{\ocircle p} $ is a $p$-times composition of the functional $\text{Com}(A,B)$, with
$\text{Com}^{\ocircle1}(A,B)=[A,B]$, 
$\text{Com}^{\ocircle2}(A,B)=[[A,B],B]$,
$\text{Com}^{\ocircle3}(A,B)=[[[A,B],B],B]$, and so on. Note that we define the nested commutator to order zero to be $Com^0(A,B)=A$.


In sum, a naive, factorized application of the polaron mapping within the EFFH method results in a {\it nonunique} effective Hamiltonian if the OQS is coupled to multiple baths through {\it noncommuting} operators. To address this issue, we employ next a non-factorized polaron mapping. 

\begin{figure*}[htbp]
\begin{centering}
\includegraphics[width=1\textwidth]{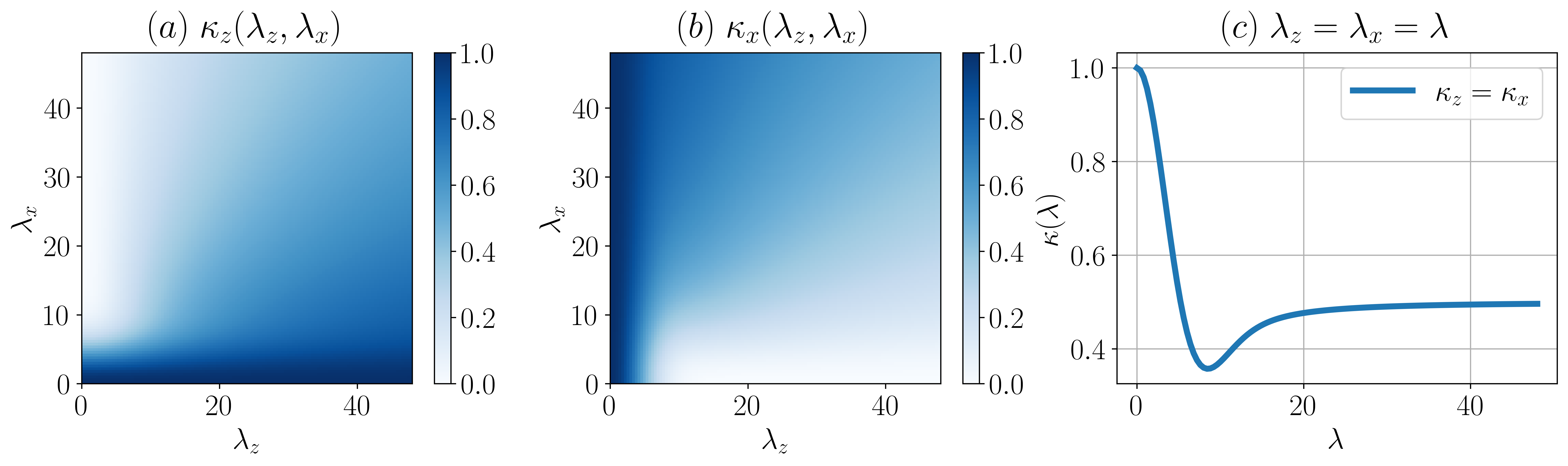} 
\par\end{centering}
\caption{\label{fig:dawson}(a)-(b) Surface plots of the suppression functions
$\kappa_{z}(\lambda_z,\lambda_x)$ and $\kappa_{x}(\lambda_z,\lambda_x)$ dressing the operators $\sigma^{z}$
and $\sigma^{x}$, respectively. (c) The suppression coefficient $\kappa(\lambda)$
for $\lambda=\lambda_{z,x}$. The frequencies of the RC  are set to be equal, $\Omega_{z}=\Omega_{x}=8$. The coupling strengths, RCs frequencies and the temperature are given in units of $\Delta$.
The $\kappa$ functions are evaluated numerically based on the integral form in Appendix \ref{app:3}.}
\end{figure*}

\vspace{3mm}
\section{Non-factorized polaron mapping: Treating noncommuting coupling operators}
\label{sec:polaron2}

We return to step {\bf B} in the EFFH procedure of 
Sec. \ref{sec:ModelMethod} and specify how it is generalized when the system is coupled to noncommuting baths. We construct the non-factorized polaron operator, which treats symmetrically the different baths. For simplicity, we keep the presentation focused on two baths only, 
\begin{equation}
\hat{U}_{P}=\exp\left[\sum_{n=1,2}\frac{\lambda_{n}}{\Omega_{n}}(\hat{a}_{n}^{\dagger}-\hat{a}_{n})\hat{S}_{n}\right].
\label{eq:UP}
\end{equation}
Trivially, it reduces to a product of two polaron transforms if $[\hat S_n,\hat S_m]=0$.

Continuing to steps {\bf B} and {\bf C} in the EFFH procedure of Sec. \ref{sec:ModelMethod}, we perform the polaron transformation and truncate the RC modes to their ground states.  We get, up to constant terms,
\begin{equation}
\begin{aligned}
\hat{H}^{\text{eff}} & =\langle0| \hat{U}_P \hat{H}_S \hat{U}_P^\dagger |0\rangle -\sum_{n}\frac{\lambda_{n}^{2}}{\Omega_{n}}\Big(\hat{S}_{n}^2\Big)^{\text{eff}}\\
 & +\sum_{n}\tilde{\sum_{k}}\left[\omega_{n,k}\hat{b}_{n,k}^{\dagger}\hat{b}_{n,k}-\frac{2\lambda_{n}f_{n,k}}{\Omega_{n}}\hat{S}_{n}^{\text{eff}}(\hat{b}_{n,k}^{\dagger}+\hat{b}_{n,k})\right],
\end{aligned}
\label{eq:Heff_UP}
\end{equation}
where for an operator $\hat{\mathcal{O}}_{S}$, its effective
operator is defined as $\hat{\mathcal{O}}_{S}^{\text{eff}}=\langle0|\hat{U}_{P}\hat{\mathcal{O}}_{S}\hat{U}_{P}^{\dagger}|0\rangle$. We delegate the derivation to Appendix \ref{app:2}.

\color{black}
Importantly, the system's Hamiltonian, 
$\langle0| \hat{U}_P \hat{H}_S \hat{U}_P^\dagger |0\rangle$,
is augmented by an additional term,  the second term in Eq. (\ref{eq:Heff_UP}),
which arises due to the system-bath coupling.
This term is responsible of bath induced phase transitions in spin lattices \cite{min2024bathengineering}, modification of topological phases in fermionic chains \cite{SSH}, and synchronization dynamics of otherwise noninteracting spins through a common bath
\cite{brenes2024bathinduced}.

We put forward the ansatz that the state of the system, the outcome of the (now, generalized) EFFH procedure, can be evaluated from the effective system Hamiltonian, 
\bea
\hat H_S^{\text{eff}} = 
\langle0| \hat{U}_P \hat{H}_S \hat{U}_P^\dagger |0\rangle  -\sum_{n}\frac{\lambda_{n}^{2}}{\Omega_{n}}\Big(\hat{S}_{n}^2\Big)^{\text{eff}}.
 \label{eq:H_MF}
\eea 
The second term in Eq. (\ref{eq:H_MF}) does not play a role when considering a single spin coupled to baths through Pauli matrices, as discussed in Sec. \ref{sec:TLS}. However, this term creates important interactions in the lattice model, as detailed in Sec. \ref{sec:lattice}.  

Equation (\ref{eq:Heff_UP}) presents an effective Hamiltonian, similar in form to equation (\ref{eq:H_original}), but with a modified system and dressed coupling constants. Both the system Hamiltonian and system-bath coupling operators are altered through the non-factorized polaron transform, followed by an RC mode truncation,
\begin{equation}
    \begin{aligned}
        \hat{H}_S &\rightarrow \hat{H}_S^{\text{eff}}, \nonumber \\
        t_{n,k} \hat{S}_n &\rightarrow \frac{2 \lambda_n}{\Omega_n} f_{n,k} \hat{S}_n^{\text{eff}}.
    \end{aligned}
\end{equation}
After the EFFH mapping, each bath from the original model, with a spectral density function $ J_n(\omega) = \sum_k t^2_{n,k} \delta(\nu_{n,k}-\omega)$, turns into
\begin{equation}
    \begin{aligned}
        J_n^{\text{eff}}(\omega)=\frac{4\lambda_n^2}{\Omega_n^2} \sum_k f^2_{n,k} \delta(\omega_{n,k}-\omega).
    \end{aligned}
\end{equation}
The next task we need to tackle is building explicitly the operators  $\hat{\mathcal{O}}_{S}^{\text{eff}}$ in Eq. (\ref{eq:Heff_UP}), which we do in momentum basis.

Reaction coordinates are harmonic modes, described by the Hamiltonian $\Omega\left(\hat{a}^{\dagger}\hat{a}+\frac{1}{2}\right)=\frac{\hat{p}^{2}}{2m}+\frac{1}{2}m\Omega^{2}\hat{x}^{2}$. In momentum space, the eigenfunctions of the harmonic Hamiltonian are \cite{HOp}
\begin{equation}
\langle p|l\rangle=\frac{e^{-p^{2}/(2m\Omega)}}{\sqrt{2^{l}l!\sqrt{\pi m\Omega}}}H_{l}\left(\frac{p}{\sqrt{m\Omega}}\right),
\end{equation}
with $l=0,1,2,...$ indexing the states and where $H_{l}(p)=(-1)^{l}e^{p^{2}} 
\frac{d^{l}}{dp^{l}}e^{-p^{2}}$ is the $l$th Hermite polynomial. The exponent of the non-factorized polaron transform includes creation and annihilation operators of the RC, now expressed as
$(\hat{a}^{\dagger}-\hat{a})=-i\sqrt{\frac{2}{m\Omega}}\hat{p}$. The resulting effective operators, with a renormalized integration variable, $\tilde{p}_n = p_n/\sqrt{m_n \Omega_n}$, 
are
\bea
&&\hat{\mathcal{O}}_{S}^{\text{eff}} 
=\langle0|\hat{U}_{P}\hat{\mathcal{O}}_{S}\hat{U}_{P}^{\dagger}|0\rangle
\nonumber\\ 
& &=\int \prod_n dp'_{n} dp_{n} \langle0|p'_{n}\rangle \langle p_{n}|0\rangle \left( \otimes_n \langle p'_{n}| \hat{U}_P \hat{\mathcal{O}}_{S} \hat{U}_P^{\dagger} \otimes_n |p_{n}\rangle \right)
 \nonumber\\
 & &=\int\hat{U}_{P}( \tilde{ \boldsymbol{p}})\hat{\mathcal{O}}_{S}\hat{U}_{P}^{\dagger}(\tilde{ \boldsymbol{p}})\prod_{n}\frac{e^{-\tilde{p}_{n}^{2}}}{\sqrt{\pi}}d \tilde{p}_{n}.
 \label{eq:O_eff}
\eea
Here,  $\otimes_n \langle p_n |$ is the tensor product of all  $p_n$ eigenstates with $n$ indexing the reaction coordinates. For a two-bath problem with two RCs, 1 and 2, this expression corresponds to
$ \otimes_n \langle p_n | =  \langle p_1,  p_2 |$. The
polaron transformation in the momentum basis is
\bea
        \hat{U}_{P}(\boldsymbol{p}) &= &\otimes_n \langle p_n | \hat{U}_{P} \otimes_n |p_{n} \rangle \nonumber \\
& = & \exp \left( -i \sqrt{2} \sum_n  \frac{p_n}{\sqrt{m_n \Omega_n}} \frac{\lambda_n}{\Omega_n} \hat{S}_n \right) \nonumber \\
\rightarrow \hat{U}_{P}(\tilde{\boldsymbol{p}}) &=& \exp \left( -i \sqrt{2} \sum_n  \tilde{p}_n \frac{\lambda_n}{\Omega_n} \hat{S}_n \right).
\label{eq:Upp}
\eea
Equation (\ref{eq:O_eff}) together with Eq. (\ref{eq:Upp}) provides a computational approach
for deriving closed-form expressions for the effective operators in Eq. (\ref{eq:Heff_UP}), completing the task of constructing an effective Hamiltonian for noncommuting coupling operators.
As a reminder, $\tilde p_n$ is a momentum variable corresponding to the $n$th reaction coordinate, which was extracted from the $n$th bath; $\hat S_n$ is a system operator coupled to that bath. $\lambda_n$ and $\Omega_n$ characterize the system-bath coupling strength and the spectrum of the bath, respectively. 

It is possible to generalize step {\bf C} of the EFFH method and retain more levels in the reaction coordinate beyond its ground state. This would require generalizing Eq. (\ref{eq:Heff_UP}) and Eq.  (\ref{eq:O_eff}). 
Studying the functional form of effective system operators, or more generally, elements of some Lie groups might give new insights into efficient computations \cite{nested_commutators}. This observation is inspired by identities of nested commutators that are derived in Appendix \ref{app:1}. They resemble known identities for operators, but instead of multiplication, they
include nested commutator functions. 

It is worth noting here that, for noncommuting operators, the ultrastrong coupling limit of the system, which is  constructed from the effective Hamiltonian, cannot be written as a sum of projectors onto the basis of the system coupling operator, as was done in the commuting case \cite{strong_limit_MFGS}.
This is because the pointer basis of either individual $\hat S_n$ cannot construct the complete solution.


\section{Cooperative thermal relaxation and decoherence in an impurity model}
\label{sec:TLS}

In this Section, we derive a closed-form expression for the effective Hamiltonian of a spin impurity 
coupled to two heat baths, where the coupling operators of the system to the baths are different Pauli matrices. 
This system might serve as a generic model for a qubit suffering both dissipation and decoherence from different independent sources. 

\subsection{Closed-form expression for the Effective Hamiltonian}

We consider a spin impurity coupled to two heat baths.
Going back to the Hamiltonian (\ref{eq:H_original}), we identify the system and its coupling operators by
\bea
\hat{H}_{S}&=& \Delta\sigma^{z},
\nonumber\\
\hat{S}_{z}&=& \sigma^{z},\,\,\,\, \hat{S}_{x}= \sigma^{x}.
\label{eq:SB}
\eea
This model is depicted in Figure \ref{fig:impurity_sys}. 
We index here the baths according to their coupling operators, $x$ and $z$.
We refer to the $x$ bath as ``dissipative'', since it allows energy exchange with the system, and to the $z$ bath as ``decoherring'', since it only allows pure decoherence when acting alone. However, it is important to note that the dissipative bath also leads to decoherence effects (see  equations of motion in Appendix \ref{app:4}), and that in the strong coupling limit, the baths cooperatively act on the system to impact both the population relaxation and decoherence rates, as we show below. 

The $x$ ($z$) bath is described by a coupling energy to the system $\lambda_x$ ($\lambda_z$) and a characteristic frequency $\Omega_{x}$ 
($\Omega_z$). These parameters appear in the RC Hamiltonian (\ref{eq:HRC}), then the effective one, (\ref{eq:Heff_UP}). 

Applying the EFFH method with the non-factorized polaron transform (\ref{eq:UP}), we obtain the effective Hamiltonian  (\ref{eq:Heff_UP})-(\ref{eq:H_MF}) with
\bea
\hat{H}_{S}^{\text{eff}}&=&\kappa_{z}\left(\lambda_z, 
\lambda_x\right)\Delta \sigma^z,
\nonumber\\
\hat{S}_{z}^{\text{eff}} & =&\kappa_{z}\left(\lambda_z, \lambda_x \right)\sigma^{z},\,\,\,\,
\hat{S}_{x}^{\text{eff}} =\kappa_{x}\left(  \lambda_z, \lambda_x
 \right)\sigma^{x}, 
\label{eq:Oeff_example_1}
\eea
see Appendix \ref{app:3} for details. 
The functions $\kappa_z(\lambda_z, \lambda_x)$ and $ \kappa_x(\lambda_z, \lambda_x)$ are displayed in Figure \ref{fig:dawson}. The functions also depend on the baths' characteristic frequencies, $\Omega_{x,z}$, but we highlight the dependency on the system-bath interaction energies.
The $\kappa_z$ and $\kappa_x$ functions dress the $\sigma^z$ and $\sigma^x$ operators, respectively, through factors that depend on {\it both} coupling energies, $\lambda_{x}$ and $\lambda_z$. 
In Appendix \ref{app:3}, we prove that 
\bea
&& 
\kappa_{x,z}(\lambda_z, \lambda_x)\leq 1,
\nonumber\\
&&\kappa_z(\lambda_z, \lambda_x)=\kappa_x(\lambda_x, \lambda_z).
\eea
Furthermore, the following limits hold
\bea
\kappa_z(0,\lambda_x)&=&e^{-2\lambda_x^2/\Omega_x^2},
\nonumber\\
\kappa_z(\lambda_z,0)&=&1, \,\,\,\,\,\,
\kappa_x(0,\lambda_x)=1.
\eea
Note that $\kappa_x(\lambda_z,0)$ is inconsequential, since in this case, there is no coupling $\hat S_x$ to dress. If the two heat baths are identical in their spectral density functions, the extracted RC parameters $\lambda_n$ and $\Omega_n$ are also identical for the two baths. 
This simplifies Eq. (\ref{eq:Oeff_example_1}) to
\bea
\hat{H}_{S}^{\text{eff}}=\kappa(\lambda)\sigma^z,\,\,\,\,
\hat{S}_{z,x}^{\text{eff}} & =\kappa(\lambda)\sigma^{z,x},
\label{eq:Oeff_example1symm}
\eea
where $\kappa(\lambda)=1-\sqrt{2}\frac{\lambda}{\Omega} F(\sqrt{2}\frac{\lambda}{\Omega})$, see Appendix \ref{app:3}. Here, $F(x)=e^{-x^{2}}\int_{0}^{x}e^{t^{2}}dt$ is the Dawson integral function. Altogether, assuming identical spectral functions for the baths, the effective Hamiltonian is given by
\bea
&&\hat{H}^{\text{eff}}
  = \kappa(\lambda) \hat{H}_{S} 
 \nonumber\\
 &&+\sum_{n=z,x}\tilde{\sum_{k}}
\left[\omega_{n,k}\hat{b}_{n,k}^{\dagger}\hat{b}_{n,k}-2\frac{\lambda }{\Omega}\kappa(\lambda) \hat{S}_{n} f_{n,k}(\hat{b}_{n,k}^{\dagger}+\hat{b}_{n,k})\right].
\nonumber\\
  \label{eq:H_eff_TLS}
\eea
In this expression, the operators of the system are dressed with a function of $\lambda$, shown in Figure \ref{fig:dawson}(c).
In the weak coupling limit, $\kappa(\lambda)$ goes to one. That is, the effect of the baths on the splitting renormalization diminishes. 
In the ultrastrong coupling limit, this function approaches the value $\frac{1}{2}$. We emphasize that in physical setups, there is no reason to assume that the coupling to the dissipation source will be identical to the coupling to the decoherence bath.
We adopt here identical parameters as a toy model to gain intuition about the functional form of $\kappa(\lambda)$.
\begin{figure*}[htbp]
\begin{centering}
\includegraphics[width=1.0\textwidth]{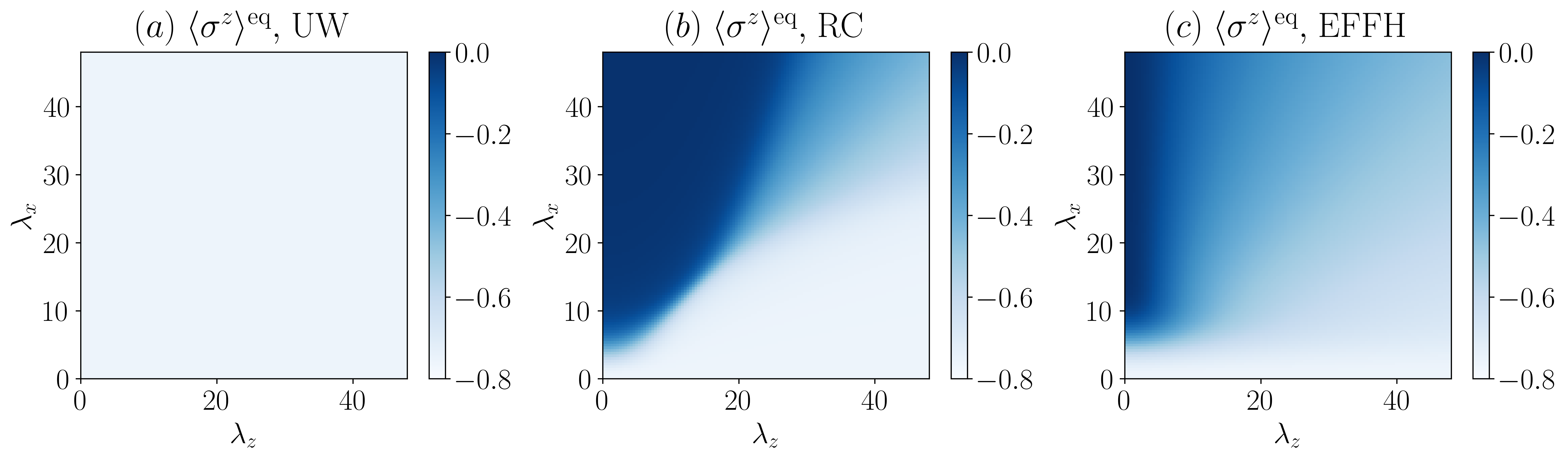}
\par\end{centering}
\caption{Surface plots of the equilibrium polarization $\langle \sigma^z \rangle$ of a single impurity spin coupled to two baths, Eq. 
(\ref{eq:H_original}) with system Hamiltonian (\ref{eq:SB}), presented as a function of the coupling parameters $\lambda_z$ and $\lambda_x$. Simulations were performed using 
(a) the original Hamiltonian (\ref{eq:SB}) assuming UW coupling, 
(b) the reaction coordinate Hamiltonian (\ref{eq:HRC}), 
and (c) the effective Hamiltonian (\ref{eq:Oeff_example_1}).
Parameters are set to $\Delta$=1, $\Omega=8$, $\gamma=0.05/\pi$, $\Lambda$=1000, $T=1$. Here and below we used $6$ levels for the RC mode, confirmed to provide converging results. The coupling strengths, RC frequencies and temperature are given in units of $\Delta$.}
\label{fig:sigmaSS}
\end{figure*}

\begin{figure}[htbp]
\begin{centering}
\includegraphics[width=0.48\textwidth]{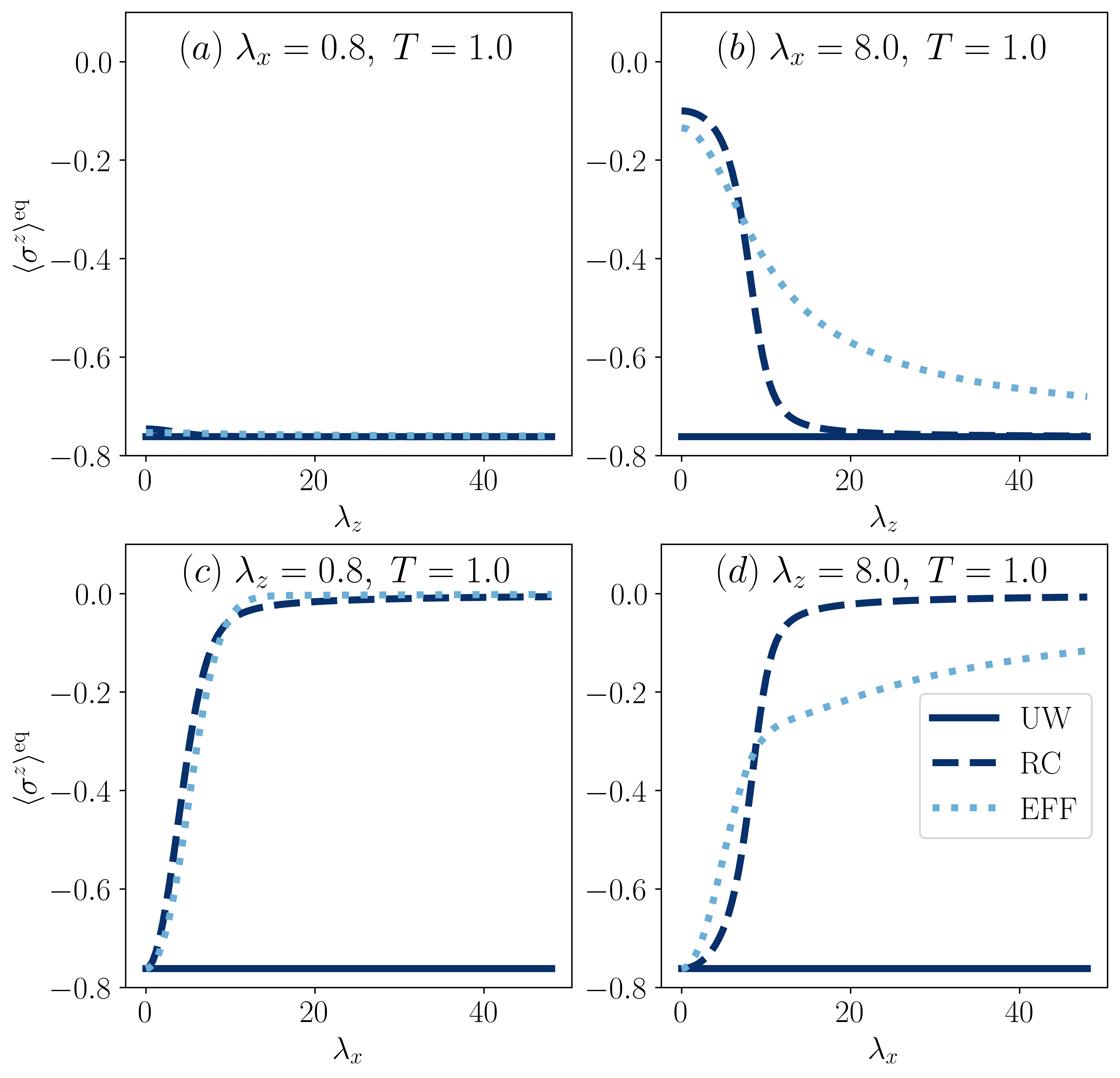} 
\par\end{centering}
\caption{Examples of the equilibrium values of the polarization plotted with respect to one
coupling parameter while keeping the other fixed, generated as cuts from Fig. \ref{fig:sigmaSS}.
(a)-(b) Polarization as a function of $\lambda_z$ for weak (a)  and strong (b) dissipative coupling $\lambda_x$. (c)-(d) Polarization as a function of $\lambda_x$
for weak (c) and strong (d) decoherence coupling $\lambda_z$. We perform calculations based on the  RC Hamiltonian (dashed), Effective Hamiltonian (dotted), and the ultraweak coupling limit (full). 
}
\label{fig:sigmaSS_plot}
\end{figure}
In Fig. \ref{fig:dawson}(a)-(b), we present  surface plots to show the dependence of $\kappa_z(\lambda_z,\lambda_x)$ and $\kappa_x(\lambda_z,\lambda_x)$
on the coupling strengths $\lambda_z$ and $\lambda_x$. 
In simulations, we set $\Omega_x=\Omega_z$, but one can easily study the more general case. 
Focusing first on Figure \ref{fig:dawson}(a), for every value of $\lambda_z$ the dressing function $\kappa_{z}(\lambda_z,\lambda_x)$ decreases with increasing $\lambda_x$. This indicates that both the system energy splitting and coupling operator to the decoherence bath are suppressed by the dissipative bath. 
The suppression of $\kappa_z$ with increasing $\lambda_x$ suggests that a dissipative bath ($x$) can slow down the effect of decoherence inflicted by another bath ($z$). However, the dynamics is more compound and this effect is not readily resolved (Appendix \ref{app:4}).
The complementary behavior of $\kappa_x$ is presented in Fig. \ref{fig:dawson}(b). Here, increasing the coupling to the decoherence bath through $\lambda_z$ reduces $\kappa_x$. This effect should lead to the suppression of thermal relaxation rate by increasing decoherence, an effect we exhibit below in Figures \ref{fig:popT} and \ref{fig:poplz}.
%

Fig. \ref{fig:dawson}(c) presents a cut along the diagonal of Fig. \ref{fig:dawson}(a)-(b), with $\lambda=\lambda_x=\lambda_z$; recall that we use here $\Omega=\Omega_x=\Omega_z$. 
Notably, we observe that the suppression coefficient $\kappa(\lambda)$ evolves {\it nonmonotonically} with $\lambda$: It starts at one in the ultraweak coupling limit, reaches a minimum of around $0.36$ for $\frac{\lambda}{\Omega}\approx 1.06$, then increases to  asymptotically approach $\frac{1}{2}$ from below at strong coupling.

We display next simulations based on the Redfield equation of motion, studying the dynamics and the steady state behavior of the system.  We use the Redfield equation based on (i) the original Hamiltonian, capturing its weak coupling limit (UW), (ii) the reaction coordinate Hamiltonian (RC) in the form of equation (\ref{eq:HRC}), and (iii) the effective Hamiltonian (EFFH), Eq. (\ref{eq:Heff_UP}).
We assume a Brownian spectral density function for the original model ($n=x,z$),
\begin{equation}
    J_{n}(\omega) = \frac{4 \gamma_n \Omega_n^2 \lambda_n^2 \omega}{ (\omega^2-\Omega_n^2)^2 + (2 \pi \gamma_n\Omega_n \omega )^2 }.
    \label{eq:Brownian}
\end{equation}
After the reaction coordinate transform, the spectral function of the residual bath becomes ohmic \cite{Anto-Sztrikacs_2021}, 
\bea J_n^{\text{RC}}(\omega) = \gamma_n \omega e^{-|\omega|/\Lambda},\eea 
which we use in RC simulations. Here, $\Lambda$ is a high frequency cutoff energy.
In the effective Hamiltonian calculations, the spectral density function is further dressed as \cite{Nick_PRX}
\bea J_n^{\text{eff}}(\omega) = \frac{4 \lambda_n^2}{\Omega_n^2} J_n^{{\text RC}}(\omega). \eea 
%
Based on previous benchmarking of the RC method against Hierarchical Equation of Motion (HEOM) simulations \cite{Anto_Sztrikacs_2023,brenes2024bathinduced}, we regard Redfield equation simulations of the RC Hamiltonian to be accurate and essentially exact when the spectral function of the model is narrow, $\gamma \ll 1$ and $\Lambda \gg 1$. These simulations construct the state of the system, which we use to study both the steady state behavior and dynamics of the spin polarization with respect to the coupling parameters, $\lambda_z$ and $\lambda_x$. 
For simplicity, in simulations we used  $\Omega_z=\Omega_x$ and  $\gamma_z=\gamma_x$. However, our theory extend beyond that.

\begin{figure*}[htbp]
\begin{centering}
\includegraphics[width=1\textwidth]{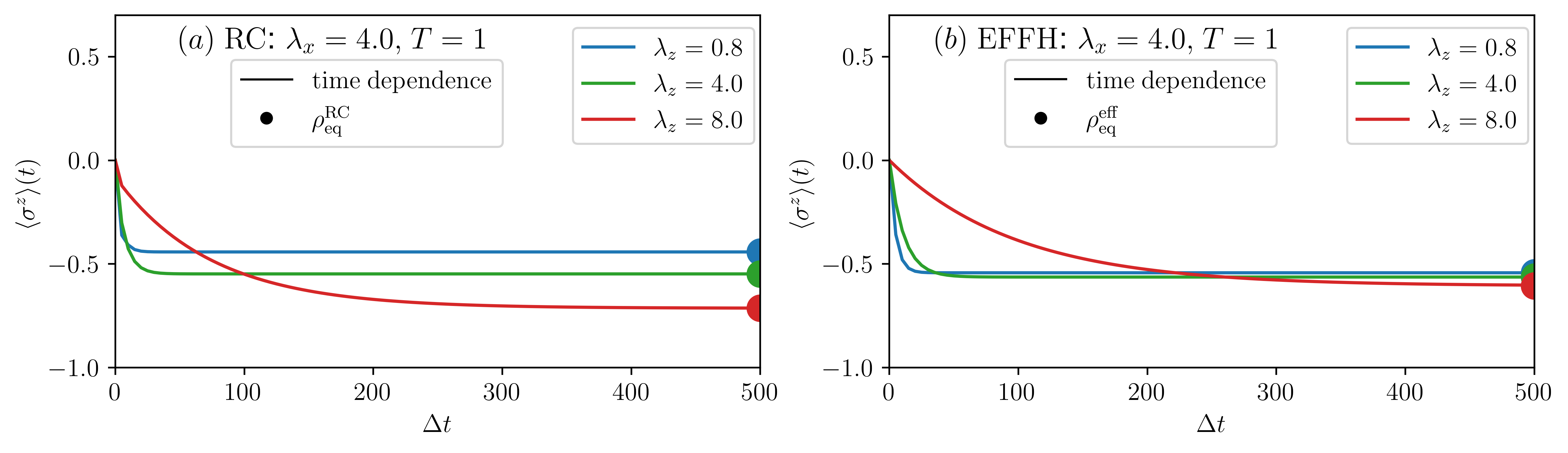}
\par\end{centering}
\caption{Confirming that the polarization dynamics reach in the long time limit a Gibbs-like equilibrium value. 
Simulations were performed with the (a) RC  and (b) the effective Hamiltonian method for different coupling energies.
The heavy circles indicate equilibrium values calculated from Eq. (\ref{eq:eqpol}). Simulation parameters are the same as in Figure \ref{fig:sigmaSS}.}
\label{fig:MFGS_check} 
\end{figure*}

\subsection{Equilibrium state}

When increasing the interaction energy of the system to the baths, the equilibrium state of the system $\rho_{\rm eq}$ generally deviates from the canonical thermal Gibbs state \cite{Thoss09, CaoFP,Miller2018, Nick_PRX, Anto_Sztrikacs_2023,Anton22, strong_limit_MFGS,
AndersAVS}. 
Nevertheless, this equilibrium state should coincide with the state of the system in the long time limit of the dynamics, generated by the appropriate dynamical equation of motion,
\bea
\lim_{t \to \infty} \rho(t) \to \rho_{\rm eq}.
\label{eq:DEQ}
\eea
Here, $\rho(t)$ is the time-dependent reduced density matrix of the system. In both reaction-coordinate and the EFFH approaches, the corresponding enlarged or effective system Hamiltonians are weakly coupled to the residual reservoirs. As such, we conjecture that the equilibrium state of the system is still a thermal Gibbs state at the inverse temperature $\beta$ of the bath. In the reaction coordinate picture, the equilibrium state of the system is
\begin{align}
\label{eq:equilibrium_rc}
\rho_{\rm eq}^{\rm RC} =  \frac{       
\Tr_{\rm RC}\left[ e^{-\beta \hat{H}_S^{\rm RC}} \right]}
{Z^{\rm RC}},
\end{align}
where $\hat{H}_S^{\rm RC}$ is the RC Hamiltonian of the system, Eq.~(\ref{eq:HRC}). $Z^{\rm RC} = \Tr[e^{-\beta \hat{H}_S^{\rm RC}}]$ and $\Tr_{\rm RC}$ denotes a partial trace over the reaction coordinate (the superscript RC on the reduced density matrix marks the method used). 
In the case of the EFFH method, the equilibrium state of the system is given by
\begin{align}
\label{eq:equilibrium_eff}
\rho_{\rm eq}^{\rm eff} = \frac{e^{-\beta \hat{H}_S^{\rm eff}}}{Z^{\rm eff}},
\end{align}
with $Z^{\rm eff} = \Tr[e^{-\beta \hat{H}_S^{\rm eff}}]$. 
The ultraweak coupling state is constructed based on $\hat H_S$ of Eq.~(\ref{eq:H_original}) assuming weak coupling. 
The different methods provide the same state at the ultraweak limit.
The expectation value of the polarization follows,
\bea
\langle {\sigma}^z \rangle^{\text{eq}} = \Tr[\rho_{\rm eq}^{\bullet} \sigma^z],
\label{eq:eqpol}
\eea
where $\bullet$ denotes adopting the state generated from the reaction-coordinate Hamiltonian, the Effective Hamiltonian of the system, or the original system's Hamiltonian. 
 

Considering the impurity spin model, 
we readily construct the corresponding equilibrium state using Eq. (\ref{eq:Oeff_example_1}),
\begin{equation}
    \begin{aligned}
        \langle \sigma^z \rangle^{\text {eq}} = -\tanh \left[  \beta \Delta \kappa_z(\lambda_z, \lambda_x) \right]. \label{eq:mag_eq_TLS}
    \end{aligned}
\end{equation}
This compact result is presented in Figures \ref{fig:sigmaSS} and \ref{fig:sigmaSS_plot} and it is compared to predictions from RC and the UW coupling limit.

In Figure \ref{fig:sigmaSS}, we present the equilibrium polarization with respect to the two coupling parameters. We compare results from the RC Hamiltonian (b), the effective Hamiltonian (c) and the UW coupling limit (a). 
We find that when both $\lambda_{x}\to0$ and $\lambda_z\to 0$, and with our choice of parameters, $\langle \sigma^z \rangle^{\text {eq}} = -\tanh(1) \approx -0.76$. This value 
corresponds to the ultraweak coupling limit shown in panel (a), where the polarization does not depend on the coupling parameters. Next, we increase $\lambda_x$ while keeping $\lambda_z=0$. In this scenario, according to both the RC and the EFFH methods (panels (b) and (c)),  
we suppress the polarization---ultimately to zero in the ultrastrong limit---reflecting the complete suppression of spin splitting.
However, when the coupling parameter $\lambda_z$ increases, we observe that the impact of the decoherring bath is to counteract the suppression of polarization, achieving e.g., $\langle \sigma^z \rangle^{\text{eq}} \approx -0.42$ for the RC model and $\langle \sigma^z \rangle^{\text{eq}} \approx -0.45$ with the EFFH at $\lambda_{x}=\lambda_z = 48$.
We comment that the effective Hamiltonian results shown in Fig. \ref{fig:sigmaSS}(c) are qualitatively correct compared to RC results of Fig. \ref{fig:sigmaSS}(b). 

Figure \ref{fig:sigmaSS_plot} presents horizontal and vertical sections of Figure \ref{fig:sigmaSS} for values of $\lambda_z$ and $\lambda_x$ corresponding to weak and strong coupling limit.
When the coupling to the dissipative bath is weak, the polariztion behaves as in the UW coupling limit, see Fig \ref{fig:sigmaSS}(a). In contrast, when the coupling is strong, the decoherring bath counteracts the impact of dissipation, recovering the UW coupling behavior once $\lambda_z\gg\lambda_x$. 
Fig.  \ref{fig:sigmaSS_plot} (c)-(d) shows that while ultrastrong  $\lambda_x$ coupling suppresses the spin polarization, the effect is being counteracted by $\lambda_z$.

To complete the analysis of the equilibrium behavior, we test Eq. (\ref{eq:DEQ}), that is, whether dynamical calculations (Appendix \ref{app:4}) approach the equilibrium state in the long time limit.
In Figure \ref{fig:MFGS_check}, we present this comparison for both the RC and the EFFH methods, and for several values of the coupling parameters, $\lambda_z$ and $\lambda_x$. In each case, we find that the dynamics, derived from the Redfield equation with the respective (RC, eff) Hamiltonian, reaches the state of equilibrium corresponding to the Gibbs state of that Hamiltonian.

We also study the equilibrium value of the coherence  $\langle {\sigma}^x \rangle^{\text{eq}} = \Tr[\hat{\rho}_{\rm eq}^{\bullet} \hat{\sigma}^x]$
as a function of coupling strengths $\lambda_z$ and $\lambda_x$ with the different methods. We observe that in our model, the equilibrium coherences are zero.
We obtain the same results by studying the long time dynamics, and by directly computing the coherences using the equilibrium state.

\subsection{Suppression of relaxation by decoherence}
\label{sec:suppression}

Our analytical analysis revealed a unique effect: The mutual suppression of coupling parameters by noncommuting operators,
see the dressing functions in Fig. \ref{fig:dawson}.
To showcase this behavior explicitly, we study the time evolution of the polarization, $\langle \sigma^z \rangle(t)$, using equations of motion as detailed in Appendix \ref{app:4}. The results are displayed in Figs. \ref{fig:popT} and \ref{fig:poplz}, with Fig. \ref{fig:short_time} zooming-in on the short-time dynamics.

In Appendix \ref{app:4}, we detail the equations of motion of the model. We show that the population (and thus the polarization) dynamics decays as $\dot \rho_{11}(t)= -\Gamma_x^{\bullet} \rho_{11}(t) +A^{\bullet}$. Here, $\bullet$ stands for the effective Hamiltonian method (eff) or the ultraweak limit (UW).
The solution for the equation of motion is
\bea
 \rho_{11}(t) = 
 \rho_{11}(0)e^{-\Gamma_x^{\bullet}t} +\frac{A^{\bullet}}{\Gamma_x^{\bullet}}\left(1-e^{-\Gamma_x^{\bullet}t}\right).
 \label{eq:Gammaxeff}
\eea
Under the effective Hamiltonian,
the relaxation rate constant depends on both $\lambda_x$ and $\lambda_z$ in a nontrivial manner, 
\bea
\Gamma_x^{\text {eff}}&=& 2 \pi \kappa_x^2(\lambda_z,\lambda_x)
J_x^{\text{eff}}(2\kappa_z\Delta)
\times [2n_B(2\kappa_z\Delta)+1],
\nonumber\\
&\approx& 
4 \pi T \kappa_x^2(\lambda_z,\lambda_x) \frac{4\lambda_x^2}{\Omega_x^2}
\gamma_x.
\label{eq:GammaXM}
\eea
For brevity, we do not explicitly note the dependence of $\kappa_z$ on both $\lambda_{x}$ and $\lambda_{x}$.
To reach the second line, we used $J_x^{\text{ eff}} (2\kappa_z \Delta)\approx\frac{4\lambda_x^2}{\Omega_x^2} \gamma_x \kappa_z \Delta$ and we assumed that $T>\Delta$. It is also required to assume that $\Omega>\Delta$ as part of the EFFH framework.
In contrast, in the ultraweak coupling limit we get
\bea
\Gamma_x^{\text {UW}}&=& 2 \pi
J_x(2\Delta) \times [2n_B(2\Delta)+1]
\nonumber\\
&\approx& 
4 \pi  T  \frac{4\lambda_x^2}{\Omega_x^2}
\gamma_x.
\label{eq:GammaXMUW}
\eea
Again, the second line was derived based on $T,\Omega>\Delta$.

Equation (\ref{eq:GammaXM}) is one of the main results of this paper. It
reveals the impact of $\lambda_z$ on the relaxation rate. In more details:
As expected, the relaxation rate scales explicitly with the coupling strength to the dissipative bath, $\lambda_x^2$. However, the additional $0\leq\kappa_x^2\leq1$ term in Eq. (\ref{eq:GammaXM}) encodes a nontrivial dependence of the relaxation on the coupling to the decoherring bath, $\lambda_z$, as presented in Fig. \ref{fig:dawson}(b). 
There, we note that for a fixed $\lambda_x$, increasing $\lambda_z$ (moving to the right) progressively reduces $\kappa_x$. 
Most critically, since $\kappa_x$ is suppressed when $\lambda_z$ increases, it indicates that $\Gamma_x^{\text{eff}}$ as a whole is {\it suppressed} when the decoherring bath couples more strongly to the system via $\lambda_z$.
The overall effect that Eq. (\ref{eq:GammaXM}) exposes {\it analytically} is a {\it suppression} or {\it slowing down} of the relaxation dynamics due to the added, independent source of decoherence to the qubit.


We now illustrate this behavior in simulations. First, in Fig. \ref{fig:popT} we show that this effect 
takes place at both (a) low  and (b) intermediate  temperatures. 
Next, in Fig. \ref{fig:poplz}, we show that the effect holds when the dissipative coupling $\lambda_x$ is  tuned up from (a) weak  to (b) intermediate. 
As can be seen from these figures,
the EFFH theory provides a qualitative correct behavior, compared to RC simulations. 

The relaxation rate arrived at from EFFH simulations precisely agrees with $\Gamma^\text{eff}_x$ of Eq. (\ref{eq:Gammaxeff}).
We also study the relaxation dynamics by fitting the RC curves in Fig. \ref{fig:poplz}(a) to an exponential function, see the Table \ref{tab:slopes}. 
The general trend observed in the Table is that increasing $\lambda_z$ with a fixed $\lambda_x$ reduces $\kappa_x$ substantially, thus suppressing $\Gamma_x^{\rm eff}$. The effective Hamiltonian however tends to overestimate the suppression phenomenon, compared to RC results. 
We also notice that the dressing function $\kappa_x$ is the dominant factor that causes the decay rate to get smaller with the increase of $\lambda_z$. 

%
\begin{table}[]
    \centering
    \begin{tabular}{|c|c|c|c|c|c|}
    \hline
Coupling & Coupling & Dressing   & Dressing  & Relax. & Relax.  \\ 
energy	& energy &  function  &  function & rate & rate \\ 
    \hspace{1mm}$\lambda_{x}$ \hspace{1mm} & \hspace{2mm} $\lambda_{z}$ \hspace{2mm} & \hspace{3mm} $\kappa_z$ \hspace{3mm} & \hspace{3mm} $\kappa_x$ \hspace{3mm} & \hspace{3mm} $\Gamma_{x}^{\text{eff}}$ \hspace{3mm} & 
   \hspace{5mm} $\Gamma_x^{\text{RC}}$ \hspace{5mm}\\ \hline
0.8	& 0.8 & 0.9803 & 0.9803 & 0.0100 & 0.0116 \\ \hline
0.8	& 4.0 & 0.9817 & 0.6079 & 0.0038 & 0.0048 \\ \hline
0.8	& 8.0 & 0.9849 & 0.1386 & 0.0002 & 0.0006 \\ \hline
    \end{tabular}
    \caption{Comparison of the numerically-fitted relaxation (Relax.) rate for the RC approach from Fig. \ref{fig:poplz}(a) with the analytically calculated relaxation rates $\Gamma_x^\text{eff}$.}
    \label{tab:slopes}
\end{table}


Trivially, the cooperative effect of dissipation suppression by deocherence is missing in the ultraweak coupling limit (depicted by dotted lines in Figs. \ref{fig:popT}-\ref{fig:poplz}). Indeed, as we show in Appendix \ref{app:4}, in the UW limit the baths act in an additive manner.

Our results here support and explain a recent qausiadiabatic path integral (QUAPI) study  \cite{Suppressing_relaxation_through_dephasing}, which reported through simulations on the suppression of spin relaxation with increased decoherence. 
While the QUAPI approach is numerically exact; it relies on a Trotter discretization and the memory time truncation within the Feynman-Vernon Influence functional \cite{Makri_PI}, it lacks an analytical insight, which our approach offers through Eqs. (\ref{eq:Gammaxeff})-(\ref{eq:GammaXM}). It is important though to note that QUAPI usually adopts Ohmic spectral functions, given the challenge to converge dynamics with a long memory time, while the EFFH method is best-suited to deal with a highly peaked bath. 
The methods are thus complementary in this respect.

In Fig. \ref{fig:short_time}, we focus on the short time dynamics of Fig. \ref{fig:poplz}. A notable observation is that early rapid dynamics is missing in the effective Hamiltonian dynamics, compared to the (more accurate) RC predictions. This is because the effective Hamiltonian is constructed such that it excludes non-Markovian effects, which the RC method captures \cite{NickNM}. 
Interestingly, this distinct early dynamics is partially responsible for the EFFH dynamics deviating from RC results throughout the whole evolution. 

In Appendix \ref{app:4}, we also investigate the dynamics of coherences while varying the couplings $\lambda_z$ and $\lambda_x$. In this case, the equations of motion demonstrate that both environments, $x$ and $z$, contribute to the decoherence dynamics, and even at weak coupling. Numerically, we observe that increasing either $\lambda_z$ or $\lambda_z$ typically results in an elevated decoherence rate. Thus, while the counter-intuitive impact of decoherence on the relaxation rate can be readily observed, the complementary phenomenon, of reduced decoherence due to enhanced relaxation, does not seem to materialize, or at least it requires specific conditions to show up.

\begin{figure*}[htbp]
\begin{centering}
\includegraphics[width=1.0\textwidth]{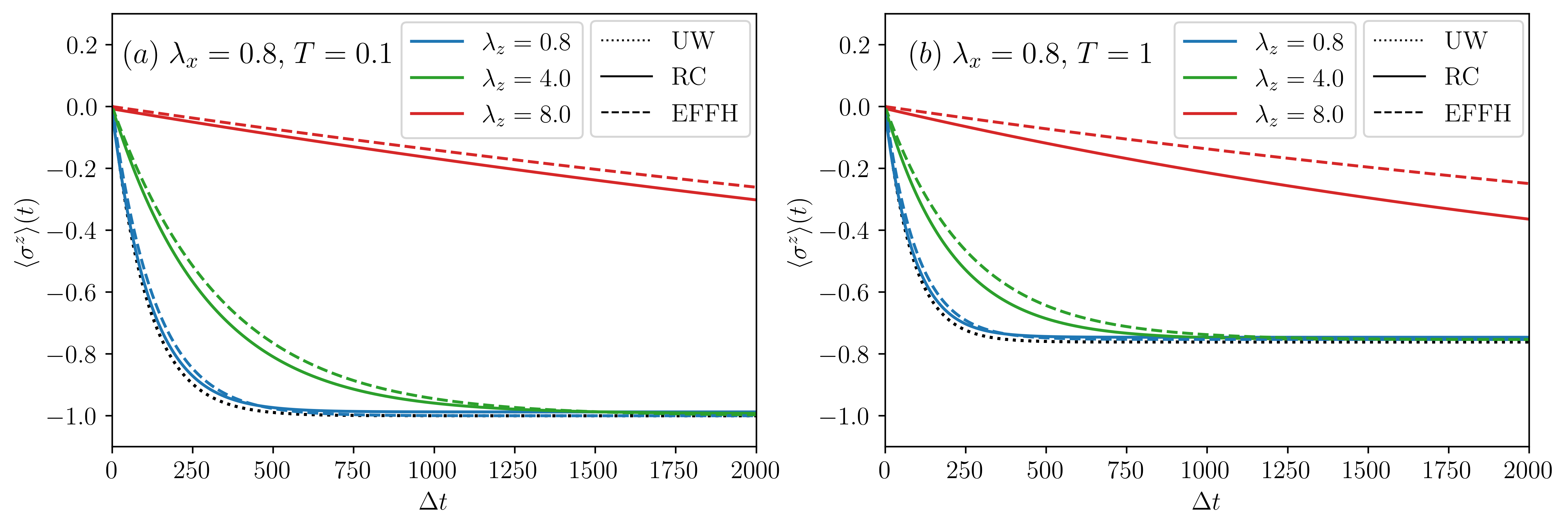} 
\par\end{centering}
\caption{\label{fig:popT}
Demonstration of the suppression of relaxation by decoheerence effect at (a) low and (b) high temperatures.
We present the dynamics of polarization $\langle \sigma^z \rangle$ from an initial state $\rho(0)=|\psi\rangle\langle\psi|$ for $|\psi\rangle=\frac{1}{\sqrt{2}}(|0\rangle+|1\rangle)$. We use a fixed dissipative coupling strength, $\lambda_x = 0.8$, but the different colored lines correspond to different values of the decoherring bath coupling, 
$\lambda_{z}$. 
(a) $T=0.1$, (b) $T=1$, where solid and dashed lines correspond to RC and the effective Hamiltonian simulations, 
respectively. The black dotted curve depicts the ultraweak coupling limit, which is independent of $\lambda_z$.
Parameters are set to $\Delta=1$, $\Omega=8$, $\gamma=0.05/\pi$, $\Lambda=1000$ with
energy parameters defined relative to $\Delta$.
}
\end{figure*}

\begin{figure*}[htbp]
\begin{centering}
\includegraphics[width=1.0\textwidth]{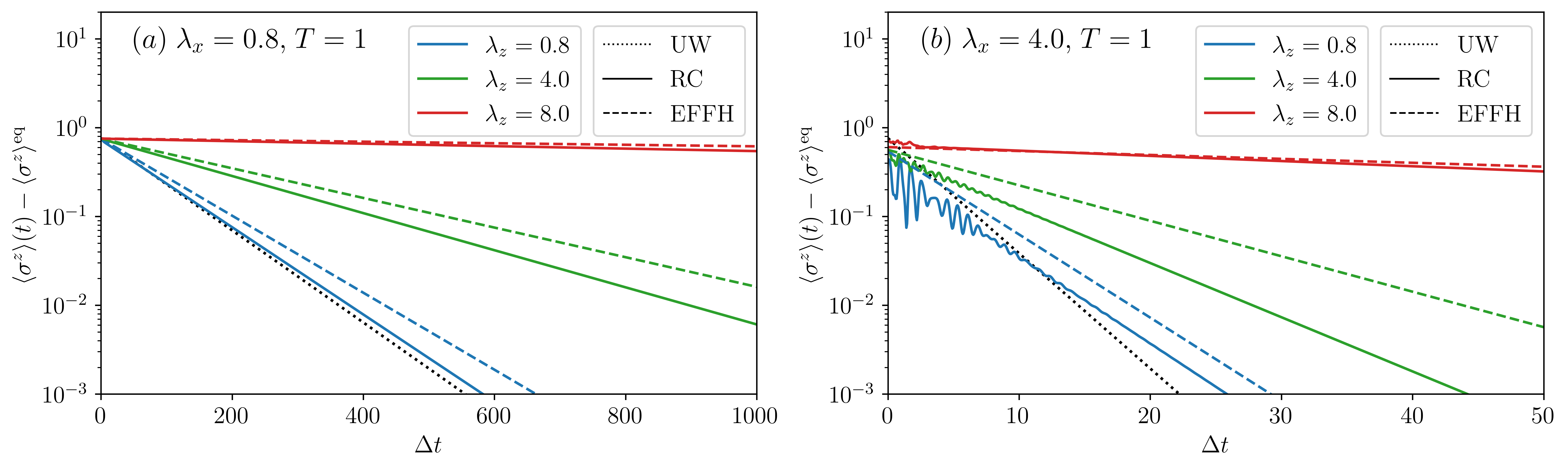} 
\par\end{centering}
\caption{
Demonstration of the suppression of relaxation by decoheerence effect at  (a) weak and (b) intermdiate dissipation strength.
We present the dynamics of polarization $\langle \sigma^z \rangle(t)$ relative to the equilibrium value,
starting with the initial condition
$\rho(0)=|\psi\rangle\langle\psi|$ with $|\psi\rangle=\frac{1}{\sqrt{2}}(|0\rangle+|1\rangle)$
for $\lambda_x=0.8$ (a) and $\lambda_x=4$ (b). The plotted value of the polarization is shifted by its equilibrium value $\langle \sigma^z \rangle^\text{eq}$ calculated from equation (\ref{eq:eqpol}).
The different colors correspond to different values of the decoherring bath coupling, $\lambda_{z}$. We compare 
RC (full) to EFFH (dashed) and ultraweak coupling (dotted) simulations. In the UW case, the polarization dynamics does not depend on $\lambda_z$.
Parameters are set to $\Delta=1$, $\Omega=8$, $\gamma=0.05/\pi$, $\Lambda=1000$, $T=1$. The coupling strengths, RC frequencies and the temperature are in units of $\Delta$.}
\label{fig:poplz}
\end{figure*}

\begin{figure*}[htbp]
\begin{centering}
\includegraphics[width=1.0\textwidth]{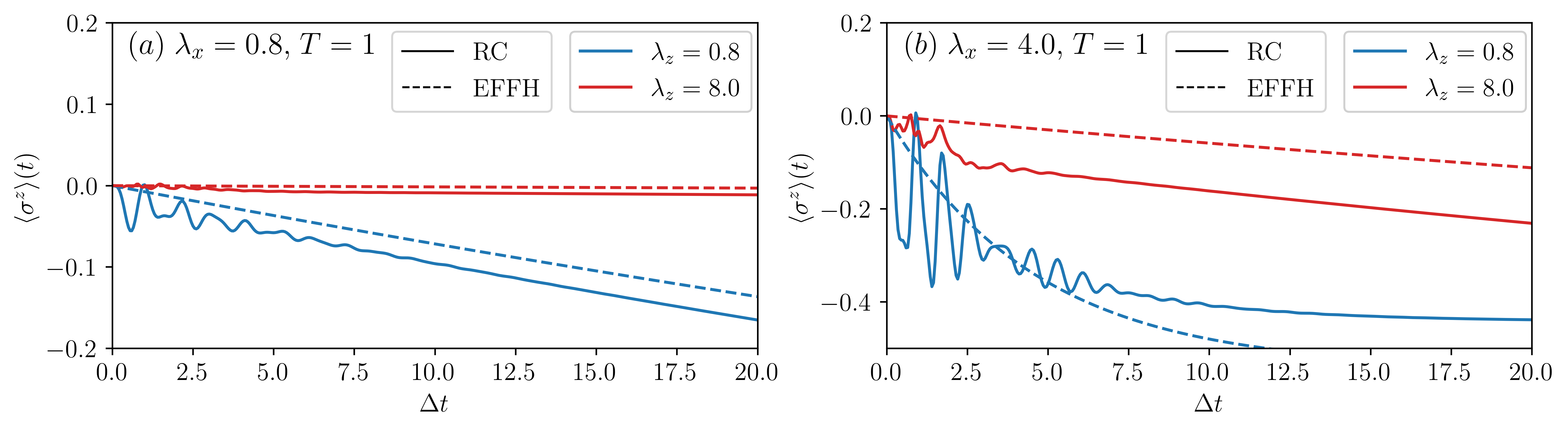} 
\par\end{centering}
\caption{
Short time dynamics of polarization $\langle \sigma^z \rangle(t)$. Simulations are the same as in Fig. \ref{fig:poplz} (omitting the $\lambda_z=4$ case, for clarity), but zooming over the short-time dynamics. The figure highlights shot time features that are resolved in the RC simulations, but not in the EFFH method, and their effect on the evolution.}
\label{fig:short_time}
\end{figure*}

\section{Bath-engineering the Kitaev XY spin chain} 
\label{sec:lattice}



We continue with the demonstration of novel behavior of systems coupled to their environments through noncommuting system's operators. While Sec. \ref{sec:TLS}  concerned with an ``impurity'' problem involving a single qubit embedded in independent environments, the present section focuses on an ensemble of spins interacting with a set of baths.
By adopting the EFFH mapping and creating an effective Hamiltonian, we exemplify the powerful idea of bath-engineering many-body lattices, previously demonstrated on variety of models
\citenum{min2024bathengineering,chain21,PhysRevB.105.L180407,weber2023quantumspinchainsbond,PhysRevLett.113.260403,PhysRevLett.129.056402,SSH}. 
Specifically, here our objective is to create the Kitaev XY spin chain \cite{Kitaev_2001,Kitaev_2006,xu2024manipulating,chain23,chain22} by designing specific system-bath couplings. 
The Kitaev XY spin chain is interesting for its potential applications in quantum computing \cite{xu2024manipulating} and in describing dynamics in real materials \cite{Morris_2021}. 
Experimentally, there have been efforts to realize the model and its variants in an optomechanical setup \cite{Slim_2024}, in solid state nanowires \cite{doi:10.1126/science.1222360}, and with Josephson junctions \cite{ren_topological_2019}. Here, rather than searching a material that realizes the Kitaev model directly, we show how to build it through the method of bath engineering. 

Our starting point is a one-dimensional spin chain with $N$ spins coupled to $N$ baths in the curious way depicted in Fig. \ref{fig:spin_chain}. The model is written in the form of Eq. (\ref{eq:H_original}),
\bea
\hat H= \hat H_S + \sum_{n=1}^N\hat H_{B,n} + \sum_{n=1}^N \hat S_n \otimes \hat B_n.
\eea
The system Hamiltonian includes an ensemble of spins, which do not interact directly. 
Each bath (index $n$) couples to an operator of the system, $\hat S_n$, which conjoins two neighboring spins,
\begin{align}
\hat{H}_{S} & =\sum_{n=1}^{N}\Delta_{n} \sigma_{n}^{z},\;\;\;\hat{S}_{n}=\begin{cases}
\sigma_{n}^{x}+\sigma_{n+1}^{x} & n\;\text{odd}\\
\sigma_{n}^{y}+\sigma_{n+1}^{y} & n\;\text{even}
\end{cases}.
\label{eq:Hs_chain}
\end{align}
As for the system-bath interaction, we use the standard form, $\hat B_n=\sum_{k} t_{n,k}(\hat c_{n,k}^{\dagger}+\hat c_{n,k})$, with the baths comprising collections of oscillators, enumerated by the index $k$, $\hat H_{B,n} = \sum_k \nu_{n,k}\hat c_{n,k}^{\dagger} \hat c_{n,k}$.
For simplicity, we assume periodic boundary conditions on the spins,
$\sigma_{N+1}^{\alpha}=\sigma_{1}^{\alpha}$.
Additionally, we take the number of spins $N$ to be even.  

In more details, the baths interact with the spins in an alternating fashion, see Fig. \ref{fig:spin_chain}: The $n$th bath interacts with two neighboring spins ($n$ and $n+1$) via their $\sigma^{y}$ operators. The $n+1$ bath interacts with spins $n+1$ and  $n+2$ via their $\sigma^{x}$ operators, and so on. Overall, $N$ baths interact with $N$ spins via alternating $\sigma^{x}$ and $\sigma^{y}$ operators, with each bath coupled to two neighboring spins. We index both the spins and baths from $1$ to $N$.  
For demonstration purposes,
we set all system-bath coupling strengths to be identical ($\lambda$) and assume that the baths are characterized by modes with identical frequencies ($\Omega$).

After the EFFH procedure, we end up with an effective Hamiltonian, Eq. (\ref{eq:Heff_UP}). The system part of it is given by
\begin{align}
\hat{H}_{S}^{\text{eff}}  =\left[ 2\kappa(\lambda) -1 \right]\sum_{n=1}^{N}\Delta_{n} \sigma_{n}^{z} -4\frac{\lambda^{2}}{\Omega} \sum_{n=1}^{N} \left(\hat{S}_{n}^{2}\right)^{\text{eff}}.
\label{eq:chain_Hseff}
\end{align}
We notice two effects derived from the environments: First, the spin splittings are dressed by the factor $2\kappa(\lambda)-1$. Second, new bond dependent two-body interaction terms are generated (Appendix \ref{app:3}), 
\bea
\sum_{n=1}^{N}\Big(\hat{S}_{n}^{2}\Big)^{\text{eff}} =\sum_{n\;\text{odd}}\xi(\lambda)\sigma_{n}^{x}\sigma_{n+1}^{x}  
+\sum_{n\;\text{even}}\xi(\lambda)\sigma_{n}^{y}\sigma_{n+1}^{y}.
\nonumber \\
\label{eq:Ssquared_chain}
\eea
Here, $\xi(\lambda)$ is a real-valued function. Thus, though we started with spins that were not directly coupled, the specially-designed couplings to the environments---through noncommuting operators of the spins---results in a system Hamiltonian with interaction terms parallel to those in the Kitaev XY spin chain. The interaction terms in Eq. (\ref{eq:chain_Hseff}) are bath-mediated, and they vanish in the $\lambda\to 0$ limit.  

We also derive expressions for the effective system operators that couple to the baths, Eq. (\ref{eq:Heff_UP}),
\begin{equation}
\hat{S}_{n}^{\text{eff}}=\begin{cases}
\kappa\left(\lambda\right)\left(\sigma_{n}^{y}+\sigma_{n+1}^{y}\right) & n\;\text{even}\\
\kappa\left(\lambda\right)\left(\sigma_{n}^{x}+\sigma_{n+1}^{x}\right) & n\;\text{odd}
\end{cases}.\label{eq:chain_Seff}
\end{equation}
Here, we retrieved the same coupling operators as we started with, but with a suppressed coupling strength.

We find that the effective model depends on three functions of $\lambda$, written explicitly in Appendix \ref{app:3}: the dressing function of the spins splitting $(2\kappa(\lambda)-1)$, the bath-generated spin interaction $\xi(\lambda)$, and the dressing function of the system-bath coupling operators, $\kappa(\lambda)$. 

\begin{figure}[htbp]
\begin{centering}
\includegraphics[width=0.45\textwidth]{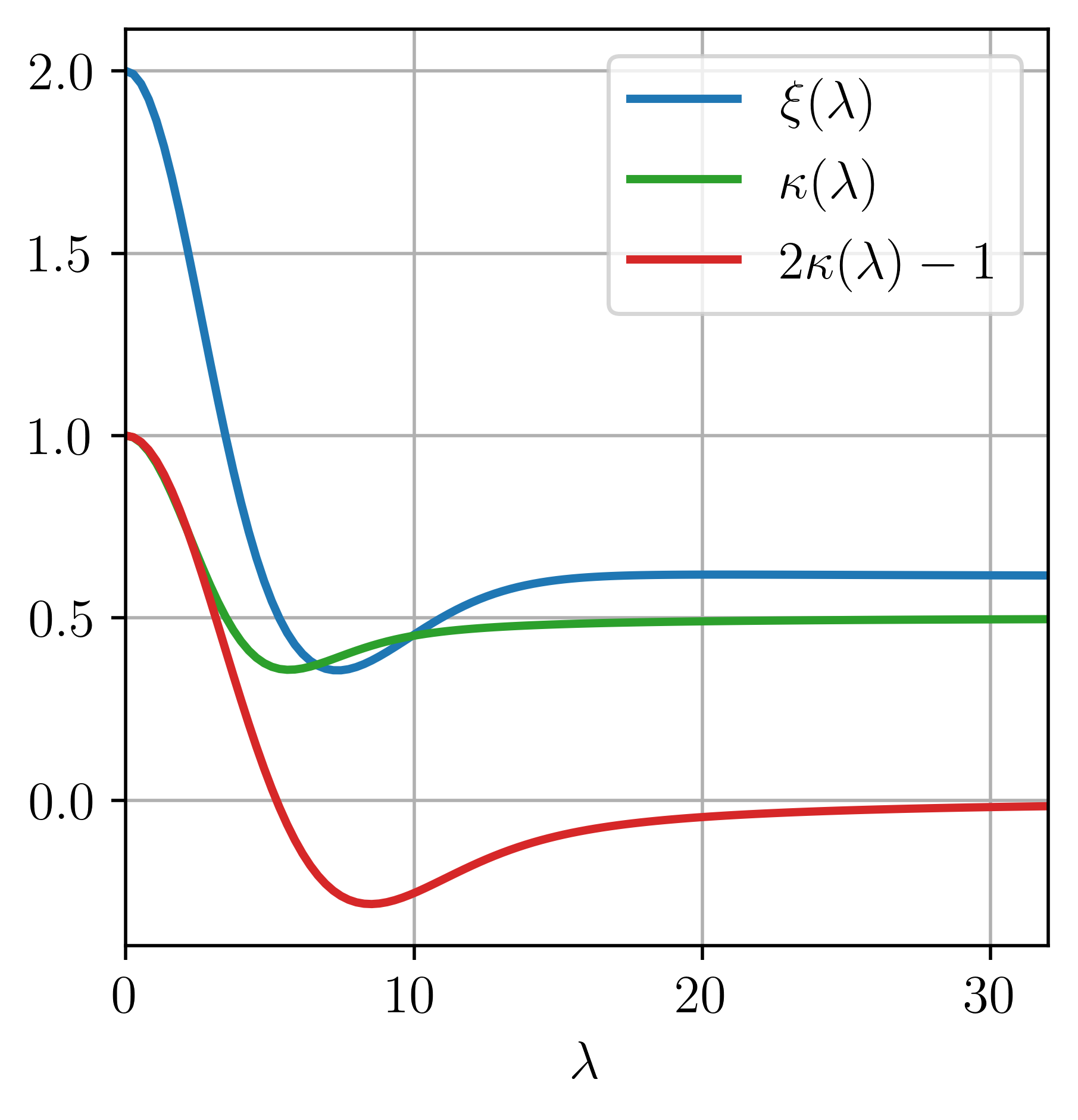} 
\par\end{centering}
\caption{\label{fig:xi_chain}
Functions defining the Kitaev XY model, Eqs. (\ref{eq:chain_Hseff})-(\ref{eq:chain_Seff}), through bath-engineering.
The functions are presented against $\lambda$.
$\xi(\lambda)$ (blue) builds the bath-induced two-body bond-dependent interaction terms. $\kappa(\lambda)$ (green) dresses the effective coupling operators. $2\kappa(\lambda)-1$  (red) dresses the splitting of each spin. }
\end{figure}

We evaluate these functions numerically and  present them in Figure \ref{fig:xi_chain}.
In the ultraweak coupling limit, we retrieve the noninteracting model. The interaction terms in the system Hamiltonian are proportional to $ \frac{\lambda^2}{\Omega}\xi\left( \lambda \right) $, going to zero as $\lambda\rightarrow 0$. 
In contrast, in the ultrastrong coupling limit, spin splittings get suppressed to zero, and two-body terms grow like $\frac{\lambda^2}{\Omega}$ times a factor $\lim_{\lambda\rightarrow\infty} \xi\left( \lambda\right) $, which
approaches a constant $\approx 0.62$. 
The latter regime corresponds to the XY Kitaev spin chain model.

It is fair to note that what we did here was to transfer the difficulty of creating directly the Kitaev XY spin chain based on spin-spin interactions to the challenge of creating a specific type of system-bath interactions, which are designed to imprint the required bond dependent interactions into the chain. Indeed, the take-away message of this discussion is not that we necessarily suggest here a simple route to engineer complex materials, rather that specifically-tailored system-bath interactions can mediate and create novel quantum materials.
In experimental realizations, the spin chain may be subjected to a transversal magnetic field \cite{Morris_2021}. This can  be feasibly treated with the effective mapping, since it only introduces new one-body terms in the system Hamiltonian, and the mapping itself can be applied to all terms in the Hamiltonian separately. 
In this case, the transverse magnetic field component wold be dressed with functions of the coupling parameters, as in equation (\ref{eq:chain_Hseff}).



\section{Conclusions}
\label{sec:summary}

In this work, we tackled the problem of open quantum systems that are coupled to multiple environments through noncommuting system operators, and at arbitrary system-bath coupling. The method that we introduced here generalized Ref. 
\cite{Nick_PRX}. 
The effective Hamiltonian that the method builds reflects both the renormalization of the system's energy parameters due to the strong couplings to baths, and the cooperative effects of the baths due to the noncommutativity of coupling operators.  
The effective system Hamiltonian only weakly couples to the residual environments, allowing us to study the system's dynamics and equilibrium properties using low-cost computational methods. 

The principal application considered in this work was of a qubit coupled to two environments, which individually were responsible for thermal relaxation and pure decoherence. The derived formula for the effective Hamiltonian exposed the effect of suppression of relaxation by decoherence, a phenomenon that was previously observed numerically without a theoretical underpinning. Simulations using the reaction coordinate technique demonstrated that the effective Hamiltonian provided qualitative correct predictions.  
As a second example, we applied our generalized EFFH theoretical framework to a spin system, where engineered bath interactions induce bond-dependent interactions as in the Kitaev XY spin chain. 
It is interesting to continue this line of work and use thermal baths to engineer e.g., the Kitaev Honeycomb lattice.  

The next technical stage for this study includes further analysis of the commutator algebra \cite{wyss2017noncommutative,doubledHilberSpace} to allow feasible analytical studies  of higher dimensional systems, beyond spins. Furthermore, beyond the analysis of the phases of magnetic materials of high spin and higher spatial dimensions, prospects include the construction of genuine quantum effects in quantum thermal machines, emerging due to the combination of noncommutativity, strong system-bath coupling, and nonequilibrium conditions. While several numerical studies addressed such setups \cite{Gelbwaser15, 
Cao_unusual_transport, Felix22,Dahai24}, we hope that our analytical tool will enable a rational approach to quantum thermal machine design. 

Inhibition of relaxation effects, or more generally, bath-mediated dynamics, by coupling the system to an additional heat bath could see varied applications in quantum information processing, quantum reaction dynamics, quantum state engineering and quantum metrology. Future work will be focused on testing and applying our method to such applications. 


\begin{acknowledgements}
We acknowledge discussions with Brett Min, Nicholas Anto-Sztrikacs and Marlon Brenes.
The work of JG was supported by the QUARMEN Erasmus Mundus Program and
the University of Toronto. DS acknowledges support from an NSERC Discovery Grant and the Canada Research Chair program.
This research has been funded in part or completely by the research project: "Quantum Software Consortium: Exploring Distributed Quantum Solutions for Canada" (QSC). QSC is financed under the National Sciences and Engineering Research Council of Canada (NSERC) Alliance Consortia Quantum Grants \#ALLRP587590-23.
\end{acknowledgements}

\begin{widetext}


\appendix
\section{Useful identities}
\renewcommand{\theequation}{A\arabic{equation}}
\setcounter{equation}{0}  
\label{app:1}

In this Appendix, we provide identities that are used in derivations throughout this manuscript. 
Let us denote by $\text{Com}^{\ocircle n}(\bullet,B)$
the $n$ times composition of the function $[\bullet,B]$. For example,
$\text{Com}^{\ocircle1}(A,B)=[A,B]$, 
$\text{Com}^{\ocircle2}(A,B)=[[A,B],B]$ and 
$\text{Com}^{\ocircle3}(A,B)=[[[A,B],B],B]$. 
Let us consider two generally noncommuting operators $A$ and $B$. It can be shown that
%
\begin{equation}
\begin{aligned}\sum_{r=1}^{k}B^{r-1}[A,B]B^{k-r} & =\sum_{r=1}^{k}B^{r-1}AB^{k-(r-1)}-B^{r}AB^{k-r}=\\
 & =\sum_{r=0}^{k-1}B^{r}AB^{k-r}-\sum_{r=1}^{k}B^{r}AB^{k-r}=\\
 & =[A,B^{k}].
\end{aligned}
\end{equation}
This relation can be used to derive
\begin{equation}
\begin{aligned}[][A,B]B^{n-r} & =B^{n-r}[A,B]+[[A,B],B^{n-r}]\\
 & =B^{n-r}[A,B]+\sum_{r'=1}^{n-r}B^{r'-1}[[A,B],B]B^{n-r-r'}.
\end{aligned}
\end{equation}
Similarly,
\begin{equation}
\begin{aligned}B^{r-1}[A,B] & =[A,B]B^{r-1}-[[A,B],B^{r-1}]\\
 & =[A,B]B^{r-1}-\sum_{r'=1}^{r-1}B^{r'-1}[[A,B],B]B^{r-1-r'}.
\end{aligned}
\end{equation}
We continue by using the above identities to prove the following relations
\begin{equation}
\begin{aligned}AB^{n} & =B^{n}A+[A,B^{n}]\\
 & =B^{n}A+\sum_{r=1}^{n}B^{r-1}[A,B]B^{n-r}\\
 & =B^{n}A+\sum_{r=1}^{n}B^{r-1}\Big(B^{n-r}[A,B]+\sum_{r'=1}^{n-r}B^{r'-1}[[A,B],B]B^{n-r-r'}\Big)\\
 & =B^{n}A+nB^{n-1}[A,B]+\sum_{r=1}^{n}\sum_{r'=1}^{n-r}B^{r+r'-2}[[A,B],B]B^{n-r-r'}\\
 & =B^{n}A+nB^{n-1}[A,B]+\sum_{r=1}^{n}\sum_{r'=1}^{n-r}B^{r+r'-2}\Big(B^{n-r-r'}[[A,B],B]+\sum_{r''=1}^{n-r-r'}B^{r''-1}[[[A,B],B],B]B^{n-r-r'-r''}\Big)\\
 & =B^{n}A+nB^{n-1}[A,B]+\frac{n(n-1)}{2}B^{n-2}[[A,B],B]+\sum_{r=1}^{n}\sum_{r'=1}^{n-r}\sum_{r''=1}^{n-r-r'}B^{r''-1}[[[A,B],B],B]B^{n-r-r'-r''}\\
 & =\sum_{j=0}^{n}\begin{pmatrix}n\\
j
\end{pmatrix}B^{n-j}\text{Com}^{\ocircle j}(A,B).
\end{aligned}
\label{eq:Com_identity}
\end{equation}
If we wish to commute $B^{n}$ past $A$ in the other direction,
\begin{equation}
\begin{aligned}B^{n}A & =AB^{n}-\sum_{r=1}^{n}\Big(B^{r-1}[A,B]\Big)B^{n-r}\\
 & =AB^{n}-\sum_{r=1}^{n}\Big([A,B]B^{r-1}-\sum_{r'=1}^{r-1}B^{r'-1}[[A,B],B]B^{r-1-r'}\Big)B^{n-r}\\
 & =AB^{n}-n[A,B]B^{n-1}+\sum_{r=1}^{n}\sum_{r'=1}^{r-1}B^{r'-1}[[A,B],B]B^{n-1-r'}\\
 & =\sum_{j=0}^{n}(-1)^{j}\begin{pmatrix}n\\
j
\end{pmatrix}\text{Com}^{\ocircle j}(A,B)B^{n-j}.
\end{aligned}
\end{equation}
As a side comment, we also note that 
\begin{equation}
\text{Com}^{\ocircle n}(AB,C)=\sum_{k=0}^{n}\frac{n!}{(n-k)!k!}\text{Com}^{\ocircle k}(A,C)\text{Com}^{\ocircle n-k}(B,C),
\end{equation}
which resembles the standard binomial theorem for a regular algebra with multiplication \cite{wyss2017noncommutative}.


\section{Derivation of the Effective Hamiltonian}
\renewcommand{\theequation}{B\arabic{equation}}
\setcounter{equation}{0}  
\label{app:2}

In this Appendix, we derive the general form of the effective Hamiltonian.
We start by writing the reaction coordinate Hamiltonian
\begin{equation}
\begin{aligned}\hat{H}_{{\rm RC}} & =\hat{H}_{S}+\sum_{n}\Big(\lambda_{n}\hat{S}_{n}(\hat{a}_{n}^{\dagger}+\hat{a}_{n})+\Omega_{n}\hat{a}_{n}^{\dagger}\hat{a}_{n}\Big)\\
 & +\sum_{n}\tilde{\sum_{k}}\Big(f_{n,k}(\hat{a}_{n}^{\dagger}+\hat{a}_{n})(\hat{b}_{n,k}^{\dagger}+\hat{b}_{n,k})+\omega_{n,k}\hat{b}_{n,k}^{\dagger}\hat{b}_{n,k}\Big).
\end{aligned}
\end{equation}
We now apply the non-factorized polaron transformation, followed by a
projection onto the ground state of both reaction coordinate spaces
and get
\begin{equation}
\begin{aligned}\hat{H}^{\text{eff}} & =\langle0|\hat{U}_{P}\hat{H}_{{\rm S}}\hat{U}_{P}^{\dagger}|0\rangle+\sum_{n}\Big(\lambda_{n}\langle0|\hat{U}_{P}\hat{S}_{n}(\hat{a}_{n}^{\dagger}+\hat{a}_{n})\hat{U}_{P}^{\dagger}|0\rangle+\Omega_{n}\langle0|\hat{U}_{P}\hat{a}_{n}^{\dagger}\hat{a}_{n}\hat{U}_{P}^{\dagger}|0\rangle\Big)\\
 & +\sum_{n}\tilde{\sum_{k}}\Big(f_{n,k}\langle0|\hat{U}_{P}(\hat{a}_{n}^{\dagger}+\hat{a}_{n})\hat{U}_{P}^{\dagger}|0\rangle(\hat{b}_{n,k}^{\dagger}+\hat{b}_{n,k})+\omega_{n,k}\hat{b}_{n,k}^{\dagger}\hat{b}_{n,k}\Big).
\end{aligned}
\end{equation}
Next, we express the effective Hamiltonian in the momentum representation.
For each of the reaction coordinates we have
\begin{equation}
\begin{aligned}\hat{a}^{\dagger}-\hat{a} & =-i\sqrt{\frac{2}{m\Omega}}\hat{p}=-i\sqrt{2}\tilde{p},\\
\hat{a}^{\dagger}+\hat{a} & =\sqrt{2m\Omega}\hat{x}=\tilde{x}=\sqrt{2}i\frac{\partial}{\partial\tilde{p}}.
\end{aligned}
\end{equation}
We also express the number operator $\hat{a}^{\dagger}\hat{a}$ in
this representation, 
\begin{equation}
\begin{aligned}
\hat{a}^{\dagger}\hat{a} & =\frac{1}{2}(\tilde{p}^{2}-\frac{\partial^{2}}{\partial\tilde{p}^{2}}-1).
\end{aligned}
\end{equation}
We express each of the terms in $\hat{H}^{\text{eff}}$ of the form
$\langle0|\hat{U}_{P}\cdots\hat{U}_{P}^{\dagger}|0\rangle$ as an
integral over momenta, corresponding to all reaction coordinates,
\begin{equation}
\langle0|\hat{U}_{P}\hat{H}_{{\rm S}}\hat{U}_{P}^{\dagger}|0\rangle=\int\prod_{n}d\tilde{p_{n}}\frac{e^{-\tilde{p}_{n}^{2}/2}}{\pi^{1/4}}\hat{U}_{P}\hat{H}_{{\rm S}}\hat{U}_{P}^{\dagger}\prod_{n}\frac{e^{-\tilde{p}_{n}^{2}/2}}{\pi^{1/4}}.
\end{equation}
The two factors $\frac{e^{-\tilde{p}_{n}^{2}/2}}{\pi^{1/4}}$ could
be combined in the above expression, but we keep them separated to
indicate that the transformed operator might be a derivative acting
on this factor. Proceeding to other terms we get
\begin{equation}
\begin{aligned}\langle0|\hat{U}_{P}(\hat{a}_{l}^{\dagger}+\hat{a}_{l})\hat{U}_{P}^{\dagger}|0\rangle & =i\sqrt{2}\int\prod_{n}d\tilde{p_{n}}\frac{e^{-\tilde{p}_{n}^{2}/2}}{\pi^{1/4}}\hat{U}_{P}\frac{\partial}{\partial\tilde{p}_{l}}\Big(\hat{U}_{P}^{\dagger}\prod_{n}\frac{e^{-\tilde{p}_{n}^{2}/2}}{\pi^{1/4}}\Big)\\
 & =i\sqrt{2}\int\prod_{n}d\tilde{p_{n}}\frac{e^{-\tilde{p}_{n}^{2}/2}}{\pi^{1/4}}\hat{U}_{P}\Big(\frac{\partial}{\partial\tilde{p}_{l}}(\hat{U}_{P}^{\dagger})\prod_{n}\frac{e^{-\tilde{p}_{n}^{2}/2}}{\pi^{1/4}}+\hat{U}_{P}^{\dagger}\frac{\partial}{\partial\tilde{p}_{l}}(\prod_{n}\frac{e^{-\tilde{p}_{n}^{2}/2}}{\pi^{1/4}})\Big)\\
 & =i\sqrt{2}\int\prod_{n}d\tilde{p_{n}}\frac{e^{-\tilde{p}_{n}^{2}/2}}{\pi^{1/4}}\hat{U}_{P}\Big(\frac{\partial}{\partial\tilde{p}_{l}}(\hat{U}_{P}^{\dagger})\prod_{n}\frac{e^{-\tilde{p}_{n}^{2}/2}}{\pi^{1/4}}-\hat{U}_{P}^{\dagger}\tilde{p}_{l}\prod_{n}\frac{e^{-\tilde{p}_{n}^{2}/2}}{\pi^{1/4}}\Big)\\
 & =-\frac{2\lambda_{l}}{\Omega_{l}}\int\prod_{n}d\tilde{p_{n}}\frac{e^{-\tilde{p}_{n}^{2}/2}}{\pi^{1/4}}\hat{U}_{P}\hat{S}_{l}\hat{U}_{P}^{\dagger}\prod_{n}\frac{e^{-\tilde{p}_{n}^{2}/2}}{\pi^{1/4}},
\end{aligned}
\end{equation}
where we used the fact that
\begin{equation}
\frac{\partial}{\partial\tilde{p}_{l}}\hat{U}_{P}^{\dagger}=i\sqrt{2}\frac{\lambda_{l}}{\Omega_{l}}\hat{S}_{l}\hat{U}_{P}^{\dagger}.
\end{equation}
For the term coupling the  system and RC space we get
\begin{equation}
\begin{aligned}\langle0|\hat{U}_{P}\hat{S}_{l}(\hat{a}_{l}^{\dagger}+\hat{a}_{l})\hat{U}_{P}^{\dagger}|0\rangle & =i\sqrt{2}\int\prod_{n}d\tilde{p_{n}}\frac{e^{-\tilde{p}_{n}^{2}/2}}{\pi^{1/4}}\hat{U}_{P}(\hat{S}_{l}\frac{\partial}{\partial\tilde{p}_{l}})\Big(\hat{U}_{P}^{\dagger}\prod_{n}\frac{e^{-\tilde{p}_{n}^{2}/2}}{\pi^{1/4}}\Big)\\
 & =-\frac{2\lambda_{l}}{\Omega_{l}}\int\prod_{n}d\tilde{p_{n}}\frac{e^{-\tilde{p}_{n}^{2}/2}}{\pi^{1/4}}\hat{U}_{P}\hat{S}_{l}^{2}\hat{U}_{P}^{\dagger}\prod_{n}\frac{e^{-\tilde{p}_{n}^{2}/2}}{\pi^{1/4}}.
\end{aligned}
\end{equation}
The term with the number operator becomes
\begin{equation}
\begin{aligned}\langle0|\hat{U}_{P}\hat{a}_{l}^{\dagger}\hat{a}_{l}\hat{U}_{P}^{\dagger}|0\rangle & =\frac{1}{2}\int\prod_{n}d\tilde{p_{n}}\frac{e^{-\tilde{p}_{n}^{2}/2}}{\pi^{1/4}}\hat{U}_{P}(\tilde{p}_{l}^{2}-\frac{\partial^{2}}{\partial\tilde{p}_{l}^{2}}-1)\Big(\hat{U}_{P}^{\dagger}\prod_{n}\frac{e^{-\tilde{p}_{n}^{2}/2}}{\pi^{1/4}}\Big)\\
 & =\int\prod_{n}d\tilde{p_{n}}\frac{e^{-\tilde{p}_{n}^{2}/2}}{\pi^{1/4}}\frac{\tilde{p}_{l}^{2}-1}{2}\prod_{n}\frac{e^{-\tilde{p}_{n}^{2}/2}}{\pi^{1/4}}\\
 & -\frac{1}{2}\int\prod_{n}d\tilde{p_{n}}\frac{e^{-\tilde{p}_{n}^{2}/2}}{\pi^{1/4}}\hat{U}_{P}\frac{\partial}{\partial\tilde{p}_{l}}\Big(i\sqrt{2}\frac{\lambda_{l}}{\Omega_{l}}\hat{S}_{l}\hat{U}_{P}^{\dagger}\prod_{n}\frac{e^{-\tilde{p}_{n}^{2}/2}}{\pi^{1/4}}-\hat{U}_{P}^{\dagger}\tilde{p}_{l}\prod_{n}\frac{e^{-\tilde{p}_{n}^{2}/2}}{\pi^{1/4}}\Big)\\
 & =\int\prod_{n}d\tilde{p_{n}}\frac{e^{-\tilde{p}_{n}^{2}/2}}{\pi^{1/4}}\frac{\tilde{p}_{l}^{2}-1}{2}\prod_{n}\frac{e^{-\tilde{p}_{n}^{2}/2}}{\pi^{1/4}}\\
 & -i\frac{\sqrt{2}}{2}\frac{\lambda_{l}}{\Omega_{l}}\int\prod_{n}d\tilde{p_{n}}\frac{e^{-\tilde{p}_{n}^{2}/2}}{\pi^{1/4}}\hat{U}_{P}\Big(\frac{\partial}{\partial\tilde{p}_{l}}(\hat{S}_{l}\hat{U}_{P}^{\dagger})\prod_{n}\frac{e^{-\tilde{p}_{n}^{2}/2}}{\pi^{1/4}}-\hat{S}_{l}\hat{U}_{P}^{\dagger}\tilde{p}_{l}\prod_{n}\frac{e^{-\tilde{p}_{n}^{2}/2}}{\pi^{1/4}}\Big)\\
 & +\frac{1}{2}\int\prod_{n}d\tilde{p_{n}}\frac{e^{-\tilde{p}_{n}^{2}/2}}{\pi^{1/4}}\hat{U}_{P}\frac{\partial}{\partial\tilde{p}_{l}}\Big(\hat{U}_{P}^{\dagger}\tilde{p}_{l}\prod_{n}\frac{e^{-\tilde{p}_{n}^{2}/2}}{\pi^{1/4}}\Big)\\
 & =\int\prod_{n}d\tilde{p_{n}}\frac{e^{-\tilde{p}_{n}^{2}/2}}{\pi^{1/4}}\frac{\tilde{p}_{l}^{2}-1}{2}\prod_{n}\frac{e^{-\tilde{p}_{n}^{2}/2}}{\pi^{1/4}}\\
 & -i\frac{\sqrt{2}}{2}\frac{\lambda_{l}}{\Omega_{l}}\int\prod_{n}d\tilde{p_{n}}\frac{e^{-\tilde{p}_{n}^{2}/2}}{\pi^{1/4}}\hat{U}_{P}\Big(i\sqrt{2}\frac{\lambda_{l}}{\Omega_{l}}\hat{S}_{l}^{2}\hat{U}_{P}^{\dagger}\prod_{n}\frac{e^{-\tilde{p}_{n}^{2}/2}}{\pi^{1/4}}-\hat{S}_{l}\hat{U}_{P}^{\dagger}\tilde{p}_{l}\prod_{n}\frac{e^{-\tilde{p}_{n}^{2}/2}}{\pi^{1/4}}\Big)\\
 & +\frac{1}{2}\int\prod_{n}d\tilde{p_{n}}\frac{e^{-\tilde{p}_{n}^{2}/2}}{\pi^{1/4}}\hat{U}_{P}\Big(i\sqrt{2}\frac{\lambda_{l}}{\Omega_{l}}\hat{S}_{l}\hat{U}_{P}^{\dagger}\tilde{p}_{l}\prod_{n}\frac{e^{-\tilde{p}_{n}^{2}/2}}{\pi^{1/4}}+\hat{U}_{P}^{\dagger}(1-\tilde{p}_{l}^{2})\prod_{n}\frac{e^{-\tilde{p}_{n}^{2}/2}}{\pi^{1/4}}\Big)\\
 & =\frac{\lambda_{l}^{2}}{\Omega_{l}^{2}}\int\prod_{n}d\tilde{p_{n}}\frac{e^{-\tilde{p}_{n}^{2}/2}}{\pi^{1/4}}\hat{U}_{P}\hat{S}_{l}^{2}\hat{U}_{P}^{\dagger}\prod_{n}\frac{e^{-\tilde{p}_{n}^{2}/2}}{\pi^{1/4}}.
\end{aligned}
\end{equation}
Thus, up to a constant, the effective Hamiltonian is equal to
\begin{equation}
\begin{aligned}\hat{H}^{\text{eff}} & =\int\hat{U}_{P}\Big(\hat{H}_{{\rm S}}-\sum_{l}\big(\frac{\lambda_{l}^{2}}{\Omega_{l}}\hat{S}_{l}^{2}\big)\Big)\hat{U}_{P}^{\dagger}\prod_{n}\frac{e^{-\tilde{p}_{n}^{2}}}{\pi^{1/2}}d\tilde{p_{n}}\\
 & +\sum_{l}\tilde{\sum_{k}}\Big(-\frac{2f_{l,k}\lambda_{l}}{\Omega_{l}}\int\hat{U}_{P}\hat{S}_{l}\hat{U}_{P}^{\dagger}\prod_{n}\frac{e^{-\tilde{p}_{n}^{2}}}{\pi^{1/2}}d\tilde{p_{n}}(\hat{b}_{l,k}^{\dagger}+\hat{b}_{l,k})+\omega_{l,k}\hat{b}_{l,k}^{\dagger}\hat{b}_{l,k}\Big)\\
 & =\Big(\hat{H}_{{\rm S}}-\sum_{l}\frac{\lambda_{l}^{2}}{\Omega_{l}}\hat{S}_{l}^{2}+\sum_{l}\tilde{\sum_{k}}\big(\omega_{l,k}\hat{b}_{l,k}^{\dagger}\hat{b}_{l,k}-\frac{2f_{l,k}\lambda_{l}}{\Omega_{l}}\hat{S}_{l}(\hat{b}_{l,k}^{\dagger}+\hat{b}_{l,k})\big)\Big)^{\text{eff}}.
\end{aligned}
\end{equation}
Here, for an operator $\mathcal{\hat{O}}$ we define the effective counterpart as
\begin{equation}
\begin{aligned}(\mathcal{\hat{O}})^{\text{eff}} & =\int\end{aligned}
\hat{U}_{P}(\tilde{\boldsymbol{p}})\mathcal{\hat{O}}\hat{U}_{P}^{\dagger}(\tilde{\boldsymbol{p}})\prod_{n}\frac{e^{-\tilde{p}_{n}^{2}}}{\pi^{1/2}}d\tilde{p_{n}}.
\end{equation}

\color{black}
\section{Derivation of the effective Hamiltonian for a spin impurity model and a spin chain}
\renewcommand{\theequation}{C\arabic{equation}}
\setcounter{equation}{0}  
\label{app:3}
In this Appendix we perform the effective mapping of the system operators $\hat{\mathcal{O}}_S\rightarrow\hat{\mathcal{O}}_S^{\text{eff}}$ for the spin impurity and spin chain models. Our goal is to find an explicit form of the effective system Hamiltonian and bath coupling operators.

\subsection{Equal couplings to different baths}

We consider 
$\hat{H}_{S}=\Delta\sigma^{z}$, $\hat{S}_{z}=\sigma^{z}$, and $\hat{S}_{x}=\sigma^{x}$,
and assume that $\frac{\lambda_{n}}{\Omega_{n}}=\frac{\lambda}{\Omega}$
for all $n$. Using 
the identify $e^{ia\hat \alpha( \cdot \vec\sigma)}= \sigma^0\cos a + i (\hat \alpha \cdot \vec{\sigma})\sin a $, with $\hat \alpha $ a unit vector,
we simplify Eq. (\ref{eq:Upp}) and get
\bea
\hat{U}_{P}(\boldsymbol{\tilde p}) & =&
\exp\Big(-i\sqrt{2}\sum_{n=z,x}\frac{\lambda_{n}}{\Omega_{n}}{\tilde p}_{n}\hat{S}_{n}\Big)
\nonumber\\
 & =&\sigma^{0}\cos(\sqrt{2}\frac{\lambda}{\Omega}\sqrt{\tilde p_{z}^{2}+\tilde p_{x}^{2}})-\frac{i\sin(\sqrt{2}\frac{\lambda}{\Omega}\sqrt{\tilde p_{z}^{2}+\tilde p_{x}^{2}})}{\sqrt{\tilde p_{z}^{2}+\tilde p_{x}^{2}}}(\sigma^{z}\tilde p_{z}+\sigma^{x}\tilde p_{x}),
\eea
where $\sigma^{0}$ is the identity matrix.  The substitution $\tilde p_{z}\rightarrow r\cos\theta,\;\tilde p_{x}\rightarrow r\sin\theta$ greatly simplifies the above expression leading to 
\begin{equation}
\begin{aligned}\hat{U}_{P}(r,\theta) =\sigma^{0}\cos(\sqrt{2}\frac{\lambda}{\Omega} r)-i\sin(\sqrt{2}\frac{\lambda}{\Omega} r)\big( \sigma^{z} \cos\theta +\sigma^{x} \sin\theta \big).
\end{aligned}
\end{equation}
We now act on the system operators $\sigma^x$ and $\sigma^z$ with the polaron transform in cylindrical coordinates  to obtain
 \begin{equation}
\begin{aligned}\hat{U}_{P}(r,\theta)\sigma^{z}\hat{U}_{P}^{\dagger}(r,\theta) & =\sigma^{z}\left[\cos^{2}\theta+\sin^{2}\theta\cos\left(2\sqrt{2}r\frac{\lambda}{\Omega}\right)\right]\\
 & +\sigma^{x}\left[\sin\left(2\theta\right)\sin^{2}(\sqrt{2}r\frac{\lambda}{\Omega})\right]\\
 & -\sigma^{y}\left[\sin\theta\sin\left(2\sqrt{2}r\frac{\lambda}{\Omega}\right)\right],
\end{aligned}
\label{sigma_z_UP}
\end{equation}
and
\begin{equation}
\begin{aligned}\hat{U}_{P}(r,\theta)\sigma^{x}\hat{U}_{P}^{\dagger}(r,\theta) & =\sigma^{x}\left[\cos^{2}\theta\cos(2\sqrt{2}r\frac{\lambda}{\Omega})+\sin^{2}\theta\right]\\
 & +\sigma^{y}\left[\cos\theta\sin(2\sqrt{2}r\frac{\lambda}{\Omega})\right]\\
 & +\sigma^{z}\left[\sin(2\theta)\sin^{2}\left(\sqrt{2}r\frac{\lambda}{\Omega}\right)\right].
\end{aligned}
\label{sigma_x_UP}
\end{equation}

Next, we wish to integrate the expression in equation (\ref{eq:O_eff})
using the new variables.  After changing the integration measure from $d\tilde{p}_z d\tilde{p}_x$ to $rdrd\theta$, and substituting the above results, we get
\begin{equation}
\hat{\mathcal{O}}_{S}^{\text{eff}}=\int_{0}^{\infty}\int_{0}^{2\pi}\hat{U}_{P}(r,\theta)\hat{\mathcal{O}}_{S}\hat{U}_{P}^{\dagger}(r,\theta)e^{-r^{2}}\frac{r}{\pi}drd\theta.
\end{equation}
We notice that both effective Pauli matrices, $(\sigma^z)^{\text{eff}}$ and $(\sigma^x)^{\text {eff}}$, will only pick up a multiplicative factor, rather than contributions from other Pauli matrices. This is because both the second and third lines of equations (\ref{sigma_z_UP}) and (\ref{sigma_x_UP}) sum up to zero after integration over $\theta$.  As a result, we are left with the task of evaluating the first lines of those equations,
\bea
(\sigma^z)^{\text{eff}} \nonumber &=& 
\sigma^z \int_{0}^{\infty}\int_{0}^{2\pi} \left[\cos^{2}\theta+\sin^{2}\theta \cos(2\sqrt{2}r\frac{\lambda}{\Omega}) \right] e^{-r^{2}}\frac{r}{\pi}drd\theta \\ &=& \sigma^z \int_0^\infty \left[ 1 + \cos(2\sqrt{2}r\frac{\lambda}{\Omega}) \right]  e^{-r^{2}} r  dr,
\eea
and similarly
\bea
(\sigma^x)^{\text{eff}} = \sigma^x \int_0^\infty \left[ 1 + \cos( 2\sqrt{2} r \frac{\lambda}{\Omega} ) \right] e^{-r^{2}} r dr.
\eea
We collect the prefactor and define it as 
\bea
\kappa(\lambda)\equiv\int_0^\infty \left[ 1 + \cos( 2\sqrt{2} r \frac{\lambda}{\Omega} ) \right] e^{-r^{2}} r dr.
\eea
Using $\int_0^{\infty}e^{-y^2}y dy= 1/2$, we can easily verify that 
$\kappa(\lambda)\xrightarrow{\lambda\to 0} 1$ and 
$\kappa(\lambda)\xrightarrow{\lambda\to \infty} 1/2$.
We can also express the above integrals using known special functions. Let us Taylor expand the cosine function around zero in the definition of $\kappa(\lambda)$ and get
\begin{equation}
    \begin{aligned}
\kappa(\lambda)&=\int_{0}^{\infty}\left(1+\sum_{n=0}^{\infty}\frac{(-1)^{n}\left(2\sqrt{2}r\frac{\lambda}{\Omega}\right)^{2n}}{(2n)!}\right)re^{-r^{2}}dr\\&=
\frac{1}{2}+\frac{1}{2}\sum_{n=0}^{\infty}\frac{(-1)^{n}}{(2n)!}\left(2\sqrt{2}\frac{\lambda}{\Omega}\right)^{2n}n!\\&=1-\frac{\sqrt{2}\frac{\lambda}{\Omega}}{2}\sum_{n=1}^{\infty}\frac{(-1)^{n-1}}{(2n)!}n!\left(2\sqrt{2}\frac{\lambda}{\Omega}\right)^{2n}\left(\sqrt{2}\frac{\lambda}{\Omega}\right)^{-1}\\&
=1-\frac{\sqrt{2}\frac{\lambda}{\Omega}}{2}\sum_{n=0}^{\infty}\frac{(-1)^{n}2^{2n+2}}{(2n+2)!}(n+1)!\left(\sqrt{2}\frac{\lambda}{\Omega}\right)^{2n+1}\\&=1-\frac{\sqrt{2}\frac{\lambda}{\Omega}}{2}\sum_{n=0}^{\infty}\frac{(-1)^{n}2^{n+1}}{(2n+1)!!}\left(\sqrt{2}\frac{\lambda}{\Omega}\right)^{2n+1}\\&
=1-\sqrt{2}\frac{\lambda}{\Omega}\sum_{n=0}^{\infty}\frac{(-1)^{n}2^{n}}{(2n+1)!!}\left(\sqrt{2}\frac{\lambda}{\Omega}\right)^{2n+1}\\&
=1-\sqrt{2}\frac{\lambda}{\Omega} F\left(\sqrt{2}\frac{\lambda}{\Omega}\right),
\end{aligned}
\end{equation}
where we used the identity $\frac{(n+1)!}{(2n+2)!}=\frac{1}{2^{n+1}(2n+1)!!}$. We denote by $F(x)=\sum_{n=0}^{\infty}\frac{(-1)^{n}2^{n}}{(2n+1)!!}x^{2n+1}=e^{-x^{2}}\int_{0}^{x}e^{t^{2}} dt$ the Dawson integral function. Then, the system operators become
\begin{equation}
(\sigma^z)^{\text{eff}}=\sigma^{z}\left[1-\sqrt{2}\frac{\lambda}{\Omega}F\left(\sqrt{2}\frac{\lambda}{\Omega}\right)\right],\;\;\;(\sigma^x)^{\text{eff}}=\sigma^{x}\left[1-\sqrt{2}\frac{\lambda}{\Omega}F\left(\sqrt{2}\frac{\lambda}{\Omega}\right)\right].
\end{equation}

\subsection{Unequal couplings}
We now let the coupling parameters to be different, $\frac{\lambda_{1}}{\Omega_{1}}\neq\frac{\lambda_{2}}{\Omega_{2}}$. We then
get through the same procedure
\begin{equation}
\begin{aligned}
\hat{U}_{P}(\tilde{\boldsymbol{p}}) & =\sigma^{0} \cos \left(\sqrt{2} \sqrt{\tilde{p}_{z}^{2}\frac{\lambda_{z}^{2}}{\Omega_{z}^{2}}+\tilde{p}_{x}^{2}\frac{\lambda_{x}^{2}}{\Omega_{x}^{2}} }\right) \\ & -i \frac{\sin(\sqrt{2} \sqrt{\tilde{p}_{z}^{2}\frac{\lambda_{z}^{2}}{\Omega_{z}^{2}}+\tilde{p}_{x}^{2}\frac{\lambda_{x}^{2}}{\Omega_{x}^{2}}})}{\sqrt{\tilde{p}_{z}^{2}\frac{\lambda_{z}^{2}}{\Omega_{z}^{2}}+\tilde{p}_{x}^{2}\frac{\lambda_{x}^{2}}{\Omega_{x}^{2}}}}\Big(\sigma^{x} \tilde{p}_{x}\frac{\lambda_{x}}{\Omega_{x}}+\sigma^{z} \tilde{p}_{z}\frac{\lambda_{z}}{\Omega_{z}}\Big).
\end{aligned}
\end{equation}
We transform to polar coordinates, $\tilde p_z= r\cos \theta$ and $\tilde p_x=r \sin \theta$, and compute the transformed system operators, 
\begin{equation}
\begin{aligned}\hat{U}_{P}(r,\theta)
\sigma^{z}\hat{U}_{P}^{\dagger}(r,\theta) & =\sigma^{z}\left(\frac{\frac{\lambda_{z}^{2}}{\Omega_z^2}\cos^{2}\theta+\frac{\lambda_{x}^{2}}{\Omega_x^2}\sin^{2}\theta\cos(2\sqrt{2}r\sqrt{\frac{\lambda_{z}^{2}}{\Omega_z^2}\cos^{2}\theta
+\frac{\lambda_{x}^{2}}{\Omega_x^2}\sin^{2}\theta})}{\frac{\lambda_{z}^{2}}{\Omega_z^2}\cos^{2}\theta+\frac{\lambda_{x}^{2}}{\Omega_x^2}\sin^{2}\theta}\right)\\
 & +\sigma^{x}\left(\frac{\frac{\lambda_{z}\lambda_{x}}{\Omega_z \Omega_x}\sin(2\theta)\sin^2(\sqrt{2}r\sqrt{\frac{\lambda_{z}^{2}}{\Omega_z^2}\cos^{2}\theta+\frac{\lambda_{x}^{2}}{\Omega_x^2}\sin^{2}\theta})}{\frac{\lambda_{z}^{2}}{\Omega_z^2}\cos^{2}\theta+\frac{\lambda_{x}^{2}}{\Omega_x^2}\sin^{2}\theta}\right)\\
 & +\sigma^{y}\left(\frac{\frac{\lambda_{x}}{\Omega_x}\sin\theta\sin(2\sqrt{2}r\sqrt{\frac{\lambda_{z}^{2}}{\Omega_z^2}\cos^{2}\theta+\frac{\lambda_{x}^{2}}{\Omega_x^2}\sin^{2}\theta})}{\sqrt{\frac{\lambda_{z}^{2}}{\Omega_z^2}\cos^{2}\theta+\frac{\lambda_{x}^{2}}{\Omega_x^2}\sin^{2}\theta}}\right),
\end{aligned}
\end{equation}
and
\begin{equation}
\begin{aligned}
\hat{U}_{P}(r,\theta)\sigma^{x}\hat{U}_{P}^{\dagger}(r,\theta) & =\sigma^{x}\left( \frac{\frac{\lambda_{x}^{2}}{\Omega_x^2}\sin^{2}\theta+\frac{\lambda_z^2}{\Omega_z^2}\cos^{2}\theta\cos(2\sqrt{2}r\sqrt{\cos^{2}(\theta)\frac{\lambda_{z}^{2}}{\Omega_z^2}+\sin^{2}(\theta)\frac{\lambda_{x}^{2}}{\Omega_x^2}})}{\cos^{2}(\theta)\frac{\lambda_{z}^{2}}{\Omega_z^2}+\sin^{2}(\theta)\frac{\lambda_{x}^{2}}{\Omega_x^2}}\right)\\ & +\sigma^{y}\left(\frac{-\frac{\lambda_{z}}{\Omega_z}\cot(\theta)\sin(2\sqrt{2}r\sin(\theta)\sqrt{\cot^{2}(\theta)\frac{\lambda_{z}^{2}}{\Omega_z^2}+\frac{\lambda_{x}^{2}}{\Omega_x^2}})}{\sqrt{\cot^{2}(\theta)\frac{\lambda_{z}^{2}}{\Omega_z^2}+\frac{\lambda_{x}^{2}}{\Omega_x^2}}} \right)\\ & +\sigma^{z}\left(\frac{\frac{\lambda_{z}\lambda_{x}}{\Omega_z \Omega_x}\sin(2\theta)\sin(\sqrt{2}r\sqrt{\cos^{2}(\theta)\frac{\lambda_{z}^{2}}{\Omega_z^2}+\sin^{2}(\theta)\frac{\lambda_{x}^{2}}{\Omega_x^2}})}{\cos^{2}(\theta)\frac{\lambda_{z}^{2}}{\Omega_z^2}+\sin^{2}(\theta)\frac{\lambda_{x}^{2}}{\Omega_x^2}}\right).
\end{aligned}
\end{equation}
We notice that both the second and third lines of the above equations amount to zero after integration over $\theta$. Altogether, we obtain dressing functions for the Pauli operators, each dependent on both $\lambda_x$ and $\lambda_z$:
\begin{equation}
\begin{aligned}
(\sigma^z)^{\text{eff}} & =\sigma^{z}\kappa_z\left(\lambda_z,\lambda_x\right),\;\;\;(\sigma^x)^{\text{eff}}= \sigma^x \kappa_x \left(\lambda_z,\lambda_x\right) =\sigma^{x}\kappa_z\left(\lambda_x,\lambda_z\right),\end{aligned}
\end{equation}
where
\begin{equation}
\kappa_z(\lambda_z,\lambda_x)=\int_{0}^{\infty}\int_{0}^{2\pi}drd\theta e^{-r^{2}}\frac{r}{\pi}\frac{ \frac{\lambda_z^2}{\Omega_z^2}\cos^{2}\theta+\frac{\lambda_x^2}{\Omega_x^2}\sin^{2}\theta\cos(2\sqrt{2}r\sqrt{ \frac{\lambda_z^2}{\Omega_z^2}\cos^{2}\theta+\frac{\lambda_x^2}{\Omega_x^2}\sin^{2}\theta })}{\frac{\lambda_z^2}{\Omega_z^2}\cos^{2}\theta +\frac{\lambda_x^2}{\Omega_x^2}\sin^{2}\theta }.
\label{eq:kappaD}
\end{equation}
Eq. (\ref{eq:kappaD}) can be numerically integrated. The result of this
integration is presented in Figure \ref{fig:dawson}.  Let us check that for $\lambda_z \rightarrow 0$ the function $\kappa(\lambda_z,\lambda_x)$ reduces to known results \cite{Nick_PRX},
\begin{equation}
\begin{aligned}
        \kappa_z(0,\lambda_x) &= \int_{0}^{\infty}\int_{0}^{2\pi}drd\theta e^{-r^{2}}\frac{r}{\pi}\cos\left(2\sqrt{2}r  \left| \sin(\theta) \frac{\lambda_x}{\Omega_x} \right| \right) \\
        &= \int_{0}^{2\pi} d\theta \left( \frac{1}{2 \pi} - \frac{\sqrt{2} \frac{\lambda_x}{\Omega_x} \sin(\theta) F\left( \frac{\sqrt{2} \sin(\theta) \frac{\lambda_x}{\Omega_x} }{\text{sgn}(\sin(\theta))} \right) }{\pi \text{sgn}(\sin(\theta)) } \right) \\
        &= \exp\left(-2\frac{\lambda_x^2}{\Omega_x^2}\right).
\end{aligned}
\end{equation}
We can use equation (\ref{eq:kappaD}) to prove that $\kappa_{z,x} \leq 1$ and that $\kappa_z(\lambda_z,\lambda_x)=\kappa_x(\lambda_x,\lambda_z)$:
\begin{equation}
    \begin{aligned}\kappa_{z}(\lambda_{z},\lambda_{x}) & \leq\int_{0}^{\infty}\int_{0}^{2\pi}drd\theta e^{-r^{2}}\frac{r}{\pi}\frac{\frac{\lambda_z^2}{\Omega_z^2}\cos^{2}\theta+\frac{\lambda_x^2}{\Omega_x^2}\sin^{2}\theta\cdot1}{\frac{\lambda_z^2}{\Omega_z^2}\cos^{2}\theta+\frac{\lambda_x^2}{\Omega_x^2}\sin^{2}\theta}\\
 & = \int_{0}^{\infty}\int_{0}^{2\pi}drd\theta e^{-r^{2}}\frac{r}{\pi}=1,
\end{aligned}
\end{equation}
\begin{equation}
    \begin{aligned}\kappa_{z}(\lambda_{x},\lambda_{z}) & =\int_{0}^{\infty}\int_{0}^{2\pi}drd\theta e^{-r^{2}}\frac{r}{\pi}\frac{\frac{\lambda_x^2}{\Omega_x^2}\cos^{2}\theta+\frac{\lambda_z^2}{\Omega_z^2}\sin^{2}\theta\cos(2\sqrt{2}r\sqrt{\frac{\lambda_x^2}{\Omega_x^2}\cos^{2}\theta+\frac{\lambda_z^2}{\Omega_z^2}\sin^{2}\theta})}{\frac{\lambda_x^2}{\Omega_x^2}\cos^{2}\theta+\frac{\lambda_z^2}{\Omega_z^2}\sin^{2}\theta}\\
 & \rightarrow\int_{0}^{\infty}\int_{\pi/2}^{-3\pi/2}drd\theta'e^{-r^{2}}\frac{r}{\pi}\frac{\frac{\lambda_x^2}{\Omega_x^2}\sin^{2}\theta'+\frac{\lambda_z^2}{\Omega_z^2}\cos^{2}\theta'\cos(2\sqrt{2}r\sqrt{\frac{\lambda_x^2}{\Omega_x^2}\sin^{2}\theta'+\frac{\lambda_z^2}{\Omega_z^2}\cos^{2}\theta'})}{\frac{\lambda_x^2}{\Omega_x^2}\sin^{2}\theta'+\frac{\lambda_z^2}{\Omega_z^2}\cos^{2}\theta'}\\
 & =\kappa_{x}(\lambda_{z},\lambda_{x}),
\end{aligned}
\end{equation}
where in the last derivation we performed the substitution $\theta'=\pi/2-\theta$.

\subsection{Kitaev XY spin chain model}
We present the calculation of the effective Hamiltonian for the spin system (\ref{eq:Hs_chain}).
We note that each term in the system's Hamiltonian, that is, $\Delta_n \sigma^z_n$, does not commute with the two neighboring coupling operators, $\hat{S}_{n-1}$, and $\hat{S}_{n}$. Furthermore, each coupling operator $\hat{S}_n$ does not commute with its neighboring coupling operators, $\hat{S}_{n-1}$, and $\hat{S}_{n+1}$. This allows us to simplify the mixed polaron transformation operator, when acting on the system Hamiltonian and bath coupling operators. The non-factorized polaron transformation that we apply on the operator $\Delta_n \sigma^z_n$ is
\begin{equation}
\hat{U}_{P,n}=\exp\left[\frac{\lambda}{\Omega}\sum_{m=n-1,n}(\hat{a}_{m}^{\dagger}-\hat{a}_{m})\hat{S}_{m}\right].
\end{equation}
In contrast, the polaron transformation of the operator $\hat{S}_n$ is given by
\begin{equation}
\hat{U}_{P,n}=\exp\left[\frac{\lambda}{\Omega}\sum_{m=n-1,n+1}(\hat{a}_{m}^{\dagger}-\hat{a}_{m})\hat{S}_{m}\right].
\end{equation}
%
Continuing to derive the effective Hamiltonian, we get
\begin{equation}
    \begin{aligned}
\hat{H}_{S}^{\text{eff}}=\sum_{n=1}^{N}\left( \Delta_n \sigma^z_n \right)^{\text{eff}},
    \end{aligned}
\end{equation}
where
\begin{equation}
\left( \Delta_n \sigma^z_n \right)^{\text{eff}}=\int\hat{U}_{P,n}(\tilde{\boldsymbol{p}})\left( \Delta_n \sigma_{n}^{z} \right) \hat{U}_{P,n}^{\dagger}(\tilde{\boldsymbol{p}})\prod_{m=n-1,n}\frac{e^{-\tilde{p}_{m}^{2}}}{\sqrt{\pi}}d\tilde{p}_{m},
\end{equation}
and
\begin{equation}
\hat{S}_{m}^{\text{eff}}=\begin{cases}
\int\hat{U}_{P,m}(\tilde{\boldsymbol{p}})(\sigma_{m}^{x}+\sigma_{m+1}^{x})\hat{U}_{P,m}^{\dagger}(\tilde{\boldsymbol{p}})\prod_{n=m-1,m+1}\frac{e^{-\tilde{p}_{n}^{2}}}{\sqrt{\pi}}d\tilde{p}_{n} & m\;\text{odd}\\
\int\hat{U}_{P,m}(\tilde{\boldsymbol{p}})(\sigma_{m}^{y}+\sigma_{m+1}^{y})\hat{U}_{P,m}(\tilde{\boldsymbol{p}})\prod_{n=m-1,m+1}\frac{e^{-\tilde{p}_{n}^{2}}}{\sqrt{\pi}}d\tilde{p}_{n} & m\;\text{even}
\end{cases},
\end{equation}
Note that the integrals are linear in the system operators, which are being transformed. This means that finding the effective Hamiltonian reduces to evaluating integrals in equation (\ref{eq:Ssquared_chain}) and
\begin{equation}
\left( \sigma_n^\alpha \right) ^{\text{eff}}=\int\hat{U}_{P,n}(\tilde{\boldsymbol{p}}) \sigma_{n}^{\alpha} \hat{U}_{P,n}^{\dagger}(\tilde{\boldsymbol{p}})\prod_{m=n-1,n}\frac{e^{-\tilde{p}_{m}^{2}}}{\sqrt{\pi}}d\tilde{p}_{m},
\end{equation}
where $n \in [1,N]$, and $\alpha \in \{x,y,z\}$. 
By relabelling the operator indices $n-1 \rightarrow 2$, and $n \rightarrow 1$,  we end up with the same integrals as in the previous example of a two-level system. Explicitly performing those integrals, after changing variables to $\tilde{p}_{n-1} = \tilde{p}_{2} \rightarrow r\cos(\theta),\; \tilde{p}_{n} = \tilde{p}_{1}\rightarrow r\sin(\theta)$, leads to equations (\ref{eq:chain_Hseff}) and (\ref{eq:chain_Seff}). 
As for the mapping of the new two-body term in the effective system Hamiltonian appearing in equation \ref{eq:chain_Hseff}, these two body terms include (besides identities)
\begin{equation}
    \begin{aligned}
        \hat{S}_{m}^{2}=\begin{cases}
2\sigma_{m}^{x}\sigma_{m+1}^{x} & m\;\text{odd}\\
2\sigma_{m}^{y}\sigma_{m+1}^{y} & m\;\text{even}
\end{cases}.\label{eq:S_squared}
    \end{aligned}
\end{equation}
Each term includes a product of two Pauli matrices, corresponding to two neighboring sites. Such terms do not commute with three coupling operators. The non-factorized polaron transform of the two-body terms is thus given by
\begin{equation}
    \begin{aligned}
        \hat{U}_{P,m}(\tilde{\boldsymbol{p}})=\exp\Big(-i\sqrt{2}\frac{\lambda}{\Omega}\sum_{n=m-1,m,m+1}\tilde{p}_{n}\hat{S}_{n}\Big),
    \end{aligned}\label{eq:UP_chain}
\end{equation}
with the effective two-body interaction term 
\begin{equation}
\Big(\hat{S}_{m}^{2}\Big)^{\text{eff}}=\begin{cases}
\int\hat{U}_{P,m}(\tilde{\boldsymbol{p}})2\sigma_{m}^{x}\sigma_{m+1}^{x}\hat{U}_{P,m}^{\dagger}(\tilde{\boldsymbol{p}})\prod_{n=m-1,m,m+1}\frac{e^{-\tilde{p}_{n}^{2}}}{\sqrt{\pi}}d\tilde{p}_{n} & m\;\text{odd}\\
\int\hat{U}_{P,m}(\tilde{\boldsymbol{p}})2\sigma_{m}^{y}\sigma_{m+1}^{y}\hat{U}_{P,m}(\tilde{\boldsymbol{p}})\prod_{n=m-1,m,m+1}\frac{e^{-\tilde{p}_{n}^{2}}}{\sqrt{\pi}}d\tilde{p}_{n} & m\;\text{even}
\end{cases}. \label{eq:S2eff_chain}
\end{equation}
Let us first consider the simple case of $N=2$. In this case,  $\hat{S}_{1}^{2}=2\sigma^x_1 \sigma^x_2$ and $\hat{S}_{2}^{2}=2\sigma^y_1 \sigma^y_2$ do not commute.
After substituting $\tilde{p}_{1} \rightarrow r\cos(\theta),\; \tilde{p}_{2} \rightarrow r\sin(\theta)$, the general polaron transform takes the form
\begin{equation}
    \begin{aligned}
     &\hat{U}_{P,1}(\tilde{\boldsymbol{p}})=\hat{U}_{P,2}(\tilde{\boldsymbol{p}})=\exp\Big(-i\sqrt{2}\frac{\lambda}{\Omega}\sum_{n=1,2}\tilde{p}_{n}\hat{S}_{n}\Big)\\&\rightarrow
        \left(
\begin{array}{cccc}
 \cos ^2\left(\frac{\sqrt{2} \lambda  r}{\Omega }\right) & -\frac{1}{2} e^{i \theta } \sin \left(\frac{2 \sqrt{2} \lambda  r}{\Omega }\right) & -\frac{1}{2} e^{i \theta } \sin \left(\frac{2 \sqrt{2} \lambda  r}{\Omega }\right) & e^{2 i \theta } \sin ^2\left(\frac{\sqrt{2} \lambda  r}{\Omega }\right) \\
 \frac{1}{2} e^{-i \theta } \sin \left(\frac{2 \sqrt{2} \lambda  r}{\Omega }\right) & \cos ^2\left(\frac{\sqrt{2} \lambda  r}{\Omega }\right) & -\sin ^2\left(\frac{\sqrt{2} \lambda  r}{\Omega }\right) & -\frac{1}{2} e^{i \theta } \sin \left(\frac{2 \sqrt{2} \lambda  r}{\Omega }\right) \\
 \frac{1}{2} e^{-i \theta } \sin \left(\frac{2 \sqrt{2} \lambda  r}{\Omega }\right) & -\sin ^2\left(\frac{\sqrt{2} \lambda  r}{\Omega }\right) & \cos ^2\left(\frac{\sqrt{2} \lambda  r}{\Omega }\right) & -\frac{1}{2} e^{i \theta } \sin \left(\frac{2 \sqrt{2} \lambda  r}{\Omega }\right) \\
 e^{-2 i \theta } \sin ^2\left(\frac{\sqrt{2} \lambda  r}{\Omega }\right) & \frac{1}{2} e^{-i \theta } \sin \left(\frac{2 \sqrt{2} \lambda  r}{\Omega }\right) & \frac{1}{2} e^{-i \theta } \sin \left(\frac{2 \sqrt{2} \lambda  r}{\Omega }\right) & \cos ^2\left(\frac{\sqrt{2} \lambda  r}{\Omega }\right) \\
\end{array}
\right).
    \end{aligned}
    \nonumber
\end{equation}
Plugging this expression into equation (\ref{eq:S2eff_chain}) and performing the integral over $\{r,\theta\}$  we obtain
\begin{equation}
    \begin{aligned}
\Big(\hat{S}_{1}^{2}\Big)^{\text{eff}} & =\kappa_{1}(\lambda)\sigma_{1}^{x}\sigma_{2}^{x}+\kappa_{2}(\lambda)\sigma_{1}^{y}\sigma_{2}^{y}+\kappa_{3}(\lambda)\sigma_{1}^{z}\sigma_{2}^{z},
\end{aligned}
\end{equation}
\begin{equation}
    \begin{aligned}
\Big(\hat{S}_{2}^{2}\Big)^{\text{eff}} & =\kappa_{2}(\lambda)\sigma_{1}^{x}\sigma_{2}^{x}+\kappa_{1}(\lambda)\sigma_{1}^{y}\sigma_{2}^{y}+\kappa_{3}(\lambda)\sigma_{1}^{z}\sigma_{2}^{z},
\end{aligned}
\end{equation}
where the dressing functions are
\begin{equation}
    \begin{aligned}\kappa_{1}(\lambda) & =2-\frac{\lambda}{\Omega\sqrt{2}}\left(2F\left(\sqrt{2}\frac{\lambda}{\Omega}\right)+3F\left(2\sqrt{2}\frac{\lambda}{\Omega}\right)\right), \\
\kappa_{2}(\lambda) & =\frac{\lambda}{\Omega\sqrt{2}}\left(2F\left(\sqrt{2}\frac{\lambda}{\Omega}\right)-F\left(2\sqrt{2}\frac{\lambda}{\Omega}\right)\right),\\
\kappa_{3}(\lambda) & =2\sqrt{2}\frac{\lambda}{\Omega} F\left(2\sqrt{2}\frac{\lambda}{\Omega}\right).
\end{aligned}
\end{equation}
Going back to the general $N>2$ model, we notice that each two body term of the form $2\sigma^{x/y}_m\sigma^{x/y}_{m+1}$ commutes with term proportional to $ \sigma^{x/y}_{m-1} $ in $\hat{S}_{m-1}$ and a term proportional to $ \sigma^{x/y}_{m+2} $ in $\hat{S}_{m+1}$. 
Thus, equation (\ref{eq:UP_chain}) simplifies to
\begin{equation}
    \begin{aligned}
        \hat{U}_{P,m}(\tilde{\boldsymbol{p}})=\begin{cases}
\exp\Big(-i\sqrt{2}\frac{\lambda}{\Omega}\big[\tilde{p}_{m}\sigma_{m}^{x}+\tilde{p}_{m-1}\sigma_{m}^{y}\big]\otimes\big[\tilde{p}_{m+1}\sigma_{m+1}^{y}+\tilde{p}_{m}\sigma_{m+1}^{x}\big]\Big) & m\;\text{odd}\\
\exp\Big(-i\sqrt{2}\frac{\lambda}{\Omega}\big[\tilde{p}_{m}\sigma_{m}^{y}+\tilde{p}_{m-1}\sigma_{m}^{x}\big]\otimes\big[\tilde{p}_{m+1}\sigma_{m+1}^{x}+\tilde{p}_{m}\sigma_{m+1}^{y}\big]\Big) & m\;\text{even}
\end{cases}.\label{eq:UP_factorized_chain}
    \end{aligned}
\end{equation}
The above formula reduces to the  $N=2$ case when substituting $\tilde{p}_{m-2}\rightarrow \tilde{p}_{m+1}$. This physically corresponds to adding correlations between baths with indices $m-2$ and $m$. 

Plugging Eq. (\ref{eq:UP_factorized_chain}) in (\ref{eq:S2eff_chain}), and with the help of a symbolic computation software, we
obtain the explicit form of the operators in equation (\ref{eq:UP_factorized_chain}),
\begin{equation}
    \begin{aligned}
        \sum_{m=1}^{N}\Big(\hat{S}_{m}^{2}\Big)^{\text{eff}}=\sum_{m\;\text{odd}}\xi\left(\lambda\right)\sigma_{m}^{x}\sigma_{m+1}^{x}+\sum_{m\;\text{even}}\xi\left(\lambda\right)\sigma_{m}^{y}\sigma_{m+1}^{y},
    \end{aligned}
\end{equation}
where the function $\xi(\lambda)$ is given by the integral
\begin{equation}
    \begin{aligned}
        \xi(\lambda)=\int_{\mathbb{R}^{3}}dp_{1}dp_{2}dp_{3}\frac{e^{-p_{1}^{2}-p_{2}^{2}-p_{3}^{2}}}{\pi^{3/2}}\left(\frac{2\left(p_{1}^{2}\cos\left(2\sqrt{2}\frac{\lambda}{\Omega}\sqrt{p_{1}^{2}+p_{2}^{2}}\right)+p_{2}^{2}\right)\left(p_{3}^{2}\cos\left(2\sqrt{2}\frac{\lambda}{\Omega}\sqrt{p_{2}^{2}+p_{3}^{2}}\right)+p_{2}^{2}\right)}{\left(p_{1}^{2}+p_{2}^{2}\right)\left(p_{2}^{2}+p_{3}^{2}\right)}\right),
    \end{aligned}
\end{equation}
which we evaluate numerically and present in Section \ref{sec:lattice}.


\section{Redfield Equation and the derivation of the population relaxation rate}

\renewcommand{\theequation}{D\arabic{equation}}
\setcounter{equation}{0}  
\label{app:4}

In this Appendix, we derive the relaxation rate for the population of the spin (and similarly, its polarization) under the influence of two heat baths, one coupled to the $\sigma^z$ operator of the system, the other to $\sigma^x$.
We apply the Redfield equation to two scenarios:
(i) The weak coupling limit of the model Eq. (\ref{eq:SB}).
(ii) The effective Hamiltonian framework, given by Eq. (\ref{eq:Oeff_example_1}). In this case, strong coupling effects are embedded within the parameters of the effective Hamiltonian, while the system is now weakly coupled to the residual baths.

The Redfield equation for the reduced density operator of a system coupled to multiple baths (indexed by $n=x,z$ here) is written as \cite{Nitzan}
\begin{equation}
    \begin{aligned}
        \frac{d}{dt} \rho_{ab}(t) &=-i\omega_{ab}\rho_{ab}(t)
-i\sum_{c}\left(\bar{V}_{ac}\rho_{cb}-\rho_{ac}\bar{V}_{cb} \right)\\ 
& -\sum_{n=z,x}\sum_{c,d}\Big(R_{ac,cd}^{n}(\omega_{dc})\rho_{db}(t)+R_{bd,dc}^{n,*}(\omega_{cd}) \rho_{ac}(t) -\big[R_{db,ac}^{n}(\omega_{ca})+R_{ca,bd}^{n,*}(\omega_{db})\big]\rho_{cd}(t) \Big),
    \end{aligned}
\end{equation}
where $\omega_{ab}=E_{a}-E_{b}$, with $E_{a}$ and $E_b$ as the eigenenergies of the system Hamiltonian.
The constants $\bar V$ are averaged over the thermal state of the baths, and they do not contribute in our model,
\begin{equation}
\bar{V}^{\text{eff}}=\left\langle \sum_{n=z,x}\tilde{\sum_{k}}\frac{2\lambda_{n}f_{n,k}}{\Omega_{n}}\hat{S}_{n}^{\text{eff}}(\hat{b}_{n,k}^{\dagger}+\hat{b}_{n,k}) \right\rangle=0; \,\,\,\,\,; 
\bar{V}^{\text{UW}} \propto \left\langle \hat c_{n,k}^{\dagger}+\hat c_{n,k} \right\rangle= 0.
\end{equation}
The elements of the dissipators are given by half Fourier transforms of bath autocorrelation functions $C_n(t)$, multiplied by the respective matrix elements,
\begin{align}
R_{ab,cd}^{n,\text{UW}}(\omega) & =
(\hat{S}_{n})_{ab}(\hat{S}_{n})_{cd} 
\int_{0}^{\infty}e^{i\omega\tau} C_{n}^{\text{UW}}(\tau)d\tau
\nonumber \\
 & =(\hat{S}_{n})_{ab}(\hat{S}_{n})_{cd}C_{n}^{\text{UW}}(\omega) ,
\end{align}
\begin{align}
R_{ab,cd}^{n,\text{eff}} & =
(\hat{S}_{n}^{\text{eff}})_{ab}(\hat{S}_{n}^{\text{eff}})_{cd} \int_{0}^{\infty}e^{i\omega\tau} C_{n}^{\text{eff}}(\tau)d\tau\nonumber \\
 & =(\hat{S}_{n})_{ab}(\hat{S}_{n})_{cd} 
 \kappa_{n}^{2}(\lambda_{z},\lambda_{x}) C_{n}^{\text{eff}}(\omega).
\end{align}
Note that the two baths are not correlated.

The full derivation of the correlation functions can be found in \cite{Nitzan}. The real part of the $n$th bath correlation function is given by
%
%
\begin{equation}
    \begin{aligned}
    \real C_{n}^{\bullet}(\omega) & =\begin{cases}
 \pi J_{n}^{\bullet}(|\omega|)n_{B}(|\omega|), & \omega<0\\
\pi J_{n}^{\bullet}(\omega)(n_{B}(\omega)+1), & \omega>0\\
\lim_{\omega\rightarrow0} \pi J_{n}^{\bullet}(\omega)n_{B}(\omega), & \omega=0
\end{cases}
\end{aligned}
\end{equation}
Here, $J_n^{\bullet}(\omega)$ is the spectral density function of the $n$th bath, with $\bullet$ standing for the method (eff, UW), and $n_B(\omega)$ the Bose-Einstein distribution function.
As for the Lamb shift, which is the imaginary part of the dissipator, we assume that it can be neglected as discussed in Refs. 
\citenum{PhysRevA.91.052108,PhysRevA.90.032114, FelixNJP, correa2024}.

Concretely, in our model, Eq. (\ref{eq:SB}) and (\ref{eq:Oeff_example_1}), the Redfield equations for the diagonal and off-diagonal elements of the reduced density matrix are given by
\bea
        \frac{d}{dt}  \rho_{11}(t) &= &-\left( R^x_{12,21}(\omega_{12}) + R^{x,*}_{12,21}(\omega_{12}) \right) \rho_{11}(t) 
 + \left( R^x_{21,12}(\omega_{21}) + R^{x,*}_{21,12}(\omega_{21}) \right) \rho_{22}(t), 
\nonumber\\
\frac{d}{dt}  \rho_{12}(t) &=&-i\omega_{12}\rho_{12}(t) 
  + \left( -R^z_{11,11}(\omega_{11}) - R^{z,*}_{22,22}(\omega_{22}) + R^z_{22,11}(\omega_{11}) + R^{z,*}_{11,22}(\omega_{22}) \right) \rho_{12}(t) 
  \nonumber\\
 &+&\left( -R^x_{12,21}(\omega_{12}) -R^{x,*}_{21,12}(\omega_{21}) \right) \rho_{12}(t) 
 + \left( R^x_{12,12}(\omega_{21}) + R^{x,*}_{21,21}(\omega_{12}) \right) \rho_{21}(t). \label{eq:refieldUW}
    \eea
Equations of motion for $\rho_{22}(t) $ and $\rho_{21}(t) $ are obtained by switching the indices $ 1 \leftrightarrow 2 $ in the above equations. 
Notably, the diagonal and off-diagonal elements of $\rho(t)$ are {\it decoupled} per the structure of the Hamiltonian without an additional secular approximation. 

Let us focus on the population dynamics. The two pairs in the dissipator combine into a full Fourier transform. In the effective Hamiltonian method, this results in
\bea
   \frac{d}{dt}  \rho_{11}(t) &=&  - \left[2 \pi  \kappa_x^2 J_x^{\text{eff}}(2\kappa_z\Delta) n_B(2\kappa_z\Delta) \right]\rho_{11}(t)
+ [2 \pi \kappa_x^2 J_x^{\text{eff}}(2\kappa_z\Delta) (n_B(2\kappa_z\Delta)+1)] \rho_{22}(t)
\nonumber\\
&=&
 - \left[2 \pi \kappa_x^2 J_x^{\text{eff}}(2\kappa_z\Delta) (2n_B(2\kappa_z\Delta)+1) \right]\rho_{11}
+ [2 \pi \kappa_x^2 J_x^{\text{eff}}(2\kappa_z\Delta) (n_B(2\kappa_z\Delta)+1)].
\label{eq:EOM11}
\eea
Note that the spectral function and the thermal occupation factor are evaluated at the renormalized spin splitting 
$\kappa_z\Delta$. For conciseness, we do not explicitly indicate the dependence of $\kappa_{x}$ and $\kappa_z$ on the two coupling parameters, $\lambda_x$ and $\lambda_z$.
Altogether, the population follows a simple dynamical equation,
\bea
 \frac{d}{dt}  \rho_{11}(t) &=& - \Gamma_x^{\text{eff}} \rho_{11}(t) + A,
\eea
where  $A$ stands for the constant term in Eq. (\ref{eq:EOM11}), and the rate constant is given by
\bea
\Gamma_x^{\text {eff}}= 2 \pi \kappa_x^2(\lambda_z,\lambda_x)  \times \frac{4\lambda_x^2}{\Omega_x^2} \gamma_x \kappa_z \Delta \times (2n_B(2\kappa_z\Delta)+1), 
\label{eq:GammaR}
\eea
which we obtained by recalling that 
$J_x^{\text{eff}}(\omega)=
\frac{4\lambda_x^2}{\Omega_x^2} \gamma_x \omega$. A reminder that $\gamma$ is a dimensionless parameter, which we inherited from the original model where it served as the width parameter of the Brownian spectral function, see Eq. (\ref{eq:Brownian}).
In this equation, we also explicitly indicate the dependence of the dressing function $\kappa_x$ on both coupling parameters, $\lambda_x$ and $\lambda_z$. Similarly, $\kappa_z$ depends on both coupling parameters, but for simplicity we suppress them from the text.

Equation (\ref{eq:GammaR}) is a central result of this study. 
It shows that the relaxation rate  depends on both the dissipative coupling $\lambda_{x}$ and the coupling to the decoheering bath, $\lambda_z$. The latter dependence is mostly embedded within 
$\kappa_x$, a function of both $\lambda_{x,z}$.
In the limit $T>\kappa_z\Delta$, the rate can be approximated as
\bea
\Gamma_x^{\text{eff}}\approx 4 \pi T \kappa_x^2 (\lambda_z,\lambda_x) \times \frac{4 \lambda_x^2}{\Omega_x^2} \gamma_x.
\label{eq:GammaxT}
\eea
%
We contrast this result with the ultraweak coupling limit, 
\bea
\Gamma_x^{\text{UW}}= 2 \pi  J_x(2\Delta)(2n_B(2\Delta)+1).
\eea
Assuming that $\Omega\gg \Delta$, and that $T>\Delta$, we simplify this result using the Brownian spectral function (\ref{eq:Brownian}) and write
\bea
\Gamma_x^{\text{UW}} \approx 4 \pi T \times
 \frac{4\lambda_x^2}{\Omega_x^2} \gamma_x. 
\label{eq:GammaUWT}
\eea
Comparing Eq. (\ref{eq:GammaxT}) to (\ref{eq:GammaUWT}) reveals the impact of strong coupling on the relaxation dynamics via the $\kappa_x^2\leq 1$ term:
Coupling the qubit to a decohering bath with a coupling parameter $\lambda_z$ {\it slows down} the relaxation dynamics by suppressing $\kappa_x$.
%
Simulations of this effect are presented in Sec. \ref{sec:suppression}.

Finally, We complement simulations of the polarization and present in Fig. \ref{fig:coh} the behavior of the coherences at different coupling strengths. Notably, the decoherence behavior is enhanced by increasing either $\lambda_z$ and $\lambda_x$, without showing a nontrivial suppression effect as for $\Gamma_x$.

\begin{figure*}[htbp]
\begin{centering}
\includegraphics[width=1.0\textwidth]{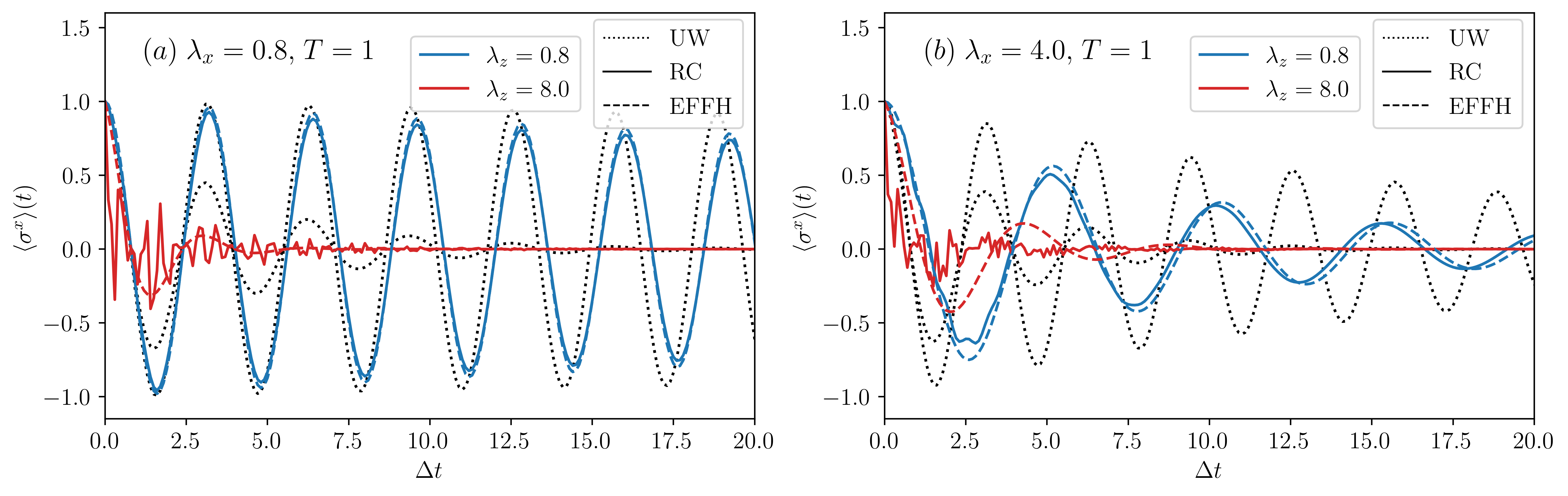} 
\par\end{centering}
\caption{\label{fig:coh}
The dynamics of coherences $\langle \sigma^x \rangle$ from an initial
state $\rho(0)=|\psi\rangle\langle\psi|$ for $|\psi\rangle=\frac{1}{\sqrt{2}}(|0\rangle+|1\rangle)$.  The different colored lines correspond
to different values of the decoherring bath coupling, $\lambda_{z}$. 
(a) $\lambda_x=0.8$, (b) $\lambda_x=4.0$, where solid and dashed lines correspond
to RC and the EFFH simulations, respectively. The dotted-black curve depicts the ultraweak coupling limit. Parameters are set to $\Delta=1$, $\Omega=8$, $\gamma=0.05/\pi$, $\Lambda=1000$, $T=1$.
}
\end{figure*}


\end{widetext}

\bibliographystyle{apsrev4-1}
\bibliography{bib}

\begin{thebibliography}{83}%
\makeatletter
\providecommand \@ifxundefined [1]{%
 \@ifx{#1\undefined}
}%
\providecommand \@ifnum [1]{%
 \ifnum #1\expandafter \@firstoftwo
 \else \expandafter \@secondoftwo
 \fi
}%
\providecommand \@ifx [1]{%
 \ifx #1\expandafter \@firstoftwo
 \else \expandafter \@secondoftwo
 \fi
}%
\providecommand \natexlab [1]{#1}%
\providecommand \enquote  [1]{``#1''}%
\providecommand \bibnamefont  [1]{#1}%
\providecommand \bibfnamefont [1]{#1}%
\providecommand \citenamefont [1]{#1}%
\providecommand \href@noop [0]{\@secondoftwo}%
\providecommand \href [0]{\begingroup \@sanitize@url \@href}%
\providecommand \@href[1]{\@@startlink{#1}\@@href}%
\providecommand \@@href[1]{\endgroup#1\@@endlink}%
\providecommand \@sanitize@url [0]{\catcode `\\12\catcode `\$12\catcode `\&12\catcode `\#12\catcode `\^12\catcode `\_12\catcode `\%12\relax}%
\providecommand \@@startlink[1]{}%
\providecommand \@@endlink[0]{}%
\providecommand \url  [0]{\begingroup\@sanitize@url \@url }%
\providecommand \@url [1]{\endgroup\@href {#1}{\urlprefix }}%
\providecommand \urlprefix  [0]{URL }%
\providecommand \Eprint [0]{\href }%
\providecommand \doibase [0]{http://dx.doi.org/}%
\providecommand \selectlanguage [0]{\@gobble}%
\providecommand \bibinfo  [0]{\@secondoftwo}%
\providecommand \bibfield  [0]{\@secondoftwo}%
\providecommand \translation [1]{[#1]}%
\providecommand \BibitemOpen [0]{}%
\providecommand \bibitemStop [0]{}%
\providecommand \bibitemNoStop [0]{.\EOS\space}%
\providecommand \EOS [0]{\spacefactor3000\relax}%
\providecommand \BibitemShut  [1]{\csname bibitem#1\endcsname}%
\let\auto@bib@innerbib\@empty
\bibitem [{\citenamefont {Witczak}\ \emph {et~al.}(2023)\citenamefont {Witczak}, \citenamefont {Chrzanowski}, \citenamefont {Sitarek}, \citenamefont {Lysien},\ and\ \citenamefont {Podhorodecki}}]{witczak_flexible_2023}%
  \BibitemOpen
  \bibfield  {author} {\bibinfo {author} {\bibfnamefont {L.}~\bibnamefont {Witczak}}, \bibinfo {author} {\bibfnamefont {M.}~\bibnamefont {Chrzanowski}}, \bibinfo {author} {\bibfnamefont {P.}~\bibnamefont {Sitarek}}, \bibinfo {author} {\bibfnamefont {M.}~\bibnamefont {Lysien}}, \ and\ \bibinfo {author} {\bibfnamefont {A.}~\bibnamefont {Podhorodecki}},\ }\href@noop {} {\bibfield  {journal} {\bibinfo  {journal} {{ACS} Omega}\ }\textbf {\bibinfo {volume} {8}},\ \bibinfo {pages} {39217} (\bibinfo {year} {2023})}\BibitemShut {NoStop}%
\bibitem [{\citenamefont {Mi}\ \emph {et~al.}(2017)\citenamefont {Mi}, \citenamefont {Cady}, \citenamefont {Zajac}, \citenamefont {Deelman},\ and\ \citenamefont {Petta}}]{Petta17}%
  \BibitemOpen
  \bibfield  {author} {\bibinfo {author} {\bibfnamefont {X.}~\bibnamefont {Mi}}, \bibinfo {author} {\bibfnamefont {J.~V.}\ \bibnamefont {Cady}}, \bibinfo {author} {\bibfnamefont {D.~M.}\ \bibnamefont {Zajac}}, \bibinfo {author} {\bibfnamefont {P.~W.}\ \bibnamefont {Deelman}}, \ and\ \bibinfo {author} {\bibfnamefont {J.~R.}\ \bibnamefont {Petta}},\ }\href {http://dx.doi.org/10.1126/science.aal2469} {\bibfield  {journal} {\bibinfo  {journal} {Science}\ }\textbf {\bibinfo {volume} {355}},\ \bibinfo {pages} {156} (\bibinfo {year} {2017})}\BibitemShut {NoStop}%
\bibitem [{\citenamefont {Bluvstein}\ \emph {et~al.}(2022)\citenamefont {Bluvstein}, \citenamefont {Levine}, \citenamefont {Semeghini}, \citenamefont {Wang}, \citenamefont {Ebadi}, \citenamefont {Kalinowski}, \citenamefont {Keesling}, \citenamefont {Maskara}, \citenamefont {Pichler}, \citenamefont {Greiner}, \citenamefont {Vuletic},\ and\ \citenamefont {Lukin}}]{Lukin22}%
  \BibitemOpen
  \bibfield  {author} {\bibinfo {author} {\bibfnamefont {D.}~\bibnamefont {Bluvstein}}, \bibinfo {author} {\bibfnamefont {H.}~\bibnamefont {Levine}}, \bibinfo {author} {\bibfnamefont {G.}~\bibnamefont {Semeghini}}, \bibinfo {author} {\bibfnamefont {T.~T.}\ \bibnamefont {Wang}}, \bibinfo {author} {\bibfnamefont {S.}~\bibnamefont {Ebadi}}, \bibinfo {author} {\bibfnamefont {M.}~\bibnamefont {Kalinowski}}, \bibinfo {author} {\bibfnamefont {A.}~\bibnamefont {Keesling}}, \bibinfo {author} {\bibfnamefont {N.}~\bibnamefont {Maskara}}, \bibinfo {author} {\bibfnamefont {H.}~\bibnamefont {Pichler}}, \bibinfo {author} {\bibfnamefont {M.}~\bibnamefont {Greiner}}, \bibinfo {author} {\bibfnamefont {V.}~\bibnamefont {Vuletic}}, \ and\ \bibinfo {author} {\bibfnamefont {M.~D.}\ \bibnamefont {Lukin}},\ }\href {http://dx.doi.org/10.1038/s41586-022-04592-6} {\bibfield  {journal} {\bibinfo  {journal} {Nature}\ }\textbf {\bibinfo {volume} {604}},\ \bibinfo {pages} {451} (\bibinfo {year} {2022})}\BibitemShut {NoStop}%
\bibitem [{\citenamefont {Senior}\ \emph {et~al.}(2020)\citenamefont {Senior}, \citenamefont {Gubaydullin}, \citenamefont {Karimi}, \citenamefont {Peltonen}, \citenamefont {Ankerhold},\ and\ \citenamefont {Pekola}}]{PekolaR}%
  \BibitemOpen
  \bibfield  {author} {\bibinfo {author} {\bibfnamefont {J.}~\bibnamefont {Senior}}, \bibinfo {author} {\bibfnamefont {A.}~\bibnamefont {Gubaydullin}}, \bibinfo {author} {\bibfnamefont {B.}~\bibnamefont {Karimi}}, \bibinfo {author} {\bibfnamefont {J.~T.}\ \bibnamefont {Peltonen}}, \bibinfo {author} {\bibfnamefont {J.}~\bibnamefont {Ankerhold}}, \ and\ \bibinfo {author} {\bibfnamefont {J.~P.}\ \bibnamefont {Pekola}},\ }\href {https://doi.org/10.1038/s42005-020-0307-5} {\bibfield  {journal} {\bibinfo  {journal} {Communications Physics}\ }\textbf {\bibinfo {volume} {3}},\ \bibinfo {pages} {40} (\bibinfo {year} {2020})}\BibitemShut {NoStop}%
\bibitem [{\citenamefont {Nielsen}\ and\ \citenamefont {Chuang}(2010)}]{Nielsen2010}%
  \BibitemOpen
  \bibfield  {author} {\bibinfo {author} {\bibfnamefont {M.~A.}\ \bibnamefont {Nielsen}}\ and\ \bibinfo {author} {\bibfnamefont {I.~A.}\ \bibnamefont {Chuang}},\ }\href@noop {} {\emph {\bibinfo {title} {Quantum Computing and Quantum Information}}},\ \bibinfo {edition} {anniversary edn.}\ ed.\ (\bibinfo  {publisher} {Cambridge University Press},\ \bibinfo {year} {2010})\BibitemShut {NoStop}%
\bibitem [{\citenamefont {Georgopoulos}\ \emph {et~al.}(2021)\citenamefont {Georgopoulos}, \citenamefont {Emary},\ and\ \citenamefont {Zuliani}}]{noise21}%
  \BibitemOpen
  \bibfield  {author} {\bibinfo {author} {\bibfnamefont {K.}~\bibnamefont {Georgopoulos}}, \bibinfo {author} {\bibfnamefont {C.}~\bibnamefont {Emary}}, \ and\ \bibinfo {author} {\bibfnamefont {P.}~\bibnamefont {Zuliani}},\ }\href {https://link.aps.org/doi/10.1103/PhysRevA.104.062432} {\bibfield  {journal} {\bibinfo  {journal} {Phys. Rev. A}\ }\textbf {\bibinfo {volume} {104}},\ \bibinfo {pages} {062432} (\bibinfo {year} {2021})}\BibitemShut {NoStop}%
\bibitem [{\citenamefont {Nazarov}(2003)}]{noise03}%
  \BibitemOpen
  \bibfield  {author} {\bibinfo {author} {\bibfnamefont {Y.}~\bibnamefont {Nazarov}},\ }\href@noop {} {\emph {\bibinfo {title} {Quantum Noise in Mesoscopic Physics}}}\ (\bibinfo  {publisher} {NATO Science Series II: Mathematics, Physics and Chemistry},\ \bibinfo {year} {2003})\BibitemShut {NoStop}%
\bibitem [{\citenamefont {Brand}\ \emph {et~al.}(2024)\citenamefont {Brand}, \citenamefont {Sinayskiy},\ and\ \citenamefont {Petruccione}}]{Francesco24}%
  \BibitemOpen
  \bibfield  {author} {\bibinfo {author} {\bibfnamefont {D.}~\bibnamefont {Brand}}, \bibinfo {author} {\bibfnamefont {I.}~\bibnamefont {Sinayskiy}}, \ and\ \bibinfo {author} {\bibfnamefont {F.}~\bibnamefont {Petruccione}},\ }\href@noop {} {\bibfield  {journal} {\bibinfo  {journal} {Scientific Reports}\ }\textbf {\bibinfo {volume} {14}},\ \bibinfo {pages} {4769} (\bibinfo {year} {2024})}\BibitemShut {NoStop}%
\bibitem [{\citenamefont {Bittner}(2006)}]{Nitzan}%
  \BibitemOpen
  \bibfield  {author} {\bibinfo {author} {\bibfnamefont {E.~R.}\ \bibnamefont {Bittner}},\ }\href {https://doi.org/10.1021/ja069771e} {\bibfield  {journal} {\bibinfo  {journal} {Journal of the American Chemical Society}\ }\textbf {\bibinfo {volume} {128}},\ \bibinfo {pages} {17156} (\bibinfo {year} {2006})}\BibitemShut {NoStop}%
\bibitem [{\citenamefont {Wilner}\ \emph {et~al.}(2013)\citenamefont {Wilner}, \citenamefont {Wang}, \citenamefont {Cohen}, \citenamefont {Thoss},\ and\ \citenamefont {Rabani}}]{MCGuy13}%
  \BibitemOpen
  \bibfield  {author} {\bibinfo {author} {\bibfnamefont {E.~Y.}\ \bibnamefont {Wilner}}, \bibinfo {author} {\bibfnamefont {H.}~\bibnamefont {Wang}}, \bibinfo {author} {\bibfnamefont {G.}~\bibnamefont {Cohen}}, \bibinfo {author} {\bibfnamefont {M.}~\bibnamefont {Thoss}}, \ and\ \bibinfo {author} {\bibfnamefont {E.}~\bibnamefont {Rabani}},\ }\href {https://link.aps.org/doi/10.1103/PhysRevB.88.045137} {\bibfield  {journal} {\bibinfo  {journal} {Phys. Rev. B}\ }\textbf {\bibinfo {volume} {88}},\ \bibinfo {pages} {045137} (\bibinfo {year} {2013})}\BibitemShut {NoStop}%
\bibitem [{\citenamefont {Simine}\ and\ \citenamefont {Segal}(2013)}]{SegalSimine13}%
  \BibitemOpen
  \bibfield  {author} {\bibinfo {author} {\bibfnamefont {L.}~\bibnamefont {Simine}}\ and\ \bibinfo {author} {\bibfnamefont {D.}~\bibnamefont {Segal}},\ }\href {https://doi.org/10.1063/1.4808108} {\bibfield  {journal} {\bibinfo  {journal} {The Journal of Chemical Physics}\ }\textbf {\bibinfo {volume} {138}},\ \bibinfo {pages} {214111} (\bibinfo {year} {2013})}\BibitemShut {NoStop}%
\bibitem [{\citenamefont {Makri}(2024)}]{makri2024quantumdynamics}%
  \BibitemOpen
  \bibfield  {author} {\bibinfo {author} {\bibfnamefont {N.}~\bibnamefont {Makri}},\ }in\ \href@noop {} {\emph {\bibinfo {booktitle} {Comprehensive Computational Chemistry (First Edition)}}},\ Vol.~\bibinfo {volume} {4},\ \bibinfo {editor} {edited by\ \bibinfo {editor} {\bibfnamefont {M.}~\bibnamefont {Yanez}}\ and\ \bibinfo {editor} {\bibfnamefont {R.~J.}\ \bibnamefont {Boyd}}}\ (\bibinfo  {publisher} {Elsevier},\ \bibinfo {year} {2024})\ pp.\ \bibinfo {pages} {293--305}\BibitemShut {NoStop}%
\bibitem [{\citenamefont {Gribben}\ \emph {et~al.}(2022)\citenamefont {Gribben}, \citenamefont {Rouse}, \citenamefont {Iles-Smith}, \citenamefont {Strathearn}, \citenamefont {Maguire}, \citenamefont {Kirton}, \citenamefont {Nazir}, \citenamefont {Gauger},\ and\ \citenamefont {Lovett}}]{PRXQuantum.3.010321}%
  \BibitemOpen
  \bibfield  {author} {\bibinfo {author} {\bibfnamefont {D.}~\bibnamefont {Gribben}}, \bibinfo {author} {\bibfnamefont {D.~M.}\ \bibnamefont {Rouse}}, \bibinfo {author} {\bibfnamefont {J.}~\bibnamefont {Iles-Smith}}, \bibinfo {author} {\bibfnamefont {A.}~\bibnamefont {Strathearn}}, \bibinfo {author} {\bibfnamefont {H.}~\bibnamefont {Maguire}}, \bibinfo {author} {\bibfnamefont {P.}~\bibnamefont {Kirton}}, \bibinfo {author} {\bibfnamefont {A.}~\bibnamefont {Nazir}}, \bibinfo {author} {\bibfnamefont {E.~M.}\ \bibnamefont {Gauger}}, \ and\ \bibinfo {author} {\bibfnamefont {B.~W.}\ \bibnamefont {Lovett}},\ }\href {https://link.aps.org/doi/10.1103/PRXQuantum.3.010321} {\bibfield  {journal} {\bibinfo  {journal} {PRX Quantum}\ }\textbf {\bibinfo {volume} {3}},\ \bibinfo {pages} {010321} (\bibinfo {year} {2022})}\BibitemShut {NoStop}%
\bibitem [{\citenamefont {Fux}\ \emph {et~al.}()\citenamefont {Fux}, \citenamefont {Fowler-Wright}, \citenamefont {Beckles}, \citenamefont {Butler}, \citenamefont {Eastham}, \citenamefont {Gribben}, \citenamefont {Keeling}, \citenamefont {Kilda}, \citenamefont {Kirton}, \citenamefont {Lawrence}, \citenamefont {Lovett}, \citenamefont {O'Neill}, \citenamefont {Strathearn},\ and\ \citenamefont {de~Wit}}]{Keeling24}%
  \BibitemOpen
  \bibfield  {author} {\bibinfo {author} {\bibfnamefont {G.~E.}\ \bibnamefont {Fux}}, \bibinfo {author} {\bibfnamefont {P.}~\bibnamefont {Fowler-Wright}}, \bibinfo {author} {\bibfnamefont {J.}~\bibnamefont {Beckles}}, \bibinfo {author} {\bibfnamefont {E.~P.}\ \bibnamefont {Butler}}, \bibinfo {author} {\bibfnamefont {P.~R.}\ \bibnamefont {Eastham}}, \bibinfo {author} {\bibfnamefont {D.}~\bibnamefont {Gribben}}, \bibinfo {author} {\bibfnamefont {J.}~\bibnamefont {Keeling}}, \bibinfo {author} {\bibfnamefont {D.}~\bibnamefont {Kilda}}, \bibinfo {author} {\bibfnamefont {P.}~\bibnamefont {Kirton}}, \bibinfo {author} {\bibfnamefont {E.~D.~C.}\ \bibnamefont {Lawrence}}, \bibinfo {author} {\bibfnamefont {B.~W.}\ \bibnamefont {Lovett}}, \bibinfo {author} {\bibfnamefont {E.}~\bibnamefont {O'Neill}}, \bibinfo {author} {\bibfnamefont {A.}~\bibnamefont {Strathearn}}, \ and\ \bibinfo {author} {\bibfnamefont {R.}~\bibnamefont {de~Wit}},\ }\href {https://arxiv.org/abs/2406.16650} {\bibinfo  {journal} {arXiv:2406.16650}\
  }\BibitemShut {NoStop}%
\bibitem [{\citenamefont {Segal}\ and\ \citenamefont {Nitzan}(2005)}]{SegalSB1}%
  \BibitemOpen
\bibfield  {journal} {  }\bibfield  {author} {\bibinfo {author} {\bibfnamefont {D.}~\bibnamefont {Segal}}\ and\ \bibinfo {author} {\bibfnamefont {A.}~\bibnamefont {Nitzan}},\ }\href {\doibase 10.1103/PhysRevLett.94.034301} {\bibfield  {journal} {\bibinfo  {journal} {Phys. Rev. Lett.}\ }\textbf {\bibinfo {volume} {94}},\ \bibinfo {pages} {034301} (\bibinfo {year} {2005})}\BibitemShut {NoStop}%
\bibitem [{\citenamefont {Segal}(2006)}]{SegalSB2}%
  \BibitemOpen
  \bibfield  {author} {\bibinfo {author} {\bibfnamefont {D.}~\bibnamefont {Segal}},\ }\href {\doibase 10.1103/PhysRevB.73.205415} {\bibfield  {journal} {\bibinfo  {journal} {Phys. Rev. B}\ }\textbf {\bibinfo {volume} {73}},\ \bibinfo {pages} {205415} (\bibinfo {year} {2006})}\BibitemShut {NoStop}%
\bibitem [{\citenamefont {Friedman}\ \emph {et~al.}(2018)\citenamefont {Friedman}, \citenamefont {Agarwalla},\ and\ \citenamefont {Segal}}]{SegalSB3}%
  \BibitemOpen
  \bibfield  {author} {\bibinfo {author} {\bibfnamefont {H.~M.}\ \bibnamefont {Friedman}}, \bibinfo {author} {\bibfnamefont {B.~K.}\ \bibnamefont {Agarwalla}}, \ and\ \bibinfo {author} {\bibfnamefont {D.}~\bibnamefont {Segal}},\ }\href {\doibase 10.1088/1367-2630/aad5fc} {\bibfield  {journal} {\bibinfo  {journal} {New Journal of Physics}\ }\textbf {\bibinfo {volume} {20}},\ \bibinfo {pages} {083026} (\bibinfo {year} {2018})}\BibitemShut {NoStop}%
\bibitem [{\citenamefont {Schinabeck}\ \emph {et~al.}(2016)\citenamefont {Schinabeck}, \citenamefont {Erpenbeck}, \citenamefont {H\"artle},\ and\ \citenamefont {Thoss}}]{HEOMThoss16}%
  \BibitemOpen
  \bibfield  {author} {\bibinfo {author} {\bibfnamefont {C.}~\bibnamefont {Schinabeck}}, \bibinfo {author} {\bibfnamefont {A.}~\bibnamefont {Erpenbeck}}, \bibinfo {author} {\bibfnamefont {R.}~\bibnamefont {H\"artle}}, \ and\ \bibinfo {author} {\bibfnamefont {M.}~\bibnamefont {Thoss}},\ }\href {https://link.aps.org/doi/10.1103/PhysRevB.94.201407} {\bibfield  {journal} {\bibinfo  {journal} {Phys. Rev. B}\ }\textbf {\bibinfo {volume} {94}},\ \bibinfo {pages} {201407} (\bibinfo {year} {2016})}\BibitemShut {NoStop}%
\bibitem [{\citenamefont {Tanimura}(2020)}]{Tanimura20}%
  \BibitemOpen
  \bibfield  {author} {\bibinfo {author} {\bibfnamefont {Y.}~\bibnamefont {Tanimura}},\ }\href {https://doi.org/10.1063/5.0011599} {\bibfield  {journal} {\bibinfo  {journal} {The Journal of Chemical Physics}\ }\textbf {\bibinfo {volume} {153}},\ \bibinfo {pages} {020901} (\bibinfo {year} {2020})}\BibitemShut {NoStop}%
\bibitem [{\citenamefont {Velizhanin}\ \emph {et~al.}(2008)\citenamefont {Velizhanin}, \citenamefont {Wang},\ and\ \citenamefont {Thoss}}]{MCTDH08}%
  \BibitemOpen
  \bibfield  {author} {\bibinfo {author} {\bibfnamefont {K.~A.}\ \bibnamefont {Velizhanin}}, \bibinfo {author} {\bibfnamefont {H.}~\bibnamefont {Wang}}, \ and\ \bibinfo {author} {\bibfnamefont {M.}~\bibnamefont {Thoss}},\ }\href {https://www.elsevier.com/locate/cplett} {\bibfield  {journal} {\bibinfo  {journal} {Chemical Physics Letters}\ }\textbf {\bibinfo {volume} {460}},\ \bibinfo {pages} {325} (\bibinfo {year} {2008})}\BibitemShut {NoStop}%
\bibitem [{\citenamefont {Wang}\ and\ \citenamefont {Thoss}(2009)}]{MCTDH09}%
  \BibitemOpen
  \bibfield  {author} {\bibinfo {author} {\bibfnamefont {H.}~\bibnamefont {Wang}}\ and\ \bibinfo {author} {\bibfnamefont {M.}~\bibnamefont {Thoss}},\ }\href {https://doi.org/10.1063/1.3173823} {\bibfield  {journal} {\bibinfo  {journal} {The Journal of Chemical Physics}\ }\textbf {\bibinfo {volume} {131}},\ \bibinfo {pages} {024114} (\bibinfo {year} {2009})}\BibitemShut {NoStop}%
\bibitem [{\citenamefont {Brenes}\ \emph {et~al.}(2020)\citenamefont {Brenes}, \citenamefont {Mendoza-Arenas}, \citenamefont {Purkayastha}, \citenamefont {Mitchison}, \citenamefont {Clark},\ and\ \citenamefont {Goold}}]{MarlonTN}%
  \BibitemOpen
  \bibfield  {author} {\bibinfo {author} {\bibfnamefont {M.}~\bibnamefont {Brenes}}, \bibinfo {author} {\bibfnamefont {J.~J.}\ \bibnamefont {Mendoza-Arenas}}, \bibinfo {author} {\bibfnamefont {A.}~\bibnamefont {Purkayastha}}, \bibinfo {author} {\bibfnamefont {M.~T.}\ \bibnamefont {Mitchison}}, \bibinfo {author} {\bibfnamefont {S.~R.}\ \bibnamefont {Clark}}, \ and\ \bibinfo {author} {\bibfnamefont {J.}~\bibnamefont {Goold}},\ }\href {https://link.aps.org/doi/10.1103/PhysRevX.10.031040} {\bibfield  {journal} {\bibinfo  {journal} {Phys. Rev. X}\ }\textbf {\bibinfo {volume} {10}},\ \bibinfo {pages} {031040} (\bibinfo {year} {2020})}\BibitemShut {NoStop}%
\bibitem [{\citenamefont {Lacerda}\ \emph {et~al.}(2023)\citenamefont {Lacerda}, \citenamefont {Purkayastha}, \citenamefont {Kewming}, \citenamefont {Landi},\ and\ \citenamefont {Goold}}]{GooldME}%
  \BibitemOpen
  \bibfield  {author} {\bibinfo {author} {\bibfnamefont {A.~M.}\ \bibnamefont {Lacerda}}, \bibinfo {author} {\bibfnamefont {A.}~\bibnamefont {Purkayastha}}, \bibinfo {author} {\bibfnamefont {M.}~\bibnamefont {Kewming}}, \bibinfo {author} {\bibfnamefont {G.~T.}\ \bibnamefont {Landi}}, \ and\ \bibinfo {author} {\bibfnamefont {J.}~\bibnamefont {Goold}},\ }\href {https://link.aps.org/doi/10.1103/PhysRevB.107.195117} {\bibfield  {journal} {\bibinfo  {journal} {Phys. Rev. B}\ }\textbf {\bibinfo {volume} {107}},\ \bibinfo {pages} {195117} (\bibinfo {year} {2023})}\BibitemShut {NoStop}%
\bibitem [{\citenamefont {Nazir}\ and\ \citenamefont {Schaller}(2018)}]{RC_termodynamics}%
  \BibitemOpen
  \bibfield  {author} {\bibinfo {author} {\bibfnamefont {A.}~\bibnamefont {Nazir}}\ and\ \bibinfo {author} {\bibfnamefont {G.}~\bibnamefont {Schaller}},\ }\enquote {\bibinfo {title} {The reaction coordinate mapping in quantum thermodynamics},}\ in\ \href {http://dx.doi.org/10.1007/978-3-319-99046-0_23} {\emph {\bibinfo {booktitle} {Thermodynamics in the Quantum Regime}}}\ (\bibinfo  {publisher} {Springer International Publishing},\ \bibinfo {year} {2018})\ pp.\ \bibinfo {pages} {551--577}\BibitemShut {NoStop}%
\bibitem [{\citenamefont {Castro~Neto}\ \emph {et~al.}(2003)\citenamefont {Castro~Neto}, \citenamefont {Novais}, \citenamefont {Borda}, \citenamefont {Zar\'and},\ and\ \citenamefont {Affleck}}]{Afflek03}%
  \BibitemOpen
  \bibfield  {author} {\bibinfo {author} {\bibfnamefont {A.~H.}\ \bibnamefont {Castro~Neto}}, \bibinfo {author} {\bibfnamefont {E.}~\bibnamefont {Novais}}, \bibinfo {author} {\bibfnamefont {L.}~\bibnamefont {Borda}}, \bibinfo {author} {\bibfnamefont {G.}~\bibnamefont {Zar\'and}}, \ and\ \bibinfo {author} {\bibfnamefont {I.}~\bibnamefont {Affleck}},\ }\href {\doibase 10.1103/PhysRevLett.91.096401} {\bibfield  {journal} {\bibinfo  {journal} {Phys. Rev. Lett.}\ }\textbf {\bibinfo {volume} {91}},\ \bibinfo {pages} {096401} (\bibinfo {year} {2003})}\BibitemShut {NoStop}%
\bibitem [{\citenamefont {Novais}\ \emph {et~al.}(2005)\citenamefont {Novais}, \citenamefont {Castro~Neto}, \citenamefont {Borda}, \citenamefont {Affleck},\ and\ \citenamefont {Zarand}}]{Zarand05}%
  \BibitemOpen
  \bibfield  {author} {\bibinfo {author} {\bibfnamefont {E.}~\bibnamefont {Novais}}, \bibinfo {author} {\bibfnamefont {A.~H.}\ \bibnamefont {Castro~Neto}}, \bibinfo {author} {\bibfnamefont {L.}~\bibnamefont {Borda}}, \bibinfo {author} {\bibfnamefont {I.}~\bibnamefont {Affleck}}, \ and\ \bibinfo {author} {\bibfnamefont {G.}~\bibnamefont {Zarand}},\ }\href {\doibase 10.1103/PhysRevB.72.014417} {\bibfield  {journal} {\bibinfo  {journal} {Phys. Rev. B}\ }\textbf {\bibinfo {volume} {72}},\ \bibinfo {pages} {014417} (\bibinfo {year} {2005})}\BibitemShut {NoStop}%
\bibitem [{\citenamefont {Guo}\ \emph {et~al.}(2012)\citenamefont {Guo}, \citenamefont {Weichselbaum}, \citenamefont {von Delft},\ and\ \citenamefont {Vojta}}]{Vojta12}%
  \BibitemOpen
  \bibfield  {author} {\bibinfo {author} {\bibfnamefont {C.}~\bibnamefont {Guo}}, \bibinfo {author} {\bibfnamefont {A.}~\bibnamefont {Weichselbaum}}, \bibinfo {author} {\bibfnamefont {J.}~\bibnamefont {von Delft}}, \ and\ \bibinfo {author} {\bibfnamefont {M.}~\bibnamefont {Vojta}},\ }\href {\doibase 10.1103/PhysRevLett.108.160401} {\bibfield  {journal} {\bibinfo  {journal} {Phys. Rev. Lett.}\ }\textbf {\bibinfo {volume} {108}},\ \bibinfo {pages} {160401} (\bibinfo {year} {2012})}\BibitemShut {NoStop}%
\bibitem [{\citenamefont {Kohler}\ \emph {et~al.}(2013)\citenamefont {Kohler}, \citenamefont {Hackl},\ and\ \citenamefont {Kehrein}}]{flow13}%
  \BibitemOpen
  \bibfield  {author} {\bibinfo {author} {\bibfnamefont {H.}~\bibnamefont {Kohler}}, \bibinfo {author} {\bibfnamefont {A.}~\bibnamefont {Hackl}}, \ and\ \bibinfo {author} {\bibfnamefont {S.}~\bibnamefont {Kehrein}},\ }\href {\doibase 10.1103/PhysRevB.88.205122} {\bibfield  {journal} {\bibinfo  {journal} {Phys. Rev. B}\ }\textbf {\bibinfo {volume} {88}},\ \bibinfo {pages} {205122} (\bibinfo {year} {2013})}\BibitemShut {NoStop}%
\bibitem [{\citenamefont {Palm}\ and\ \citenamefont {Nalbach}(2018)}]{non-commuting_PI}%
  \BibitemOpen
  \bibfield  {author} {\bibinfo {author} {\bibfnamefont {T.}~\bibnamefont {Palm}}\ and\ \bibinfo {author} {\bibfnamefont {P.}~\bibnamefont {Nalbach}},\ }\href {https://doi.org/10.1063/1.5051652} {\bibfield  {journal} {\bibinfo  {journal} {The Journal of Chemical Physics}\ }\textbf {\bibinfo {volume} {149}},\ \bibinfo {pages} {214103} (\bibinfo {year} {2018})}\BibitemShut {NoStop}%
\bibitem [{\citenamefont {Palm}\ and\ \citenamefont {Nalbach}(2019)}]{Nalbach19}%
  \BibitemOpen
  \bibfield  {author} {\bibinfo {author} {\bibfnamefont {T.}~\bibnamefont {Palm}}\ and\ \bibinfo {author} {\bibfnamefont {P.}~\bibnamefont {Nalbach}},\ }\href {\doibase 10.1063/1.5098467} {\bibfield  {journal} {\bibinfo  {journal} {The Journal of Chemical Physics}\ }\textbf {\bibinfo {volume} {150}},\ \bibinfo {pages} {234108} (\bibinfo {year} {2019})}\BibitemShut {NoStop}%
\bibitem [{\citenamefont {Richter}\ and\ \citenamefont {Hughes}(2022)}]{Richter_2022}%
  \BibitemOpen
  \bibfield  {author} {\bibinfo {author} {\bibfnamefont {M.}~\bibnamefont {Richter}}\ and\ \bibinfo {author} {\bibfnamefont {S.}~\bibnamefont {Hughes}},\ }\href {https://link.aps.org/doi/10.1103/PhysRevLett.128.167403} {\bibfield  {journal} {\bibinfo  {journal} {Phys. Rev. Lett.}\ }\textbf {\bibinfo {volume} {128}},\ \bibinfo {pages} {167403} (\bibinfo {year} {2022})}\BibitemShut {NoStop}%
\bibitem [{\citenamefont {Duan}\ \emph {et~al.}(2020)\citenamefont {Duan}, \citenamefont {Hsieh}, \citenamefont {Liu}, \citenamefont {Wu},\ and\ \citenamefont {Cao}}]{Cao_unusual_transport}%
  \BibitemOpen
  \bibfield  {author} {\bibinfo {author} {\bibfnamefont {C.}~\bibnamefont {Duan}}, \bibinfo {author} {\bibfnamefont {C.-Y.}\ \bibnamefont {Hsieh}}, \bibinfo {author} {\bibfnamefont {J.}~\bibnamefont {Liu}}, \bibinfo {author} {\bibfnamefont {J.}~\bibnamefont {Wu}}, \ and\ \bibinfo {author} {\bibfnamefont {J.}~\bibnamefont {Cao}},\ }\href {https://doi.org/10.1021/acs.jpclett.0c00985} {\bibfield  {journal} {\bibinfo  {journal} {The Journal of Physical Chemistry Letters}\ }\textbf {\bibinfo {volume} {11}},\ \bibinfo {pages} {4080} (\bibinfo {year} {2020})}\BibitemShut {NoStop}%
\bibitem [{\citenamefont {Maguire}\ \emph {et~al.}(2019)\citenamefont {Maguire}, \citenamefont {Iles-Smith},\ and\ \citenamefont {Nazir}}]{AhsanNC19}%
  \BibitemOpen
  \bibfield  {author} {\bibinfo {author} {\bibfnamefont {H.}~\bibnamefont {Maguire}}, \bibinfo {author} {\bibfnamefont {J.}~\bibnamefont {Iles-Smith}}, \ and\ \bibinfo {author} {\bibfnamefont {A.}~\bibnamefont {Nazir}},\ }\href {\doibase 10.1103/PhysRevLett.123.093601} {\bibfield  {journal} {\bibinfo  {journal} {Phys. Rev. Lett.}\ }\textbf {\bibinfo {volume} {123}},\ \bibinfo {pages} {093601} (\bibinfo {year} {2019})}\BibitemShut {NoStop}%
\bibitem [{\citenamefont {Zhang}\ \emph {et~al.}(2024)\citenamefont {Zhang}, \citenamefont {Cao},\ and\ \citenamefont {He}}]{Dahai24}%
  \BibitemOpen
  \bibfield  {author} {\bibinfo {author} {\bibfnamefont {X.}~\bibnamefont {Zhang}}, \bibinfo {author} {\bibfnamefont {X.}~\bibnamefont {Cao}}, \ and\ \bibinfo {author} {\bibfnamefont {D.}~\bibnamefont {He}},\ }\href {https://link.aps.org/doi/10.1103/PhysRevB.109.245415} {\bibfield  {journal} {\bibinfo  {journal} {Phys. Rev. B}\ }\textbf {\bibinfo {volume} {109}},\ \bibinfo {pages} {245415} (\bibinfo {year} {2024})}\BibitemShut {NoStop}%
\bibitem [{\citenamefont {Schaller}\ \emph {et~al.}(2016)\citenamefont {Schaller}, \citenamefont {Giusteri},\ and\ \citenamefont {Celardo}}]{Schaller16}%
  \BibitemOpen
  \bibfield  {author} {\bibinfo {author} {\bibfnamefont {G.}~\bibnamefont {Schaller}}, \bibinfo {author} {\bibfnamefont {G.~G.}\ \bibnamefont {Giusteri}}, \ and\ \bibinfo {author} {\bibfnamefont {G.~L.}\ \bibnamefont {Celardo}},\ }\href {\doibase 10.1103/PhysRevE.94.032135} {\bibfield  {journal} {\bibinfo  {journal} {Phys. Rev. E}\ }\textbf {\bibinfo {volume} {94}},\ \bibinfo {pages} {032135} (\bibinfo {year} {2016})}\BibitemShut {NoStop}%
\bibitem [{\citenamefont {Hogg}\ \emph {et~al.}(2024)\citenamefont {Hogg}, \citenamefont {Cerisola}, \citenamefont {Cresser}, \citenamefont {Horsley},\ and\ \citenamefont {Anders}}]{Janet3B}%
  \BibitemOpen
  \bibfield  {author} {\bibinfo {author} {\bibfnamefont {C.~R.}\ \bibnamefont {Hogg}}, \bibinfo {author} {\bibfnamefont {F.}~\bibnamefont {Cerisola}}, \bibinfo {author} {\bibfnamefont {J.~D.}\ \bibnamefont {Cresser}}, \bibinfo {author} {\bibfnamefont {S.~A.~R.}\ \bibnamefont {Horsley}}, \ and\ \bibinfo {author} {\bibfnamefont {J.}~\bibnamefont {Anders}},\ }\href {https://doi.org/10.22331/q-2024-05-23-1357} {\bibfield  {journal} {\bibinfo  {journal} {{Quantum}}\ }\textbf {\bibinfo {volume} {8}},\ \bibinfo {pages} {1357} (\bibinfo {year} {2024})}\BibitemShut {NoStop}%
\bibitem [{\citenamefont {Palm}\ and\ \citenamefont {Nalbach}(2021)}]{Suppressing_relaxation_through_dephasing}%
  \BibitemOpen
  \bibfield  {author} {\bibinfo {author} {\bibfnamefont {T.}~\bibnamefont {Palm}}\ and\ \bibinfo {author} {\bibfnamefont {P.}~\bibnamefont {Nalbach}},\ }\href {https://link.aps.org/doi/10.1103/PhysRevA.103.022206} {\bibfield  {journal} {\bibinfo  {journal} {Phys. Rev. A}\ }\textbf {\bibinfo {volume} {103}},\ \bibinfo {pages} {022206} (\bibinfo {year} {2021})}\BibitemShut {NoStop}%
\bibitem [{\citenamefont {Anto-Sztrikacs}\ \emph {et~al.}(2023{\natexlab{a}})\citenamefont {Anto-Sztrikacs}, \citenamefont {Nazir},\ and\ \citenamefont {Segal}}]{Nick_PRX}%
  \BibitemOpen
  \bibfield  {author} {\bibinfo {author} {\bibfnamefont {N.}~\bibnamefont {Anto-Sztrikacs}}, \bibinfo {author} {\bibfnamefont {A.}~\bibnamefont {Nazir}}, \ and\ \bibinfo {author} {\bibfnamefont {D.}~\bibnamefont {Segal}},\ }\href {https://link.aps.org/doi/10.1103/PRXQuantum.4.020307} {\bibfield  {journal} {\bibinfo  {journal} {PRX Quantum}\ }\textbf {\bibinfo {volume} {4}},\ \bibinfo {pages} {020307} (\bibinfo {year} {2023}{\natexlab{a}})}\BibitemShut {NoStop}%
\bibitem [{\citenamefont {Brenes}\ \emph {et~al.}(2024)\citenamefont {Brenes}, \citenamefont {Min}, \citenamefont {Anto-Sztrikacs}, \citenamefont {Bar-Gill},\ and\ \citenamefont {Segal}}]{brenes2024bathinduced}%
  \BibitemOpen
  \bibfield  {author} {\bibinfo {author} {\bibfnamefont {M.}~\bibnamefont {Brenes}}, \bibinfo {author} {\bibfnamefont {B.}~\bibnamefont {Min}}, \bibinfo {author} {\bibfnamefont {N.}~\bibnamefont {Anto-Sztrikacs}}, \bibinfo {author} {\bibfnamefont {N.}~\bibnamefont {Bar-Gill}}, \ and\ \bibinfo {author} {\bibfnamefont {D.}~\bibnamefont {Segal}},\ }\href {http://dx.doi.org/10.1063/5.0207028} {\bibfield  {journal} {\bibinfo  {journal} {The Journal of Chemical Physics}\ }\textbf {\bibinfo {volume} {160}},\ \bibinfo {pages} {244106} (\bibinfo {year} {2024})}\BibitemShut {NoStop}%
\bibitem [{\citenamefont {Anto-Sztrikacs}\ \emph {et~al.}(2023{\natexlab{b}})\citenamefont {Anto-Sztrikacs}, \citenamefont {Min}, \citenamefont {Brenes},\ and\ \citenamefont {Segal}}]{Anto_Sztrikacs_2023}%
  \BibitemOpen
  \bibfield  {author} {\bibinfo {author} {\bibfnamefont {N.}~\bibnamefont {Anto-Sztrikacs}}, \bibinfo {author} {\bibfnamefont {B.}~\bibnamefont {Min}}, \bibinfo {author} {\bibfnamefont {M.}~\bibnamefont {Brenes}}, \ and\ \bibinfo {author} {\bibfnamefont {D.}~\bibnamefont {Segal}},\ }\href {https://link.aps.org/doi/10.1103/PhysRevB.108.115437} {\bibfield  {journal} {\bibinfo  {journal} {Phys. Rev. B}\ }\textbf {\bibinfo {volume} {108}},\ \bibinfo {pages} {115437} (\bibinfo {year} {2023}{\natexlab{b}})}\BibitemShut {NoStop}%
\bibitem [{\citenamefont {Anto-Sztrikacs}\ and\ \citenamefont {Segal}(2021{\natexlab{a}})}]{Anto-Sztrikacs_2021}%
  \BibitemOpen
  \bibfield  {author} {\bibinfo {author} {\bibfnamefont {N.}~\bibnamefont {Anto-Sztrikacs}}\ and\ \bibinfo {author} {\bibfnamefont {D.}~\bibnamefont {Segal}},\ }\href {https://dx.doi.org/10.1088/1367-2630/ac02df} {\bibfield  {journal} {\bibinfo  {journal} {New Journal of Physics}\ }\textbf {\bibinfo {volume} {23}},\ \bibinfo {pages} {063036} (\bibinfo {year} {2021}{\natexlab{a}})}\BibitemShut {NoStop}%
\bibitem [{\citenamefont {Strasberg}\ \emph {et~al.}(2016)\citenamefont {Strasberg}, \citenamefont {Schaller}, \citenamefont {Lambert},\ and\ \citenamefont {Brandes}}]{non-equalibrium_strong_coupling}%
  \BibitemOpen
  \bibfield  {author} {\bibinfo {author} {\bibfnamefont {P.}~\bibnamefont {Strasberg}}, \bibinfo {author} {\bibfnamefont {G.}~\bibnamefont {Schaller}}, \bibinfo {author} {\bibfnamefont {N.}~\bibnamefont {Lambert}}, \ and\ \bibinfo {author} {\bibfnamefont {T.}~\bibnamefont {Brandes}},\ }\href {http://dx.doi.org/10.1088/1367-2630/18/7/073007} {\bibfield  {journal} {\bibinfo  {journal} {New Journal of Physics}\ }\textbf {\bibinfo {volume} {18}},\ \bibinfo {pages} {073007} (\bibinfo {year} {2016})}\BibitemShut {NoStop}%
\bibitem [{\citenamefont {Min}\ \emph {et~al.}(2024)\citenamefont {Min}, \citenamefont {Anto-Sztrikacs}, \citenamefont {Brenes},\ and\ \citenamefont {Segal}}]{min2024bathengineering}%
  \BibitemOpen
  \bibfield  {author} {\bibinfo {author} {\bibfnamefont {B.}~\bibnamefont {Min}}, \bibinfo {author} {\bibfnamefont {N.}~\bibnamefont {Anto-Sztrikacs}}, \bibinfo {author} {\bibfnamefont {M.}~\bibnamefont {Brenes}}, \ and\ \bibinfo {author} {\bibfnamefont {D.}~\bibnamefont {Segal}},\ }\href {https://link.aps.org/doi/10.1103/PhysRevLett.132.266701} {\bibfield  {journal} {\bibinfo  {journal} {Phys. Rev. Lett.}\ }\textbf {\bibinfo {volume} {132}},\ \bibinfo {pages} {266701} (\bibinfo {year} {2024})}\BibitemShut {NoStop}%
\bibitem [{\citenamefont {Tamascelli}(2020)}]{TEDOPA}%
  \BibitemOpen
  \bibfield  {author} {\bibinfo {author} {\bibfnamefont {D.}~\bibnamefont {Tamascelli}},\ }\href {http://dx.doi.org/10.3390/e22111320} {\bibfield  {journal} {\bibinfo  {journal} {Entropy}\ }\textbf {\bibinfo {volume} {22}},\ \bibinfo {pages} {1320} (\bibinfo {year} {2020})}\BibitemShut {NoStop}%
\bibitem [{\citenamefont {Makri}(1995)}]{Makri_PI}%
  \BibitemOpen
  \bibfield  {author} {\bibinfo {author} {\bibfnamefont {N.}~\bibnamefont {Makri}},\ }\href {https://doi.org/10.1063/1.531046} {\bibfield  {journal} {\bibinfo  {journal} {Journal of Mathematical Physics}\ }\textbf {\bibinfo {volume} {36}},\ \bibinfo {pages} {2430} (\bibinfo {year} {1995})}\BibitemShut {NoStop}%
\bibitem [{\citenamefont {Dani}\ and\ \citenamefont {Makri}(2021)}]{10.1063/5.0066891}%
  \BibitemOpen
  \bibfield  {author} {\bibinfo {author} {\bibfnamefont {R.}~\bibnamefont {Dani}}\ and\ \bibinfo {author} {\bibfnamefont {N.}~\bibnamefont {Makri}},\ }\href {https://doi.org/10.1063/5.0066891} {\bibfield  {journal} {\bibinfo  {journal} {The Journal of Chemical Physics}\ }\textbf {\bibinfo {volume} {155}},\ \bibinfo {pages} {234705} (\bibinfo {year} {2021})}\BibitemShut {NoStop}%
\bibitem [{\citenamefont {Kundu}\ and\ \citenamefont {Makri}(2023{\natexlab{a}})}]{PATHSUM}%
  \BibitemOpen
  \bibfield  {author} {\bibinfo {author} {\bibfnamefont {S.}~\bibnamefont {Kundu}}\ and\ \bibinfo {author} {\bibfnamefont {N.}~\bibnamefont {Makri}},\ }\href {https://doi.org/10.1063/5.0151748} {\bibfield  {journal} {\bibinfo  {journal} {The Journal of Chemical Physics}\ }\textbf {\bibinfo {volume} {158}},\ \bibinfo {pages} {224801} (\bibinfo {year} {2023}{\natexlab{a}})}\BibitemShut {NoStop}%
\bibitem [{\citenamefont {Gelin}\ and\ \citenamefont {Thoss}(2009)}]{Thoss09}%
  \BibitemOpen
  \bibfield  {author} {\bibinfo {author} {\bibfnamefont {M.~F.}\ \bibnamefont {Gelin}}\ and\ \bibinfo {author} {\bibfnamefont {M.}~\bibnamefont {Thoss}},\ }\href {https://link.aps.org/doi/10.1103/PhysRevE.79.051121} {\bibfield  {journal} {\bibinfo  {journal} {Phys. Rev. E}\ }\textbf {\bibinfo {volume} {79}},\ \bibinfo {pages} {051121} (\bibinfo {year} {2009})}\BibitemShut {NoStop}%
\bibitem [{\citenamefont {Xu}\ and\ \citenamefont {Cao}(2016)}]{CaoFP}%
  \BibitemOpen
  \bibfield  {author} {\bibinfo {author} {\bibfnamefont {D.}~\bibnamefont {Xu}}\ and\ \bibinfo {author} {\bibfnamefont {J.}~\bibnamefont {Cao}},\ }\href {https://doi.org/10.1007/s11467-016-0540-2} {\bibfield  {journal} {\bibinfo  {journal} {Frontiers of Physics}\ }\textbf {\bibinfo {volume} {11}},\ \bibinfo {pages} {110308} (\bibinfo {year} {2016})}\BibitemShut {NoStop}%
\bibitem [{\citenamefont {Miller}(2018)}]{Miller2018}%
  \BibitemOpen
  \bibfield  {author} {\bibinfo {author} {\bibfnamefont {H.~J.~D.}\ \bibnamefont {Miller}},\ }\enquote {\bibinfo {title} {Hamiltonian of mean force for strongly-coupled systems},}\ in\ \href {\doibase 10.1007/978-3-319-99046-0_22} {\emph {\bibinfo {booktitle} {Thermodynamics in the Quantum Regime: Fundamental Aspects and New Directions}}},\ \bibinfo {editor} {edited by\ \bibinfo {editor} {\bibfnamefont {F.}~\bibnamefont {Binder}}, \bibinfo {editor} {\bibfnamefont {L.~A.}\ \bibnamefont {Correa}}, \bibinfo {editor} {\bibfnamefont {C.}~\bibnamefont {Gogolin}}, \bibinfo {editor} {\bibfnamefont {J.}~\bibnamefont {Anders}}, \ and\ \bibinfo {editor} {\bibfnamefont {G.}~\bibnamefont {Adesso}}}\ (\bibinfo  {publisher} {Springer International Publishing},\ \bibinfo {address} {Cham},\ \bibinfo {year} {2018})\ pp.\ \bibinfo {pages} {531--549}\BibitemShut {NoStop}%
\bibitem [{\citenamefont {Timofeev}\ and\ \citenamefont {Trushechkin}(2022)}]{Anton22}%
  \BibitemOpen
  \bibfield  {author} {\bibinfo {author} {\bibfnamefont {G.~M.}\ \bibnamefont {Timofeev}}\ and\ \bibinfo {author} {\bibfnamefont {A.~S.}\ \bibnamefont {Trushechkin}},\ }\href {https://doi.org/10.1142/S0217751X22430217} {\bibfield  {journal} {\bibinfo  {journal} {International Journal of Modern Physics A}\ }\textbf {\bibinfo {volume} {37}},\ \bibinfo {pages} {2243021} (\bibinfo {year} {2022})}\BibitemShut {NoStop}%
\bibitem [{\citenamefont {Cresser}\ and\ \citenamefont {Anders}(2021)}]{strong_limit_MFGS}%
  \BibitemOpen
  \bibfield  {author} {\bibinfo {author} {\bibfnamefont {J.~D.}\ \bibnamefont {Cresser}}\ and\ \bibinfo {author} {\bibfnamefont {J.}~\bibnamefont {Anders}},\ }\href {https://link.aps.org/doi/10.1103/PhysRevLett.127.250601} {\bibfield  {journal} {\bibinfo  {journal} {Phys. Rev. Lett.}\ }\textbf {\bibinfo {volume} {127}},\ \bibinfo {pages} {250601} (\bibinfo {year} {2021})}\BibitemShut {NoStop}%
\bibitem [{\citenamefont {Trushechkin}\ \emph {et~al.}(2022)\citenamefont {Trushechkin}, \citenamefont {Merkli}, \citenamefont {Cresser},\ and\ \citenamefont {Anders}}]{AndersAVS}%
  \BibitemOpen
  \bibfield  {author} {\bibinfo {author} {\bibfnamefont {A.~S.}\ \bibnamefont {Trushechkin}}, \bibinfo {author} {\bibfnamefont {M.}~\bibnamefont {Merkli}}, \bibinfo {author} {\bibfnamefont {J.~D.}\ \bibnamefont {Cresser}}, \ and\ \bibinfo {author} {\bibfnamefont {J.}~\bibnamefont {Anders}},\ }\href {https://doi.org/10.1116/5.0073853} {\bibfield  {journal} {\bibinfo  {journal} {AVS Quantum Science}\ }\textbf {\bibinfo {volume} {4}},\ \bibinfo {pages} {012301} (\bibinfo {year} {2022})}\BibitemShut {NoStop}%
\bibitem [{\citenamefont {Hughes}\ \emph {et~al.}(2009{\natexlab{a}})\citenamefont {Hughes}, \citenamefont {Christ},\ and\ \citenamefont {Burghardt}}]{Irene1}%
  \BibitemOpen
  \bibfield  {author} {\bibinfo {author} {\bibfnamefont {K.~H.}\ \bibnamefont {Hughes}}, \bibinfo {author} {\bibfnamefont {C.~D.}\ \bibnamefont {Christ}}, \ and\ \bibinfo {author} {\bibfnamefont {I.}~\bibnamefont {Burghardt}},\ }\href {https://doi.org/10.1063/1.3159671} {\bibfield  {journal} {\bibinfo  {journal} {The Journal of Chemical Physics}\ }\textbf {\bibinfo {volume} {131}},\ \bibinfo {pages} {024109} (\bibinfo {year} {2009}{\natexlab{a}})}\BibitemShut {NoStop}%
\bibitem [{\citenamefont {Hughes}\ \emph {et~al.}(2009{\natexlab{b}})\citenamefont {Hughes}, \citenamefont {Christ},\ and\ \citenamefont {Burghardt}}]{Irene2}%
  \BibitemOpen
  \bibfield  {author} {\bibinfo {author} {\bibfnamefont {K.~H.}\ \bibnamefont {Hughes}}, \bibinfo {author} {\bibfnamefont {C.~D.}\ \bibnamefont {Christ}}, \ and\ \bibinfo {author} {\bibfnamefont {I.}~\bibnamefont {Burghardt}},\ }\href {https://doi.org/10.1063/1.3226343} {\bibfield  {journal} {\bibinfo  {journal} {The Journal of Chemical Physics}\ }\textbf {\bibinfo {volume} {131}},\ \bibinfo {pages} {124108} (\bibinfo {year} {2009}{\natexlab{b}})}\BibitemShut {NoStop}%
\bibitem [{\citenamefont {Hsieh}\ \emph {et~al.}(2019)\citenamefont {Hsieh}, \citenamefont {Liu}, \citenamefont {Duan},\ and\ \citenamefont {Cao}}]{variational_polaron}%
  \BibitemOpen
  \bibfield  {author} {\bibinfo {author} {\bibfnamefont {C.}~\bibnamefont {Hsieh}}, \bibinfo {author} {\bibfnamefont {J.}~\bibnamefont {Liu}}, \bibinfo {author} {\bibfnamefont {C.}~\bibnamefont {Duan}}, \ and\ \bibinfo {author} {\bibfnamefont {J.}~\bibnamefont {Cao}},\ }\href {https://doi.org/10.1021/acs.jpcc.9b05607} {\bibfield  {journal} {\bibinfo  {journal} {The Journal of Physical Chemistry C}\ }\textbf {\bibinfo {volume} {123}},\ \bibinfo {pages} {17196} (\bibinfo {year} {2019})}\BibitemShut {NoStop}%
\bibitem [{\citenamefont {Kundu}\ and\ \citenamefont {Makri}(2023{\natexlab{b}})}]{Makri23}%
  \BibitemOpen
  \bibfield  {author} {\bibinfo {author} {\bibfnamefont {S.}~\bibnamefont {Kundu}}\ and\ \bibinfo {author} {\bibfnamefont {N.}~\bibnamefont {Makri}},\ }\href {https://doi.org/10.1063/5.0151748} {\bibfield  {journal} {\bibinfo  {journal} {The Journal of Chemical Physics}\ }\textbf {\bibinfo {volume} {158}},\ \bibinfo {pages} {224801} (\bibinfo {year} {2023}{\natexlab{b}})}\BibitemShut {NoStop}%
\bibitem [{\citenamefont {Min}\ \emph {et~al.}()\citenamefont {Min}, \citenamefont {Agarwal},\ and\ \citenamefont {Segal}}]{SSH}%
  \BibitemOpen
  \bibfield  {author} {\bibinfo {author} {\bibfnamefont {B.}~\bibnamefont {Min}}, \bibinfo {author} {\bibfnamefont {K.}~\bibnamefont {Agarwal}}, \ and\ \bibinfo {author} {\bibfnamefont {D.}~\bibnamefont {Segal}},\ }\href@noop {} {\bibinfo  {journal} {arXiv:2406.13878}\ }\BibitemShut {NoStop}%
\bibitem [{\citenamefont {Ahmed}\ and\ \citenamefont {Hussein}(2019)}]{HOp}%
  \BibitemOpen
\bibfield  {journal} {  }\bibfield  {author} {\bibinfo {author} {\bibfnamefont {A.~K.}\ \bibnamefont {Ahmed}}\ and\ \bibinfo {author} {\bibfnamefont {H.~G.}\ \bibnamefont {Hussein}},\ }\href {https://doi.org/10.1063/1.5138510} {\bibfield  {journal} {\bibinfo  {journal} {AIP Conference Proceedings}\ }\textbf {\bibinfo {volume} {2190}},\ \bibinfo {pages} {020024} (\bibinfo {year} {2019})}\BibitemShut {NoStop}%
\bibitem [{\citenamefont {Oteo}(1991)}]{nested_commutators}%
  \BibitemOpen
  \bibfield  {author} {\bibinfo {author} {\bibfnamefont {J.~A.}\ \bibnamefont {Oteo}},\ }\href {https://doi.org/10.1063/1.529428} {\bibfield  {journal} {\bibinfo  {journal} {Journal of Mathematical Physics}\ }\textbf {\bibinfo {volume} {32}},\ \bibinfo {pages} {419} (\bibinfo {year} {1991})}\BibitemShut {NoStop}%
\bibitem [{\citenamefont {Anto-Sztrikacs}\ and\ \citenamefont {Segal}(2021{\natexlab{b}})}]{NickNM}%
  \BibitemOpen
  \bibfield  {author} {\bibinfo {author} {\bibfnamefont {N.}~\bibnamefont {Anto-Sztrikacs}}\ and\ \bibinfo {author} {\bibfnamefont {D.}~\bibnamefont {Segal}},\ }\href {https://link.aps.org/doi/10.1103/PhysRevA.104.052617} {\bibfield  {journal} {\bibinfo  {journal} {Phys. Rev. A}\ }\textbf {\bibinfo {volume} {104}},\ \bibinfo {pages} {052617} (\bibinfo {year} {2021}{\natexlab{b}})}\BibitemShut {NoStop}%
\bibitem [{\citenamefont {De~Filippis}\ \emph {et~al.}(2021)\citenamefont {De~Filippis}, \citenamefont {de~Candia}, \citenamefont {Mishchenko}, \citenamefont {Cangemi}, \citenamefont {Nocera}, \citenamefont {Mishchenko}, \citenamefont {Sassetti}, \citenamefont {Fazio}, \citenamefont {Nagaosa},\ and\ \citenamefont {Cataudella}}]{chain21}%
  \BibitemOpen
  \bibfield  {author} {\bibinfo {author} {\bibfnamefont {G.}~\bibnamefont {De~Filippis}}, \bibinfo {author} {\bibfnamefont {A.}~\bibnamefont {de~Candia}}, \bibinfo {author} {\bibfnamefont {A.~S.}\ \bibnamefont {Mishchenko}}, \bibinfo {author} {\bibfnamefont {L.~M.}\ \bibnamefont {Cangemi}}, \bibinfo {author} {\bibfnamefont {A.}~\bibnamefont {Nocera}}, \bibinfo {author} {\bibfnamefont {P.~A.}\ \bibnamefont {Mishchenko}}, \bibinfo {author} {\bibfnamefont {M.}~\bibnamefont {Sassetti}}, \bibinfo {author} {\bibfnamefont {R.}~\bibnamefont {Fazio}}, \bibinfo {author} {\bibfnamefont {N.}~\bibnamefont {Nagaosa}}, \ and\ \bibinfo {author} {\bibfnamefont {V.}~\bibnamefont {Cataudella}},\ }\href {https://link.aps.org/doi/10.1103/PhysRevB.104.L060410} {\bibfield  {journal} {\bibinfo  {journal} {Phys. Rev. B}\ }\textbf {\bibinfo {volume} {104}},\ \bibinfo {pages} {L060410} (\bibinfo {year} {2021})}\BibitemShut {NoStop}%
\bibitem [{\citenamefont {Butcher}\ \emph {et~al.}(2022)\citenamefont {Butcher}, \citenamefont {Pixley},\ and\ \citenamefont {Nevidomskyy}}]{PhysRevB.105.L180407}%
  \BibitemOpen
  \bibfield  {author} {\bibinfo {author} {\bibfnamefont {M.~W.}\ \bibnamefont {Butcher}}, \bibinfo {author} {\bibfnamefont {J.~H.}\ \bibnamefont {Pixley}}, \ and\ \bibinfo {author} {\bibfnamefont {A.~H.}\ \bibnamefont {Nevidomskyy}},\ }\href {https://link.aps.org/doi/10.1103/PhysRevB.105.L180407} {\bibfield  {journal} {\bibinfo  {journal} {Phys. Rev. B}\ }\textbf {\bibinfo {volume} {105}},\ \bibinfo {pages} {L180407} (\bibinfo {year} {2022})}\BibitemShut {NoStop}%
\bibitem [{\citenamefont {Weber}()}]{weber2023quantumspinchainsbond}%
  \BibitemOpen
  \bibfield  {author} {\bibinfo {author} {\bibfnamefont {M.}~\bibnamefont {Weber}},\ }\href {https://arxiv.org/abs/2310.11525} {\bibinfo  {journal} {arXiv:2310.11525}\ }\BibitemShut {NoStop}%
\bibitem [{\citenamefont {Cai}\ \emph {et~al.}(2014)\citenamefont {Cai}, \citenamefont {Schollw\"ock},\ and\ \citenamefont {Pollet}}]{PhysRevLett.113.260403}%
  \BibitemOpen
\bibfield  {journal} {  }\bibfield  {author} {\bibinfo {author} {\bibfnamefont {Z.}~\bibnamefont {Cai}}, \bibinfo {author} {\bibfnamefont {U.}~\bibnamefont {Schollw\"ock}}, \ and\ \bibinfo {author} {\bibfnamefont {L.}~\bibnamefont {Pollet}},\ }\href {https://link.aps.org/doi/10.1103/PhysRevLett.113.260403} {\bibfield  {journal} {\bibinfo  {journal} {Phys. Rev. Lett.}\ }\textbf {\bibinfo {volume} {113}},\ \bibinfo {pages} {260403} (\bibinfo {year} {2014})}\BibitemShut {NoStop}%
\bibitem [{\citenamefont {Weber}\ \emph {et~al.}(2022)\citenamefont {Weber}, \citenamefont {Luitz},\ and\ \citenamefont {Assaad}}]{PhysRevLett.129.056402}%
  \BibitemOpen
  \bibfield  {author} {\bibinfo {author} {\bibfnamefont {M.}~\bibnamefont {Weber}}, \bibinfo {author} {\bibfnamefont {D.~J.}\ \bibnamefont {Luitz}}, \ and\ \bibinfo {author} {\bibfnamefont {F.~F.}\ \bibnamefont {Assaad}},\ }\href {https://link.aps.org/doi/10.1103/PhysRevLett.129.056402} {\bibfield  {journal} {\bibinfo  {journal} {Phys. Rev. Lett.}\ }\textbf {\bibinfo {volume} {129}},\ \bibinfo {pages} {056402} (\bibinfo {year} {2022})}\BibitemShut {NoStop}%
\bibitem [{\citenamefont {Kitaev}(2001)}]{Kitaev_2001}%
  \BibitemOpen
  \bibfield  {author} {\bibinfo {author} {\bibfnamefont {A.~Y.}\ \bibnamefont {Kitaev}},\ }\href {http://dx.doi.org/10.1070/1063-7869/44/10S/S29} {\bibfield  {journal} {\bibinfo  {journal} {Physics-Uspekhi}\ }\textbf {\bibinfo {volume} {44}},\ \bibinfo {pages} {131} (\bibinfo {year} {2001})}\BibitemShut {NoStop}%
\bibitem [{\citenamefont {Kitaev}(2006)}]{Kitaev_2006}%
  \BibitemOpen
  \bibfield  {author} {\bibinfo {author} {\bibfnamefont {A.}~\bibnamefont {Kitaev}},\ }\href {http://dx.doi.org/10.1016/j.aop.2005.10.005} {\bibfield  {journal} {\bibinfo  {journal} {Annals of Physics}\ }\textbf {\bibinfo {volume} {321}},\ \bibinfo {pages} {2} (\bibinfo {year} {2006})}\BibitemShut {NoStop}%
\bibitem [{\citenamefont {Xu}\ and\ \citenamefont {Kee}()}]{xu2024manipulating}%
  \BibitemOpen
  \bibfield  {author} {\bibinfo {author} {\bibfnamefont {H.}~\bibnamefont {Xu}}\ and\ \bibinfo {author} {\bibfnamefont {H.-Y.}\ \bibnamefont {Kee}},\ }\href@noop {} {\bibinfo  {journal} {arXiv:2404.16099}\ }\BibitemShut {NoStop}%
\bibitem [{\citenamefont {Laurell}\ \emph {et~al.}(2023)\citenamefont {Laurell}, \citenamefont {Alvarez},\ and\ \citenamefont {Dagotto}}]{chain23}%
  \BibitemOpen
\bibfield  {journal} {  }\bibfield  {author} {\bibinfo {author} {\bibfnamefont {P.}~\bibnamefont {Laurell}}, \bibinfo {author} {\bibfnamefont {G.}~\bibnamefont {Alvarez}}, \ and\ \bibinfo {author} {\bibfnamefont {E.}~\bibnamefont {Dagotto}},\ }\href {https://link.aps.org/doi/10.1103/PhysRevB.107.104414} {\bibfield  {journal} {\bibinfo  {journal} {Phys. Rev. B}\ }\textbf {\bibinfo {volume} {107}},\ \bibinfo {pages} {104414} (\bibinfo {year} {2023})}\BibitemShut {NoStop}%
\bibitem [{\citenamefont {Gordon}\ and\ \citenamefont {Kee}(2022)}]{chain22}%
  \BibitemOpen
  \bibfield  {author} {\bibinfo {author} {\bibfnamefont {J.~S.}\ \bibnamefont {Gordon}}\ and\ \bibinfo {author} {\bibfnamefont {H.-Y.}\ \bibnamefont {Kee}},\ }\href {https://link.aps.org/doi/10.1103/PhysRevResearch.4.013205} {\bibfield  {journal} {\bibinfo  {journal} {Phys. Rev. Res.}\ }\textbf {\bibinfo {volume} {4}},\ \bibinfo {pages} {013205} (\bibinfo {year} {2022})}\BibitemShut {NoStop}%
\bibitem [{\citenamefont {Morris}\ \emph {et~al.}(2021)\citenamefont {Morris}, \citenamefont {Desai}, \citenamefont {Viirok}, \citenamefont {Huvonen}, \citenamefont {Nagel}, \citenamefont {Room}, \citenamefont {Krizan}, \citenamefont {Cava}, \citenamefont {McQueen}, \citenamefont {Koohpayeh}, \citenamefont {Kaul},\ and\ \citenamefont {Armitage}}]{Morris_2021}%
  \BibitemOpen
  \bibfield  {author} {\bibinfo {author} {\bibfnamefont {C.~M.}\ \bibnamefont {Morris}}, \bibinfo {author} {\bibfnamefont {N.}~\bibnamefont {Desai}}, \bibinfo {author} {\bibfnamefont {J.}~\bibnamefont {Viirok}}, \bibinfo {author} {\bibfnamefont {D.}~\bibnamefont {Huvonen}}, \bibinfo {author} {\bibfnamefont {U.}~\bibnamefont {Nagel}}, \bibinfo {author} {\bibfnamefont {T.}~\bibnamefont {Room}}, \bibinfo {author} {\bibfnamefont {J.~W.}\ \bibnamefont {Krizan}}, \bibinfo {author} {\bibfnamefont {R.~J.}\ \bibnamefont {Cava}}, \bibinfo {author} {\bibfnamefont {T.~M.}\ \bibnamefont {McQueen}}, \bibinfo {author} {\bibfnamefont {S.~M.}\ \bibnamefont {Koohpayeh}}, \bibinfo {author} {\bibfnamefont {R.~K.}\ \bibnamefont {Kaul}}, \ and\ \bibinfo {author} {\bibfnamefont {N.~P.}\ \bibnamefont {Armitage}},\ }\href {http://dx.doi.org/10.1038/s41567-021-01208-0} {\bibfield  {journal} {\bibinfo  {journal} {Nature Physics}\ }\textbf {\bibinfo {volume} {17}},\ \bibinfo {pages} {832} (\bibinfo {year} {2021})}\BibitemShut {NoStop}%
\bibitem [{\citenamefont {Slim}\ \emph {et~al.}(2024)\citenamefont {Slim}, \citenamefont {Wanjura}, \citenamefont {Brunelli}, \citenamefont {del Pino}, \citenamefont {Nunnenkamp},\ and\ \citenamefont {Verhagen}}]{Slim_2024}%
  \BibitemOpen
  \bibfield  {author} {\bibinfo {author} {\bibfnamefont {J.~J.}\ \bibnamefont {Slim}}, \bibinfo {author} {\bibfnamefont {C.~C.}\ \bibnamefont {Wanjura}}, \bibinfo {author} {\bibfnamefont {M.}~\bibnamefont {Brunelli}}, \bibinfo {author} {\bibfnamefont {J.}~\bibnamefont {del Pino}}, \bibinfo {author} {\bibfnamefont {A.}~\bibnamefont {Nunnenkamp}}, \ and\ \bibinfo {author} {\bibfnamefont {E.}~\bibnamefont {Verhagen}},\ }\href {http://dx.doi.org/10.1038/s41586-024-07174-w} {\bibfield  {journal} {\bibinfo  {journal} {Nature}\ }\textbf {\bibinfo {volume} {627}},\ \bibinfo {pages} {767} (\bibinfo {year} {2024})}\BibitemShut {NoStop}%
\bibitem [{\citenamefont {Mourik}\ \emph {et~al.}(2012)\citenamefont {Mourik}, \citenamefont {Zuo}, \citenamefont {Frolov}, \citenamefont {Plissard}, \citenamefont {Bakkers},\ and\ \citenamefont {Kouwenhoven}}]{doi:10.1126/science.1222360}%
  \BibitemOpen
  \bibfield  {author} {\bibinfo {author} {\bibfnamefont {V.}~\bibnamefont {Mourik}}, \bibinfo {author} {\bibfnamefont {K.}~\bibnamefont {Zuo}}, \bibinfo {author} {\bibfnamefont {S.~M.}\ \bibnamefont {Frolov}}, \bibinfo {author} {\bibfnamefont {S.~R.}\ \bibnamefont {Plissard}}, \bibinfo {author} {\bibfnamefont {E.~P. A.~M.}\ \bibnamefont {Bakkers}}, \ and\ \bibinfo {author} {\bibfnamefont {L.~P.}\ \bibnamefont {Kouwenhoven}},\ }\href {https://www.science.org/doi/abs/10.1126/science.1222360} {\bibfield  {journal} {\bibinfo  {journal} {Science}\ }\textbf {\bibinfo {volume} {336}},\ \bibinfo {pages} {1003} (\bibinfo {year} {2012})}\BibitemShut {NoStop}%
\bibitem [{\citenamefont {Ren}\ \emph {et~al.}(2019)\citenamefont {Ren}, \citenamefont {Pientka}, \citenamefont {Hart}, \citenamefont {Pierce}, \citenamefont {Kosowsky}, \citenamefont {Lunczer}, \citenamefont {Schlereth}, \citenamefont {Scharf}, \citenamefont {Hankiewicz}, \citenamefont {Molenkamp}, \citenamefont {Halperin},\ and\ \citenamefont {Yacoby}}]{ren_topological_2019}%
  \BibitemOpen
  \bibfield  {author} {\bibinfo {author} {\bibfnamefont {H.}~\bibnamefont {Ren}}, \bibinfo {author} {\bibfnamefont {F.}~\bibnamefont {Pientka}}, \bibinfo {author} {\bibfnamefont {S.}~\bibnamefont {Hart}}, \bibinfo {author} {\bibfnamefont {A.~T.}\ \bibnamefont {Pierce}}, \bibinfo {author} {\bibfnamefont {M.}~\bibnamefont {Kosowsky}}, \bibinfo {author} {\bibfnamefont {L.}~\bibnamefont {Lunczer}}, \bibinfo {author} {\bibfnamefont {R.}~\bibnamefont {Schlereth}}, \bibinfo {author} {\bibfnamefont {B.}~\bibnamefont {Scharf}}, \bibinfo {author} {\bibfnamefont {E.~M.}\ \bibnamefont {Hankiewicz}}, \bibinfo {author} {\bibfnamefont {L.~W.}\ \bibnamefont {Molenkamp}}, \bibinfo {author} {\bibfnamefont {B.~I.}\ \bibnamefont {Halperin}}, \ and\ \bibinfo {author} {\bibfnamefont {A.}~\bibnamefont {Yacoby}},\ }\href {https://doi.org/10.1038/s41586-019-1148-9} {\bibfield  {journal} {\bibinfo  {journal} {Nature}\ }\textbf {\bibinfo {volume} {569}},\ \bibinfo {pages} {93} (\bibinfo {year} {2019})}\BibitemShut {NoStop}%
\bibitem [{\citenamefont {Wyss}()}]{wyss2017noncommutative}%
  \BibitemOpen
  \bibfield  {author} {\bibinfo {author} {\bibfnamefont {W.}~\bibnamefont {Wyss}},\ }\href@noop {} {\bibinfo  {journal} {arXiv:1707.03861}\ }\BibitemShut {NoStop}%
\bibitem [{\citenamefont {Am-Shallem}\ \emph {et~al.}()\citenamefont {Am-Shallem}, \citenamefont {Levy}, \citenamefont {Schaefer},\ and\ \citenamefont {Kosloff}}]{doubledHilberSpace}%
  \BibitemOpen
\bibfield  {journal} {  }\bibfield  {author} {\bibinfo {author} {\bibfnamefont {M.}~\bibnamefont {Am-Shallem}}, \bibinfo {author} {\bibfnamefont {A.}~\bibnamefont {Levy}}, \bibinfo {author} {\bibfnamefont {I.}~\bibnamefont {Schaefer}}, \ and\ \bibinfo {author} {\bibfnamefont {R.}~\bibnamefont {Kosloff}},\ }\href@noop {} {\bibinfo  {journal} {arXiv:1510.08634}\ }\BibitemShut {NoStop}%
\bibitem [{\citenamefont {Gelbwaser-Klimovsky}\ and\ \citenamefont {Aspuru-Guzik}(2015)}]{Gelbwaser15}%
  \BibitemOpen
\bibfield  {journal} {  }\bibfield  {author} {\bibinfo {author} {\bibfnamefont {D.}~\bibnamefont {Gelbwaser-Klimovsky}}\ and\ \bibinfo {author} {\bibfnamefont {A.}~\bibnamefont {Aspuru-Guzik}},\ }\href {https://doi.org/10.1021/acs.jpclett.5b01404} {\bibfield  {journal} {\bibinfo  {journal} {The Journal of Physical Chemistry Letters}\ }\textbf {\bibinfo {volume} {6}},\ \bibinfo {pages} {3477} (\bibinfo {year} {2015})}\BibitemShut {NoStop}%
\bibitem [{\citenamefont {Anto-Sztrikacs}\ \emph {et~al.}(2022)\citenamefont {Anto-Sztrikacs}, \citenamefont {Ivander},\ and\ \citenamefont {Segal}}]{Felix22}%
  \BibitemOpen
  \bibfield  {author} {\bibinfo {author} {\bibfnamefont {N.}~\bibnamefont {Anto-Sztrikacs}}, \bibinfo {author} {\bibfnamefont {F.}~\bibnamefont {Ivander}}, \ and\ \bibinfo {author} {\bibfnamefont {D.}~\bibnamefont {Segal}},\ }\href {https://doi.org/10.1063/5.0091133} {\bibfield  {journal} {\bibinfo  {journal} {The Journal of Chemical Physics}\ }\textbf {\bibinfo {volume} {156}},\ \bibinfo {pages} {214107} (\bibinfo {year} {2022})}\BibitemShut {NoStop}%
\bibitem [{\citenamefont {Sch\"onleber}\ \emph {et~al.}(2015)\citenamefont {Sch\"onleber}, \citenamefont {Croy},\ and\ \citenamefont {Eisfeld}}]{PhysRevA.91.052108}%
  \BibitemOpen
  \bibfield  {author} {\bibinfo {author} {\bibfnamefont {D.~W.}\ \bibnamefont {Sch\"onleber}}, \bibinfo {author} {\bibfnamefont {A.}~\bibnamefont {Croy}}, \ and\ \bibinfo {author} {\bibfnamefont {A.}~\bibnamefont {Eisfeld}},\ }\href {https://link.aps.org/doi/10.1103/PhysRevA.91.052108} {\bibfield  {journal} {\bibinfo  {journal} {Phys. Rev. A}\ }\textbf {\bibinfo {volume} {91}},\ \bibinfo {pages} {052108} (\bibinfo {year} {2015})}\BibitemShut {NoStop}%
\bibitem [{\citenamefont {Iles-Smith}\ \emph {et~al.}(2014)\citenamefont {Iles-Smith}, \citenamefont {Lambert},\ and\ \citenamefont {Nazir}}]{PhysRevA.90.032114}%
  \BibitemOpen
  \bibfield  {author} {\bibinfo {author} {\bibfnamefont {J.}~\bibnamefont {Iles-Smith}}, \bibinfo {author} {\bibfnamefont {N.}~\bibnamefont {Lambert}}, \ and\ \bibinfo {author} {\bibfnamefont {A.}~\bibnamefont {Nazir}},\ }\href {https://link.aps.org/doi/10.1103/PhysRevA.90.032114} {\bibfield  {journal} {\bibinfo  {journal} {Phys. Rev. A}\ }\textbf {\bibinfo {volume} {90}},\ \bibinfo {pages} {032114} (\bibinfo {year} {2014})}\BibitemShut {NoStop}%
\bibitem [{\citenamefont {Ivander}\ \emph {et~al.}(2022)\citenamefont {Ivander}, \citenamefont {Anto-Sztrikacs},\ and\ \citenamefont {Segal}}]{FelixNJP}%
  \BibitemOpen
  \bibfield  {author} {\bibinfo {author} {\bibfnamefont {F.}~\bibnamefont {Ivander}}, \bibinfo {author} {\bibfnamefont {N.}~\bibnamefont {Anto-Sztrikacs}}, \ and\ \bibinfo {author} {\bibfnamefont {D.}~\bibnamefont {Segal}},\ }\href {http://dx.doi.org/10.1088/1367-2630/ac9498} {\bibfield  {journal} {\bibinfo  {journal} {New Journal of Physics}\ }\textbf {\bibinfo {volume} {24}},\ \bibinfo {pages} {103010} (\bibinfo {year} {2022})}\BibitemShut {NoStop}%
\bibitem [{\citenamefont {Correa}\ and\ \citenamefont {Glatthard}()}]{correa2024}%
  \BibitemOpen
  \bibfield  {author} {\bibinfo {author} {\bibfnamefont {L.~A.}\ \bibnamefont {Correa}}\ and\ \bibinfo {author} {\bibfnamefont {J.}~\bibnamefont {Glatthard}},\ }\href {https://arxiv.org/abs/2305.08941} {\bibinfo  {journal} {arXiv:2305.08941}\ }\BibitemShut {NoStop}%
\end{thebibliography}%

\end{document}